\pdfoutput=1
\documentclass[cernpreprint,USenglish]{na61doc}

\usepackage{amsmath}% loads amstext, amsbsy, amsopn but not amssymb
\usepackage{enumerate}
\usepackage{placeins}
\usepackage{appendix}
\usepackage[T1]{fontenc}
\usepackage[utf8]{inputenc}
\usepackage{lmodern}
\usepackage{chngcntr}
\usepackage{microtype}
\usepackage{color}
\usepackage{epstopdf}
\usepackage{colortbl}
\usepackage{tabularx}
\usepackage{varwidth}

\usepackage{subcaption}
\usepackage{booktabs}
\usepackage{url}
\usepackage{bbding}
\usepackage{tikz}
\usetikzlibrary{patterns}
\usetikzlibrary{plotmarks}
\usepackage{pgfplots}

\usepackage{hyperref}

\newcommand{\eV}{\ensuremath{\mbox{e\kern-0.1em V}}\xspace}
\newcommand{\GeV}{\ensuremath{\mbox{Ge\kern-0.1em V}}\xspace}
\newcommand{\MeV}{\ensuremath{\mbox{Me\kern-0.1em V}}\xspace}
\newcommand{\GeVc}{\ensuremath{\mbox{Ge\kern-0.1em V}\!/\!c}\xspace}
\newcommand{\GeVcc}{\ensuremath{\mbox{Ge\kern-0.1em V}\!/\!c^2}\xspace}
\newcommand{\AGeV}{\ensuremath{A\,\mbox{Ge\kern-0.1em V}}\xspace}
\newcommand{\AGeVc}{\ensuremath{A\,\mbox{Ge\kern-0.1em V}\!/\!c}\xspace}
\newcommand{\MeVc}{\ensuremath{\mbox{Me\kern-0.1em V}/c}\xspace}

\newcommand{\UrqmdLong}{{\scshape U}r{\scshape qmd1.3.1}\xspace}
\newcommand{\Urqmd}{{\scshape U}r{\scshape qmd}\xspace}
\newcommand{\Hijing}{{\scshape Hijing}\xspace}

\newcommand{\GeantThree}{{\scshape Geant3}\xspace}

\newcommand{\Epos}{{\scshape Epos}\xspace}
\newcommand{\EposLong}{{\scshape Epos1.99}\xspace}

\newcommand{\NASixtyOne}{NA61\slash SHINE\xspace}%this seems to work properly to me. aa
\newcommand{\CernVM}{\textsc{Cern\-\kern-0.05emVM}\xspace}

\newcommand{\pt}{$p_{\text{T}}$}
\newcommand{\pl}{$p_{\text{L}}$}
\newcommand{\y}{$y$}

\newcommand{\piNeg}{$\pi^{-}$}
\newcommand{\plab}{$p_{\text{lab}}$}
\newcommand{\dEdx}{$\text{d}E/\text{d}x$}

\newcommand{\dndptLong}{$\text{d}n/\text{d}p_\text{T}$}
\newcommand{\dndyLong}{$\text{d}n/\text{d}y$}
\newcommand{\dndydpT}{$\frac{\text{d}^{2}n}{\text{d}y\text{d}p_\text{T}}$}

\newcommand{\coordinate}[1]{{\fontfamily{lmss}\selectfont#1}}

\newcommand{\pbar}{\ensuremath{\overline{p}}\xspace}

\newcommand{\pp}{\mbox{\textit{p}+\textit{p}}\xspace}
\newcommand{\NN}{\mbox{\textit{N}+\textit{N}}\xspace}

\definecolor{kBlue+2}{RGB}{0,0,153}
\definecolor{kGray}{RGB}{204,204,204}
\definecolor{kRed+1}{RGB}{204,0,0}
\definecolor{kBlue}{RGB}{0,0,204}
\definecolor{kGreen}{RGB}{0,153,0}
\definecolor{kRed}{RGB}{204,0,0}
\definecolor{kBlack}{RGB}{0,0,0}
\definecolor{kRed+2}{RGB}{153,0,0}
\definecolor{redShading}{RGB}{229,127,127}
\definecolor{kBlue+1}{RGB}{0,0,204}
\definecolor{kYellow+1}{RGB}{204,204,0}
\definecolor{kGreen+2}{RGB}{0,153,0}
\definecolor{kAzure+1}{RGB}{51,153,255}
\definecolor{kBlue+1}{RGB}{127,127,229}

\definecolor{color13}{RGB}{0,0,204}
\definecolor{color19}{RGB}{51,153,255}
\definecolor{color30}{RGB}{0,153,0}
\definecolor{color40}{RGB}{204,204,0}
\definecolor{color75}{RGB}{255,102,0}
\definecolor{color150}{RGB}{204,0,0}
\definecolor{color150Comparison}{RGB}{204,0,0}
\definecolor{color75Comparison}{RGB}{204,204,0}
\definecolor{color40Comparison}{RGB}{0,153,0}
\definecolor{color30Comparison}{RGB}{51,153,255}
\definecolor{color19Comparison}{RGB}{0,0,204}
\definecolor{colorArSc}{RGB}{255,102,51}
\definecolor{colorPP}{RGB}{0,0,204}
\definecolor{colorNN}{RGB}{0,0,204}
\definecolor{colorNNWorld}{RGB}{153,153,255}
\definecolor{colorBeBe}{RGB}{0,153,0}
\definecolor{colorXeLa}{RGB}{204,0,255}
\definecolor{colorPbPb}{RGB}{204,0,0}
\definecolor{colorAuAu}{RGB}{204,0,0}

\newcommand{\dashedLine}{\tikz[baseline=-0.5ex]\draw [thick,dashed] (0,0) -- (0.5,0);}
\newcommand{\looseDashedLine}{\tikz[baseline=-0.5ex]\draw [thick,loosely dashed] (0,0) -- (0.5,0);}
\newcommand{\solidLine}{\tikz[baseline=-0.5ex]\draw [thick] (0,0) -- (0.5,0);}
\newcommand{\dottedLine}{\tikz[baseline=-0.5ex]\draw [thick,dotted] (0,0) -- (0.5,0);}
\newcommand{\markerThirteen}{\CircleSolid}
\newcommand{\markerNineteen}{\TriangleDown}
\newcommand{\markerThirty}{\FourStar}
\newcommand{\markerFourty}{\Plus}
\newcommand{\markerSeventyFife}{\TriangleUp}
\newcommand{\markerOneFifty}{\SquareSolid}
\newcommand{\markerNineteenComparison}{\CircleSolid}
\newcommand{\markerThirtyComparison}{\TriangleDown}
\newcommand{\markerFourtyComparison}{\Plus}
\newcommand{\markerSeventyFifeComparison}{\TriangleUp}
\newcommand{\markerOneFiftyComparison}{\SquareSolid}
\newcommand{\markerArSc}{\TriangleDown}

\newcommand{\markerNN}{\CircleSolid}
\newcommand{\markerPP}{\CircleSolid}
\newcommand{\markerPbPb}{\SquareSolid}
\newcommand{\markerAuAu}{\Square}
\newcommand{\markerNNWorld}{\CircleShadow}
\newcommand{\markerBeBe}{\DiamondSolid}

\ShineTitle{
Spectra and mean multiplicities of \piNeg~in \textit{central} ${}^{40}$Ar+${}^{45}$Sc collisions at 13$A$, 19$A$, 30$A$, 40$A$, 75$A$ and 150\AGeVc beam momenta measured by the \NASixtyOne spectrometer at the CERN SPS
}

\PreprintIdNumber{CERN-EP-2021-010}

\ShineJournal{Eur. Phys. J. C}

\ShineAbstract{The physics goal of the strong interaction program of the \NASixtyOne experiment at the CERN Super Proton Synchrotron (SPS) is to study the phase diagram of hadronic matter by a scan of particle production in collisions of nuclei with various sizes at a set of energies covering the SPS energy range. This paper presents differential inclusive spectra of transverse momentum, transverse mass and rapidity of \piNeg~mesons produced in \textit{central} ${}^{40}$Ar+${}^{45}$Sc collisions at beam momenta of 13$A$, 19$A$, 30$A$, 40$A$, 75$A$ and 150\AGeVc. Energy and system size dependence of parameters of these distributions -- mean transverse mass, the inverse slope parameter of transverse mass spectra, width of the rapidity distribution and mean multiplicity -- are presented and discussed. Furthermore, the dependence of the ratio of the mean number of produced pions to the mean number of wounded nucleons on the collision energy was derived. The results are compared to predictions of several models.}

\begin{document}

\maketitle

%***********************************************************************************

%\linenumbers

\newpage
{\Large The \NASixtyOne Collaboration}
%(the EB ageed to update the author-list! adding N.Abgrall, KEK group, selected T2K beam group members, G.Mills)
\bigskip
%\begin{sloppypar}
% based on XML DB with time Wed Jan 13 16:04:07 2021
% Authors in alphabetical order.

\noindent
A.~Acharya$^{\,9}$,
H.~Adhikary$^{\,9}$,
K.K.~Allison$^{\,25}$,
E.V.~Andronov$^{\,21}$,
T.~Anti\'ci\'c$^{\,3}$,
V.~Babkin$^{\,19}$,
M.~Baszczyk$^{\,13}$,
S.~Bhosale$^{\,10}$,
A.~Blondel$^{\,4}$,
M.~Bogomilov$^{\,2}$
% ...

%\end{sloppypar}
% based on XML DB with time Wed Jan 13 16:04:07 2021
% Institutes in alphabetical order.

\noindent
% ...
$^{2}$~Faculty of Physics, University of Sofia, Sofia, Bulgaria\\
$^{3}$~Ru{\dj}er Bo\v{s}kovi\'c Institute, Zagreb, Croatia\\
$^{4}$~LPNHE, University of Paris VI and VII, Paris, France\\
% ...
$^{9}$~Jan Kochanowski University in Kielce, Poland\\
$^{10}$~Institute of Nuclear Physics, Polish Academy of Sciences, Cracow, Poland\\
% ...
$^{13}$~AGH - University of Science and Technology, Cracow, Poland\\
% ...
$^{19}$~Joint Institute for Nuclear Research, Dubna, Russia\\
% ...
$^{21}$~St. Petersburg State University, St. Petersburg, Russia\\
% ...
$^{25}$~University of Colorado, Boulder, USA\\
% ...

% sections of text

\section{Introduction}

This paper presents measurements of the \NASixtyOne experiment on spectra and mean multiplicities of $\pi^{-}$ mesons produced in central  ${}^{40}$Ar+${}^{45}$Sc collisions at beam momenta of 13$A$, 19$A$, 30$A$, 40$A$, 75$A$ and 150\AGeVc. These studies form part of the strong interactions program of \NASixtyOne~\cite{Abgrall:2014fa,Antoniou:2006mh} investigating the properties of the onset of deconfinement and searching for the possible existence of a critical point in the phase diagram of strongly interacting matter. The program is mainly motivated by the observation of rapid changes of hadron production properties in central Pb+Pb collisions at about 30\AGeVc by the NA49 experiment~\cite{Afanasiev:2002mx,Alt:2007aa}, namely a sharp peak in the Kaon to pion ratio ("horn"), the start of a plateau in the inverse slope parameter for Kaons ("step"), and a steepening of the increase of pion production per wounded nucleon with increasing collision energy ("kink"). These findings were predicted as signals of the onset of deconfinement~\cite{Gazdzicki:1998vd}. They were recently confirmed by the RHIC beam energy program~\cite{Adamczyk:2017iwn} and the interpretation is supported by the LHC results (see Ref.~\cite{Rustamov:2012np} and references therein). 

The goals of the \NASixtyOne strong interaction program are pursued experimentally by a two dimensional scan in collision energy and nuclear mass number of colliding nuclei.  This allows to explore systematically the phase diagram of strongly interacting matter~\cite{Antoniou:2006mh}. In particular, the analysis of the existing data within the framework of statistical models suggests that by increasing collision energy one increases temperature and decreases baryon chemical potential  of strongly interacting matter at freeze-out~\cite{Becattini:2005xt}, whereas by increasing nuclear mass number of the colliding nuclei one decreases the temperature~\cite{Alt:2007uj,Becattini:2005xt,Vovchenko:2015idt}.

Within this program \NASixtyOne recorded data on \pp, Be+Be, Ar+Sc, Xe+La and Pb+Pb collisions. Further high statistics measurements of Pb+Pb collisions are planned with an upgraded detector starting in 2022~\cite{PbAddendum}. Results on $\pi^-$ spectra and multiplicities have already been published from \pp interactions~\cite{Abgrall:2013pp_pim,Aduszkiewicz:2017sei} and $^7$Be+$^9$Be collisions~\cite{Acharya:2020cyb, Acharya:2020hpt}. The latter provide the basic reference of a light isospin zero system for the study of dense matter effects in collisions of heavier nuclei.

In this paper the so-called $h^{-}$ method is used for determining $\pi^{-}$ production since it provides the largest phase space coverage. This procedure utilizes the fact that negatively charged particles are predominantly $\pi^{-}$ mesons with a small admixture (of order 10\%) of $K^-$ mesons and anti-protons which can be reliably subtracted.

The paper is organized as follows: after this introduction the experiment is briefly described in Sec.~2. The analysis procedure is discussed in Sec.~3. Section~4 presents the results of the analysis. In Sec.~5 the new measurements are compared to model calculations. The relevance of the new results for the study of the onset of deconfinement is discussed in Sec.~6. A summary and outlook in Sec.~7 closes the paper.

The following variables and definitions are used in this paper. The particle rapidity $y$ is calculated in the nucleon-nucleon collision center of mass system (cms), $y=0.5\ln{[(E+cp_{\text{L}})/(E-cp_{\text{L}})]}$, where $E$ and \pl~are the particle energy and longitudinal momentum, respectively. The transverse component of the momentum is denoted as \pt,~and the transverse mass $m_{\text{T}}$ is defined as $m_{\text{T}} = \sqrt{m^2 + (cp_{\text{T}})^2}$, where $m$ is the particle mass in GeV. The momentum in the laboratory frame is denoted \plab~and the collision energy per nucleon pair in the center of mass by $\sqrt{s_{NN}}$.

The Ar+Sc collisions are selected by requiring a low value of the forward energy - the energy emitted into the region  populated by projectile spectators. These collisions are referred to as \textit{central} collisions and a selection of collisions based on the forward energy is called a \textit{centrality} selection. Although for Ar+Sc collisions the forward energy is not tightly correlated with the impact parameter of the collision, the terms \textit{central} and \textit{centrality} are adopted following the convention widely used in heavy-ion physics.

\clearpage

\section{\NASixtyOne detector}

\begin{figure*}[h]
  \centering
    \includegraphics[width=\textwidth]{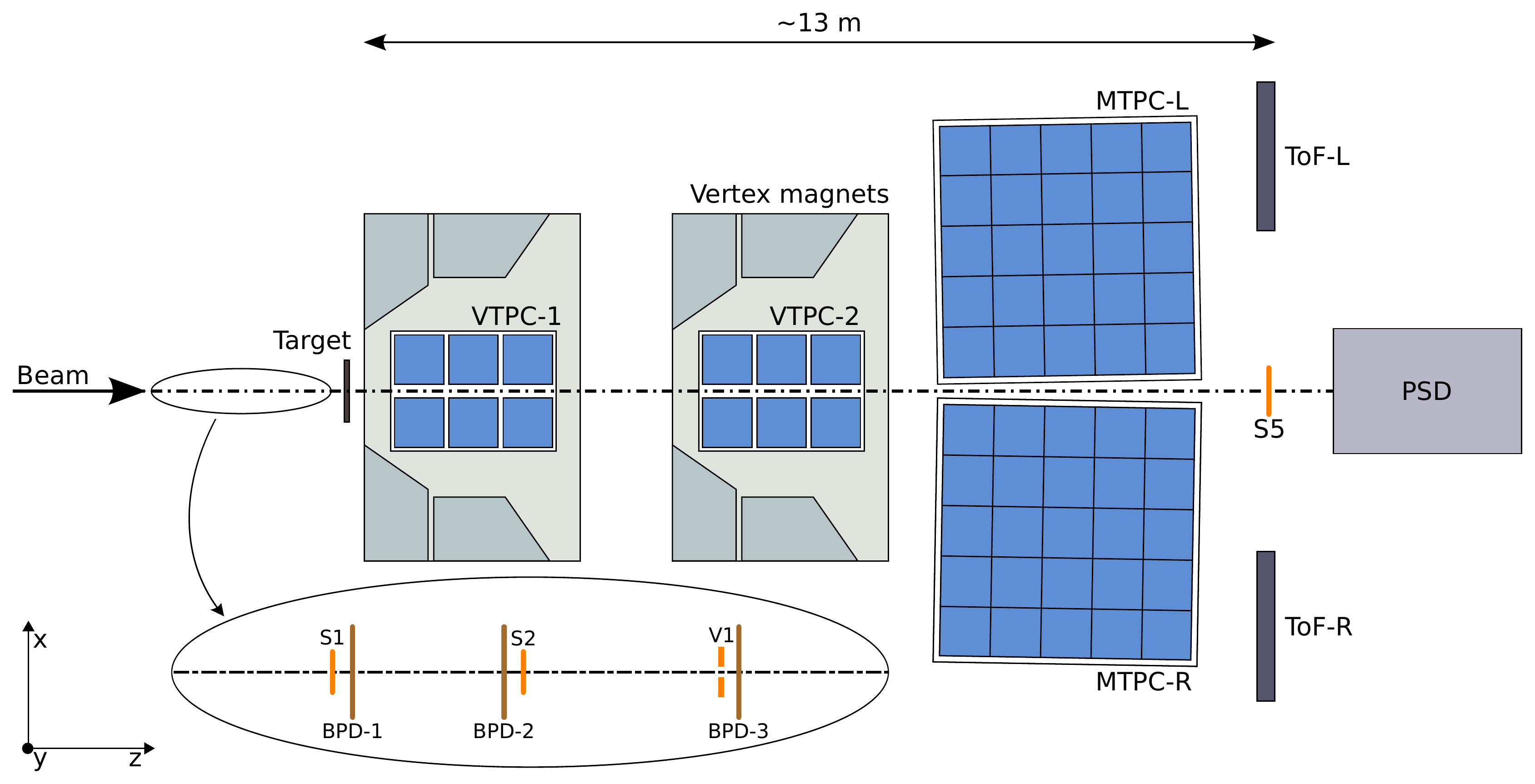}
  \caption{The schematic layout of the \NASixtyOne experiment at the CERN SPS ~\cite{Abgrall:2014fa} showing the components used for the Ar+Sc energy scan (horizontal cut, not to scale). The trigger detector configuration upstream of the target is shown in the inset. Alignment of the chosen coordinate system is shown on the plot; its origin lies in the middle of VTPC-2, on the beam axis. The nominal beam direction is along the \coordinate{z}-axis. The magnetic field bends charged particle trajectories in the \coordinate{x}--\coordinate{z} (horizontal) plane. The drift direction in the TPCs is along the  (vertical) \coordinate{y}-axis.}
  \label{fig:setup}
\end{figure*}

The \NASixtyOne detector (see Fig.~\ref{fig:setup}) is a large-acceptance hadron spectrometer situated in the North Area H2 beam-line of the CERN SPS~\cite{Abgrall:2014fa}. The main components of the detection system used in the analysis are four large volume Time Projection Chambers (TPC). Two of them, called Vertex TPCs (VTPC), are located downstream of the target inside superconducting magnets with maximum combined bending power of 9~Tm. The magnetic field was scaled down in proportion to the beam momentum in order to obtain similar phase space acceptance at all energies. The main TPCs (MTPC) and two walls of pixel Time-of-Flight (ToF-L/R) detectors are placed symmetrically on either side of the beamline downstream of the magnets. The TPCs are filled with Ar:CO$_{2}$ gas mixtures in proportions 90:10 for the VTPCs and 95:5 for the MTPCs. The Projectile Spectator Detector (PSD) is positioned 20.5 m (16.7 m) downstream of the MTPCs at beam momenta of 75$A$ and 150\AGeVc (13$A$, 19$A$, 30$A$, 40\AGeVc), centered in the transverse plane on the deflected position of the beam. Moreover a brass cylinder of 10 cm (30$A$ -- 150\AGeVc) or 5 cm (19\textit{A} GeV/c) length and 5 cm diameter (degrader) was placed in front of the center of the PSD in order to reduce electronic saturation effects and shower leakage from the downstream side.

Primary beams of fully ionized ${}^{40}$Ar nuclei were extracted from the SPS accelerator at beam momenta of 13$A$, 19$A$, 30$A$, 40$A$, 75$A$ and 150\AGeVc. Two scintillation counters, S1 and S2, provide beam definition, together with a veto counter V1 with a 1~cm diameter hole, which defines the beam before the target. The S1 counter provides also the timing reference (start time for all counters). Beam particles are selected by the trigger system requiring the coincidence T1 = $\textrm{S1} \wedge\textrm{S2} \wedge\overline{\textrm{V1}}$. Individual beam particle trajectories are precisely measured by the three beam position detectors (BPDs) placed upstream of the target~\cite{Abgrall:2014fa}. Collimators in the beam line were adjusted to obtain beam rates of $\approx 10^4$/s during the $\approx$ 10 s spill and a cycle time of 32.4 s.

The target was a stack of 2.5 x 2.5~cm$^2$ area and 1 mm thick $^{45}$Sc plates of 6~mm total thickness placed $\approx$~80~cm upstream of VTPC-1. Impurities due to other isotopes and elements were measured to be 0.3~\%. Their influence on the pion multiplicity was estimated to be an increase by less than 0.2~\% caused by the admixture of heavier elements~\cite{Banas:2018sak}. No correction was applied for this negligible contamination. Data were taken with target inserted (denoted~I) and target removed (denoted~R). 

Interactions in the target are selected with the trigger system by requiring an incoming $^{40}$Ar ion and a signal below that of beam ions from S5, a small 2~cm diameter scintillation counter placed on the beam trajectory behind the MTPCs. This minimum bias trigger is based on the breakup of the beam ion due to interactions in and downstream of the target. In addition, \textit{central} collisions were selected by requiring an energy signal below a set threshold from the 16 central modules of the PSD which measure mainly the energy carried by projectile spectators. The cut was set to retain only the events with the $\approx$~30\% smallest energies in the PSD. The event trigger condition thus was T2 = T1$\wedge\overline{\textrm{S5}}\wedge\overline{\textrm{PSD}}$. The statistics of recorded events are summarized in Table~\ref{tab:statbeam}.

\begin{table}[!htbp]
	\caption{Basic beam properties and number of events recorded and used in the analysis of ${}^{40}$Ar+${}^{45}$Sc  interactions at incident momenta of 13$A$, 19$A$, 30$A$, 40$A$, 75$A$ and 150\AGeVc.}
	\vspace{0.5cm}
	\centering
	\begin{tabular}{c | c | c | c | c}
  	$p_{beam}$ (\GeVc) & $\sqrt{s_{NN}}$ (\GeV) & \parbox[][1.5cm][c]{2cm}{\centering Recorded event triggers} & \parbox[][1.5cm][c]{2cm}{\centering Number of selected events} & \parbox[][1.5cm][c]{3.5cm}{\centering Fraction of background events after selection cuts} \\
    \hline
    13  & 5.1 & $3.0\cdot10^{6}$ & $243637$ & $1.26\cdot10^{-3}$ \\
    19  & 6.1 & $3.7\cdot10^{6}$ & $250249$ & $1.46\cdot10^{-3}$ \\
    30  & 7.6 & $4.8\cdot10^{6}$ & $431816$ & $1.11\cdot10^{-3}$ \\
    40  & 8.8 & $8.9\cdot10^{6}$ & $634001$ & $1.39\cdot10^{-3}$ \\
    75  & 11.9 & $4.4\cdot10^{6}$ & $556047$ & $4.44\cdot10^{-4}$ \\
    150 & 16.8 & $0.99\cdot10^{6}$ & $133953$ & $1.08\cdot10^{-4}$
  	\end{tabular}
	\label{tab:statbeam}
\end{table}
\clearpage

\section{Analysis procedure}

This section starts with a brief overview of the data analysis procedure and the applied corrections. It also defines to which class of particles the final results correspond. A description of the calibration and the track and vertex reconstruction procedure can be found in Ref.~\cite{Abgrall:2013pp_pim}.

The analysis procedure consists of the following steps:
\begin{enumerate}[(i)]
  \item application of event and track selection criteria,
  \item determination of spectra of negatively charged hadrons using the selected events and tracks,
  \item evaluation of corrections to the spectra based on experimental data and simulations,
  \item calculation of the corrected spectra and its parameters,
  \item calculation of statistical and systematic uncertainties.
\end{enumerate}

Corrections for the following biases were evaluated and applied when significant:
\begin{enumerate}[(i)]
 \item contribution from off-target interactions,
 \item procedure of selecting \textit{central} collisions,
 \item geometrical acceptance,
 \item contribution of particles other than \emph{primary} (see below) negatively charged pions produced in Ar+Sc interactions,
 \item losses of produced negatively charged pions due to their decays and secondary interactions.
\end{enumerate}

Correction (i) was found to be negligible and was therefore not applied. Corrections (ii)-(v) are estimated by simulations. Events were generated with the \EposLong model (version CRMC 1.5.3)~\cite{Werner:2005jf, Pierog:2009zt, Pierog:2018}, passed through detector simulation employing the \GeantThree package~\cite{Geant3} and then reconstructed by the standard \NASixtyOne program chain. Event selection in the simulation was based on the number of projectile spectator nucleons which is available from the \EposLong model.

The final results refer to $\pi^{-}$ produced in \textit{central} Ar+Sc interactions by strong interaction processes and in electromagnetic decays of produced hadrons. Such hadrons are referred to as \emph{primary} hadrons. The definition of \textit{central} collisions is given in Sec.~\ref{sec:centrality}.

\subsection{\textit{Central} collisions}
\label{sec:centrality}

A short description of the procedure defining \textit{central} collisions is given below. For more details see Ref.~\cite{h-bebe:2020}.

Final results presented in this paper refer to the 5\% of Ar+Sc collisions with the lowest value of the forward energy $E_\text{F}$ (\textit{central} collisions). The quantity $E_\text{F}$ is defined as the total energy in the laboratory system of all particles produced in Ar+Sc collisions via strong and electromagnetic processes in the forward momentum region defined by the acceptance map in Ref.~\cite{PSD_acceptance}. Final results on \textit{central} collisions, derived using this procedure, allow a precise comparison with predictions of models without any additional information about the \NASixtyOne setup and used magnetic field. Using this definition the mean number of wounded nucleons $\langle W \rangle$ was calculated within the Wounded Nucleon Model~(WNM)~\cite{Bialas:1976ed} as implemented in \Epos.

\begin{figure}[!htbp]
       \centering 
       \begin{minipage}[m]{0.45\textwidth}
       	\centering 
       \includegraphics[width=0.6\textwidth]{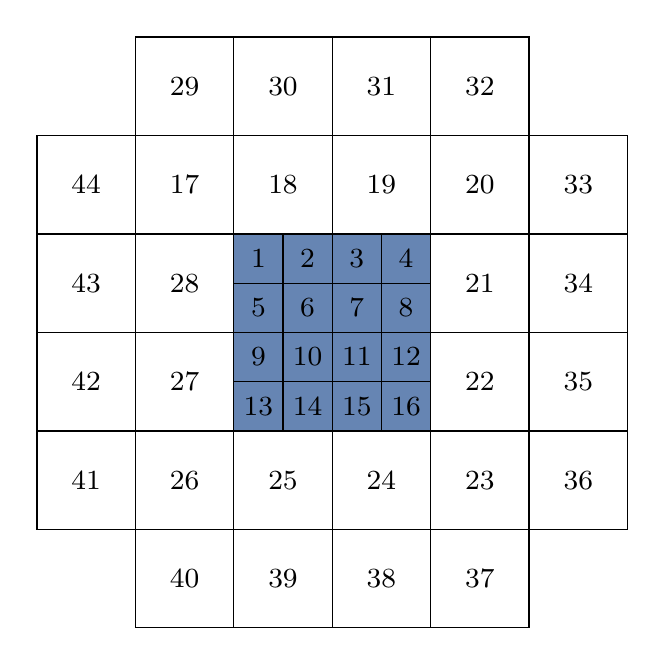}\\
       T2 trigger
   		\end{minipage}
       \begin{minipage}[m]{0.45\textwidth}
       	\centering 
       \includegraphics[width=0.6\textwidth]{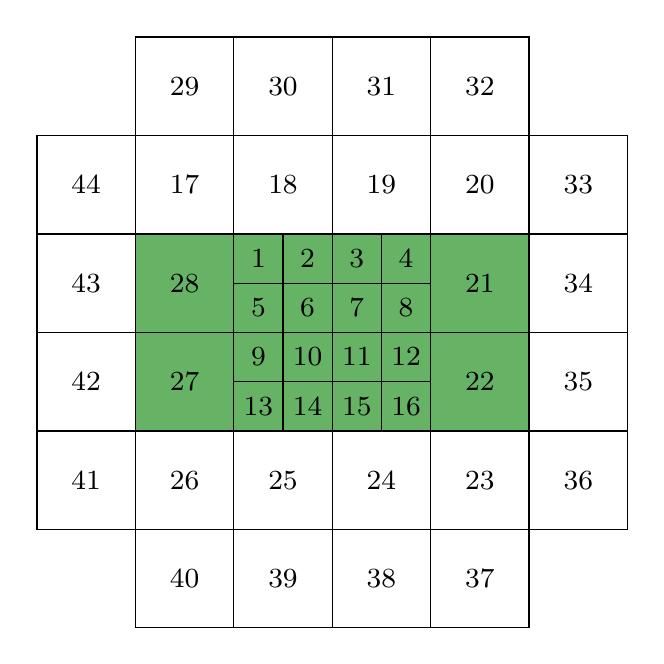}\\
       150\AGeVc
   		\end{minipage}\\\vspace{0.5cm}
       \begin{minipage}[m]{0.45\textwidth}
       	\centering 
       \includegraphics[width=0.6\textwidth]{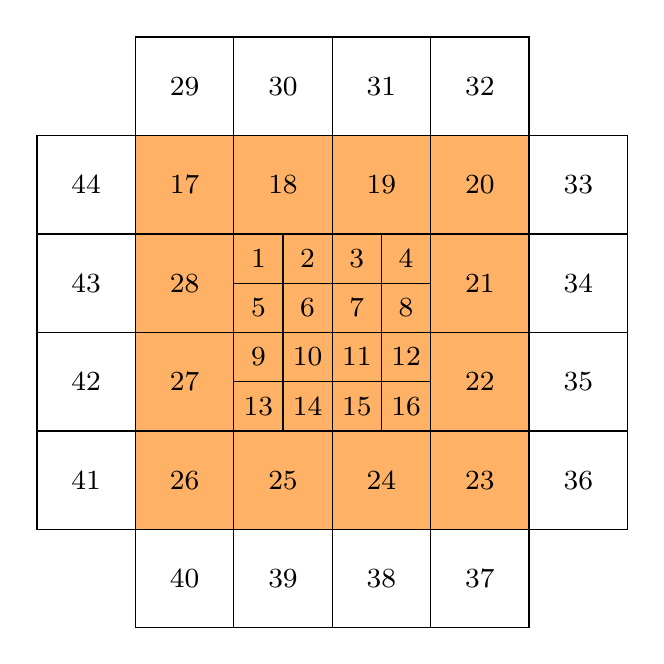}\\
       75\textit{A}, 40\textit{A}, 30\AGeVc
   		\end{minipage}
       \begin{minipage}[m]{0.45\textwidth}
       	\centering 
       \includegraphics[width=0.6\textwidth]{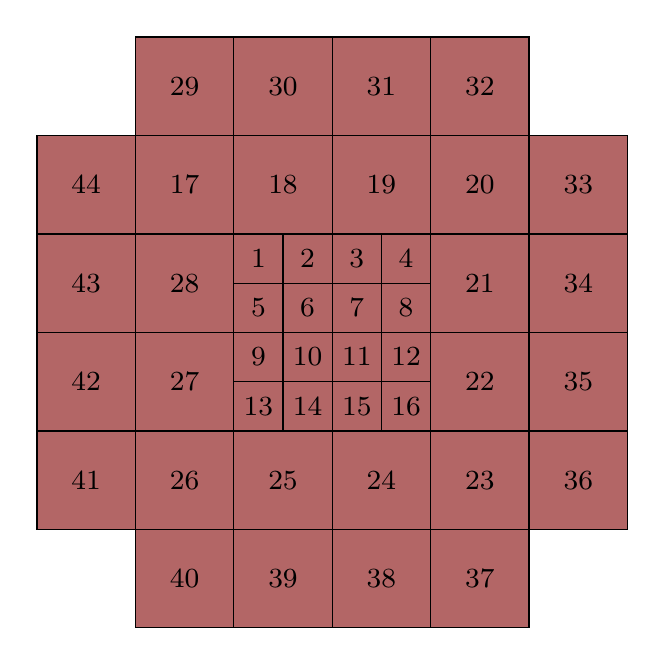}\\
       19\textit{A}, 13\AGeVc
       \end{minipage}
       \caption{Schematic diagrams indicating by shading the PSD modules used in the online and offline event selection. The trigger is derived from the energy in the central 16 modules (1-16) in blue color. Determination of the PSD energy $E_\text{PSD}$ uses the green (150\AGeVc), orange (75\textit{A}, 40\textit{A}, 30\AGeVc) or all modules (19\textit{A}, 13\AGeVc) at the respective beam momenta.}
\label{fig:PSDAllModuleSelections}
\end{figure}

For analysis of the data the event selection was based on the $\approx$~5\% of collisions with the lowest value of the energy $E_\text{PSD}$ measured by a subset of PSD modules (see Fig.~\ref{fig:PSDAllModuleSelections}) in order to optimize the sensitivity to projectile spectators. The acceptance resulting from the definition of the forward energy $E_\text{F}$ corresponds closely to the acceptance of this subset of PSD modules. 

\begin{figure}[!htbp]
	\centering
	\includegraphics[width=0.48\textwidth]{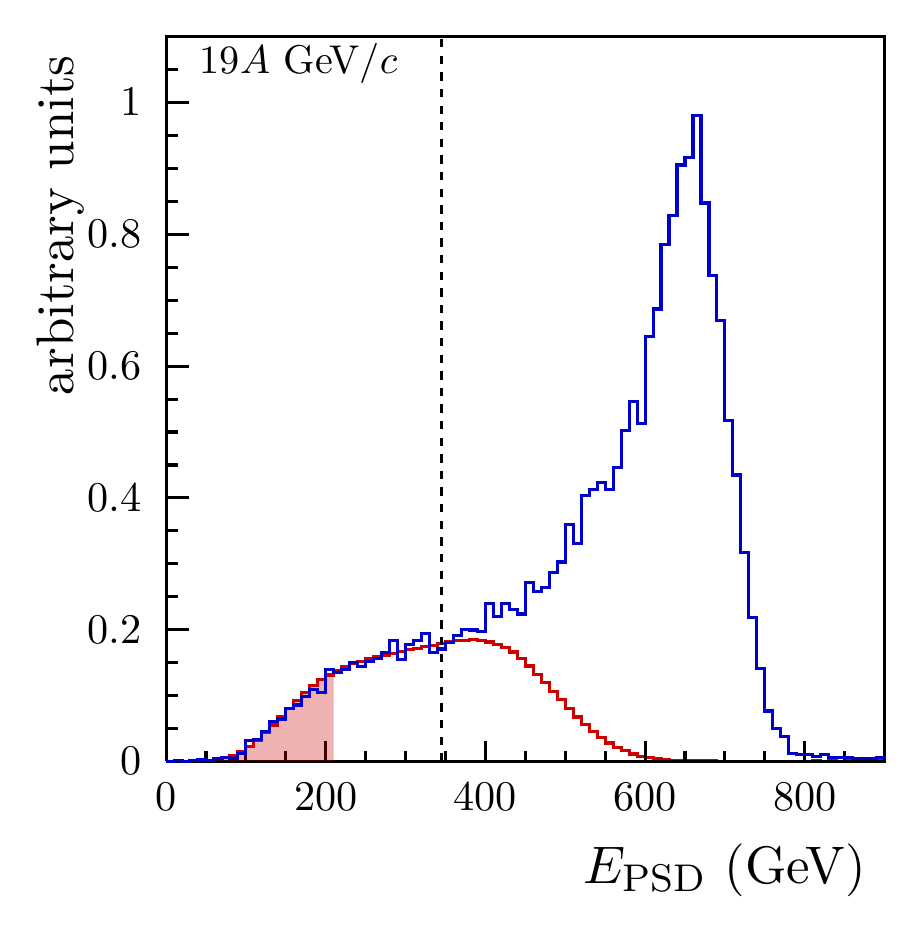}
	\includegraphics[width=0.48\textwidth]{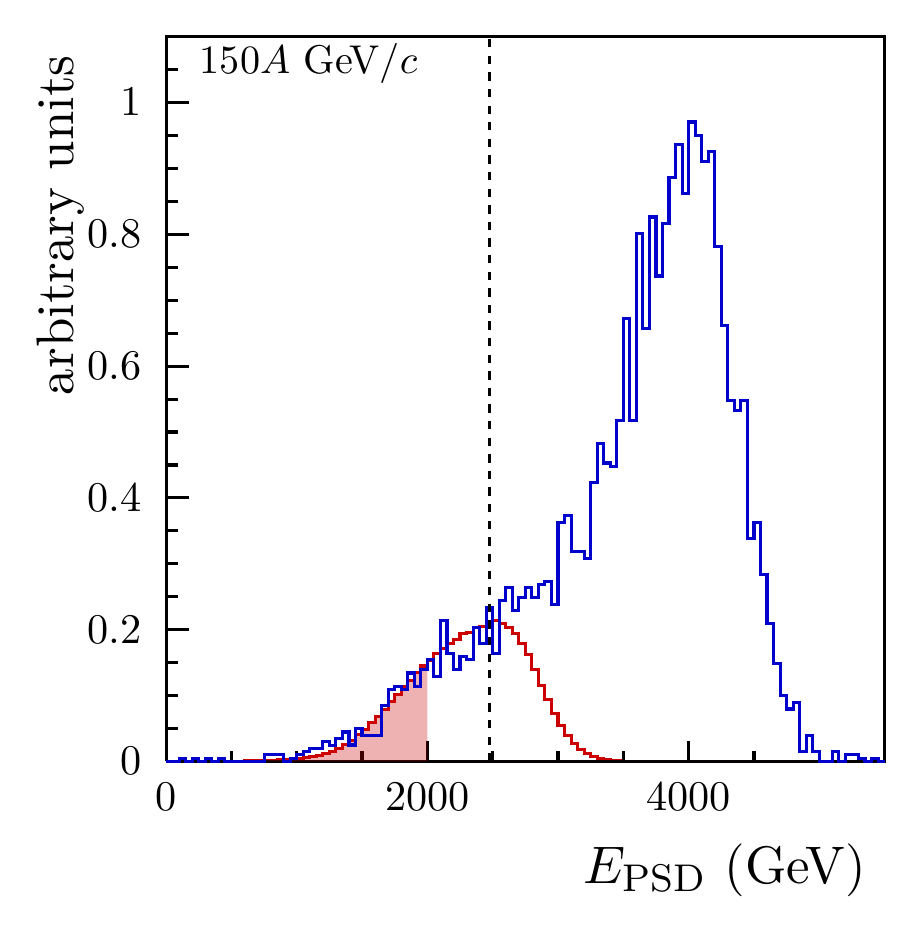}
	{\textcolor{kBlue+2}\solidLine} T1 trigger \hspace{.5cm}\textcolor{kRed+1}{\solidLine} T2 trigger (scaled) \hspace{.5cm} \textcolor{redShading}{\tiny\SquareSolid} 5\% centrality \hspace{.5cm} {\textcolor{kBlack}\dashedLine} Normalization region\\
	\caption{Event centrality selection using the energy $E_\text{PSD}$ measured by the PSD calorimeter. Distributions are shown of measured $E_\text{PSD}$ for minimum-bias selected (blue histograms) and T2 selected (red histograms) events for 19$A$~(left plot) and 150\AGeVc~(right plot) beam momenta. Histograms are normalized to agree in the overlap region (from the beginning of the distribution to the black dashed line). The T2 trigger was set to accept $\approx$~30\% of the inelastic cross section. The accepted region corresponding to the 5\% collisions with the smallest $E_\text{PSD}$ is indicated by shading.}
\label{fig:PSDEnergy_cent}
\end{figure}

Online event selection by the hardware trigger (T2) used a threshold on the sum of electronic signals from the 16 central modules of the PSD set to accept $\approx$~30\% of the inelastic interactions. Measured distributions of $E_\text{PSD}$ for minimum-bias and T2 trigger selected events, calculated in the offline analysis, are shown in Fig.~\ref{fig:PSDEnergy_cent} at beam momenta of 19\AGeVc and 150\AGeVc, respectively. The accepted region corresponding to the 5\% most \textit{central} collisions is indicated by shading. The minimum-bias distribution was obtained using the data from the beam trigger T1 with offline selection of events by requiring an event vertex in the target region. A properly normalized spectrum for target removed events was subtracted.

The forward energy $E_\text{F}$ cannot be measured directly. However, both $E_\text{F}$ and $E_\text{PSD}$ can be obtained from simulations using the \EposLong (version CRMC 1.5.3)~\cite{Werner:2005jf, Pierog:2009zt, Pierog:2018} model. A global factor $c_\text{cent}$ (listed in Table~\ref{tab:w}) was then calculated as the ratio of mean negatively charged pion multiplicities obtained with the two selection procedures in the 5\% most \textit{central} events. The resulting factors $c_\text{cent}$ range from 1.002 to 1.005 and correspond to only a small correction compared to the systematic uncertainties of the measured particle multiplicities. A possible dependence of the scaling factor on rapidity and transverse momentum was neglected. The corrections ($c_\text{cent}$) are negligibly small compared to the systematic uncertainties of the measured particle multiplicities and are therefore not applied in the calculation of $\pi^-$ yields and neglected in the quoted systematic uncertainties.

\begin{figure}[!htbp]
  \centering
     \includegraphics[width=0.49\textwidth]{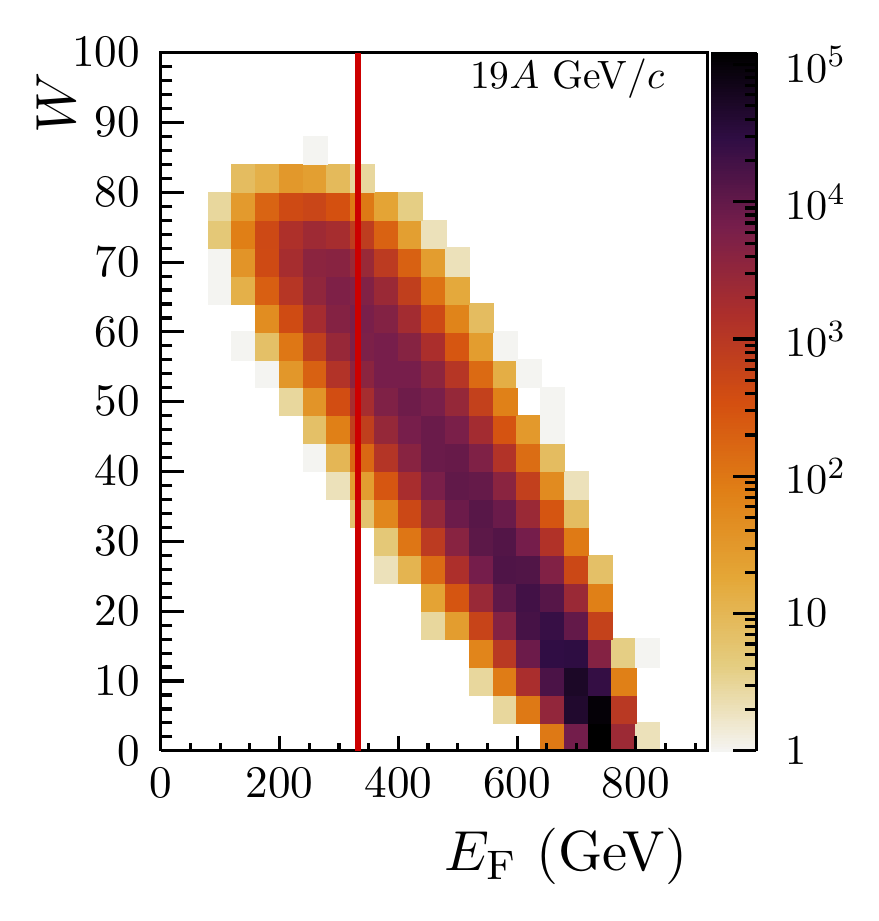}
     \includegraphics[width=0.49\textwidth]{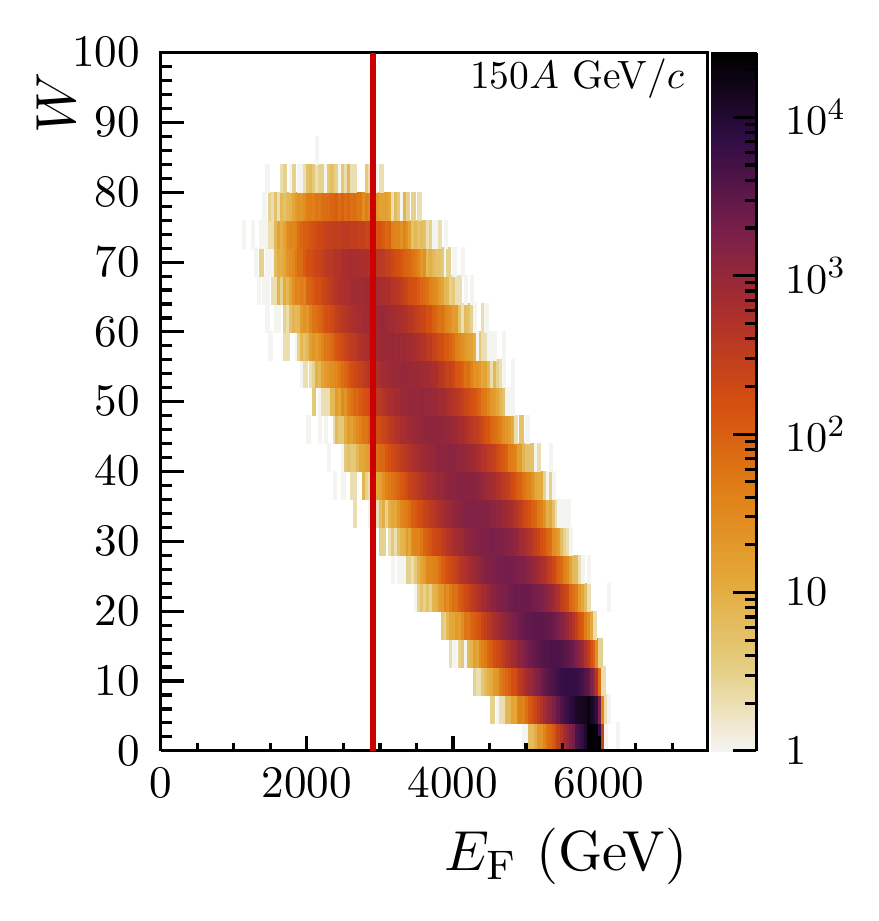}\\
  \caption{Distributions of $W$ versus $E_{F}$ for all inelastic collisions at 19$A$ (\textit{left}) and 150\AGeVc (\textit{right}) beam momenta calculated from the \EposLong model. The vertical red lines show the value of the cut on $E_\text{F}$ for selecting the 5\% most \textit{central} collisions.}
  \label{fig:woundedDistribution}
\end{figure}

Finally, events generated with the \Epos code with its implementation of the Wounded Nucleon Model~\cite{Pierog:2018} were used to estimate the average number of wounded nucleons $\langle W \rangle$ for the 5\% of events with the smallest number of spectator nucleons and with the smallest value of $E_\text{F}$. For the latter selection the average impact parameter $\langle b \rangle$ was obtained as well. Results are listed in Table~\ref{tab:w}. Example distributions of events in the $W-E_\text{F}$ plane for 19$A$ and 150~\AGeVc beam momenta are shown in Fig.~\ref{fig:woundedDistribution}. These distributions are quite broad and emphasize the importance of proper simulation of the \textit{centrality} selection when comparing model calculations with the experimental results. For comparison $\langle W \rangle$ was also calculated from the GLISSANDO model which uses a different implementation of the Wounded Nucleon Model~\cite{Broniowski:2007nz}. The resulting pion multiplicities, also listed in Table~\ref{tab:w}, differ by about 5\%. This uncertainty is not shown in the plots of the results.

\begin{table*}
 \caption{Average number of wounded nucleons $\langle W \rangle$ in the 5\% most \textit{central} Ar+Sc collisions estimated from simulations using the \Epos~\cite{Werner:2005jf, Pierog:2009zt, Pierog:2018} and GLISSANDO~\cite{Broniowski:2007nz} models. In the \Epos~WNM case the average impact parameter $\langle b\rangle$ is presented as well. The values of $\sigma$ denote the widths of the distributions of $W$ and $b$. Results from \Epos~WNM and Glissando are for \textit{centrality} selection using the smallest number of spectators, whereas the \Epos~$E_\text{F}$ results are obtained using the forward energy $E_\text{F}$ within the acceptance map in Ref.~\cite{PSD_acceptance}. The last line presents numerical values of the $c_\text{cent}$ factor.}
 \vspace{0.5cm}
 \centering
 \footnotesize
 \begin{tabular}{l|ccccccc}
  Momentum (\AGeVc) & & 13 & 19 & 30 & 40 & 75 & 150\\ 
  \hline
  \Epos WNM & $\langle W \rangle$ & $68.0$ & $68.0$ & $67.9$ & $68.0$ & $68.0$ & $68.1$\\
  & $\sigma$ & $3.7$ & $3.7$ & $3.7$ & $3.8$ & $3.7$ & $3.8$ \\
  GLISSANDO & $\langle W \rangle$ & $67.9$ & $68.2$ & $68.3$ & $68.5$ & $68.7$ & $69.1$\\
  & $\sigma$ & $4.9$ & $4.8$ & $4.8$ & $4.8$ & $4.8$ & $4.6$ \\
  \Epos $E_\text{F}$ & $\langle W \rangle$ & $65.7$ & $65.4$ & $65.1$ & $65.0$ & $65.0$ & $65.0$\\
  & $\sigma$ & $6.0$ & $6.2$ & $6.4$ & $6.5$ & $6.6$ & $6.7$ \\
  & $\langle b \rangle$ & $1.82$ & $1.95$ & $2.00$ & $2.09$ & $2.23$ & $2.08$\\
  & $\sigma$ & $0.79$ & $0.84$ & $0.86$ & $0.89$ & $0.94$ & $0.81$ \\
 \hline
  & $c_{cent}$ & $1.005$ & $1.005$ & $1.002$ & $1.003$ & $1.005$ & $1.002$ \\
 \end{tabular}
 \label{tab:w}
\end{table*}

\subsection{Event and track selection}\label{sec:cuts}

\subsubsection{Event selection}

\begin{figure}[!htbp]
\centering
    \includegraphics[width=0.8\textwidth]{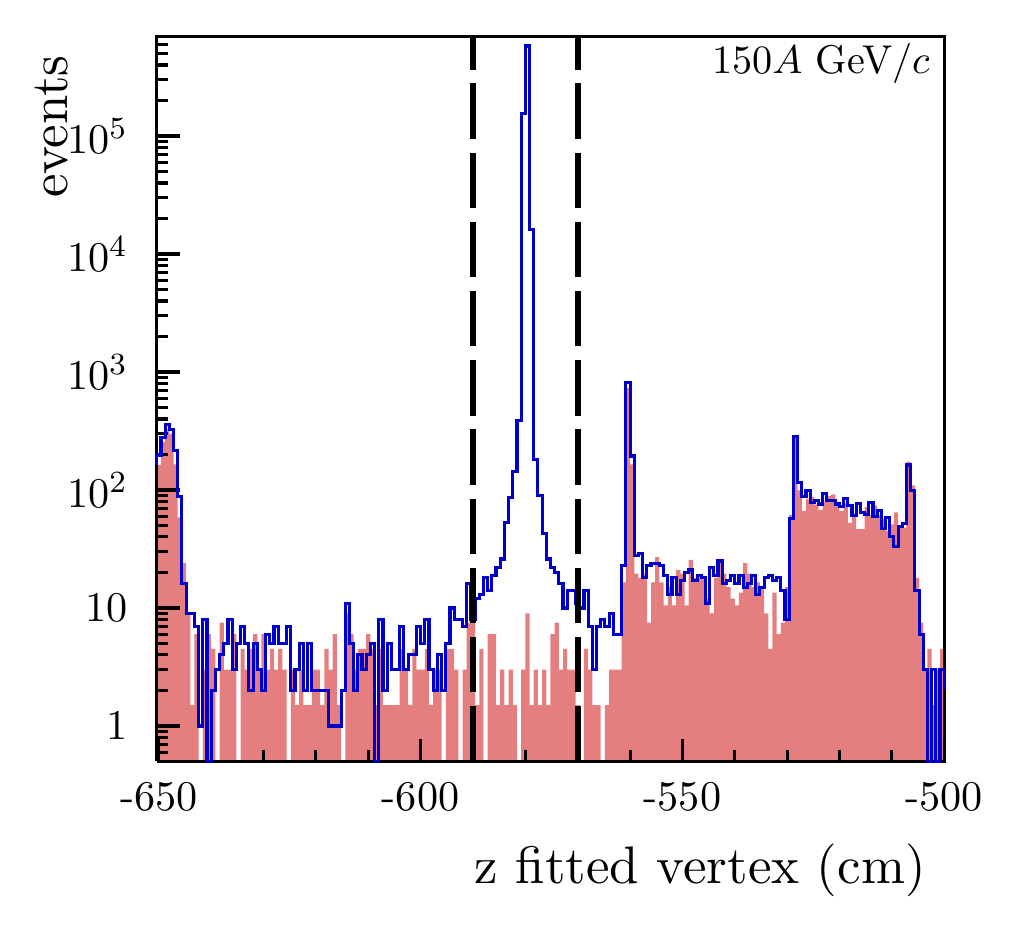}\\
	\textcolor{kBlue+1}{\solidLine} Target inserted \hspace{.5cm}\textcolor{redShading}{\tiny\SquareSolid} Target removed$\times 1.50123$ \hspace{.5cm} \textcolor{kBlack}{\dashedLine} Acceptance region
	\caption{Distribution of fitted vertex \coordinate{z} coordinate for T1 triggered events of Ar+Sc interactions at 150\AGeVc with target inserted and target removed (shaded histogram). The distribution for the data recorded with the Sc target removed was divided by a factor of $N_\text{I}/N_\text{R}$, where $N_\text{I}$ and $N_\text{R}$ are the numbers of events with Sc target inserted and removed, respectively. Vertical dashed lines show the acceptance region.}
	\label{fig:zVertex}
\end{figure}

Central Ar+Sc events were selected using the following criteria:
\begin{enumerate}[(i)]

    \item no offtime beam particle detected within a time window of $\pm$4$~\mu$s around the trigger particle,
    \item beam particle trajectory measured in at least three planes out of four of BPD-1 and BPD-2 and in both planes of BPD-3,
    \item a well reconstructed interaction vertex with \coordinate{z} position (fitted using the beam trajectory and TPC tracks) not farther away than 10~cm from the center of the Sc target (see Fig.~\ref{fig:zVertex}),
    \item an upper cut on the energy $E_\text{PSD}$ in order to select the 5~\% collisions with the lowest $E_\text{PSD}$.
\end{enumerate}

The reconstructed vertex distribution for target-removed events satisfying the above event cuts is shown in Fig.~\ref{fig:zVertex} by the shaded histogram. The latter was normalized to the same integral number of incident beams as used for the target-inserted data. One finds that within the acceptance cut region as indicated by vertical dashed lines in Fig.~\ref{fig:zVertex} the fraction of background events is smaller than $2\cdot10^{-3}$ and was therefore neglected.

The event statistics after applying the selection criteria are summarized in Table~\ref{tab:statbeam}.

\subsubsection{Track selection}

\begin{figure}[!htbp]
\centering{
    \includegraphics[width=0.45\textwidth]{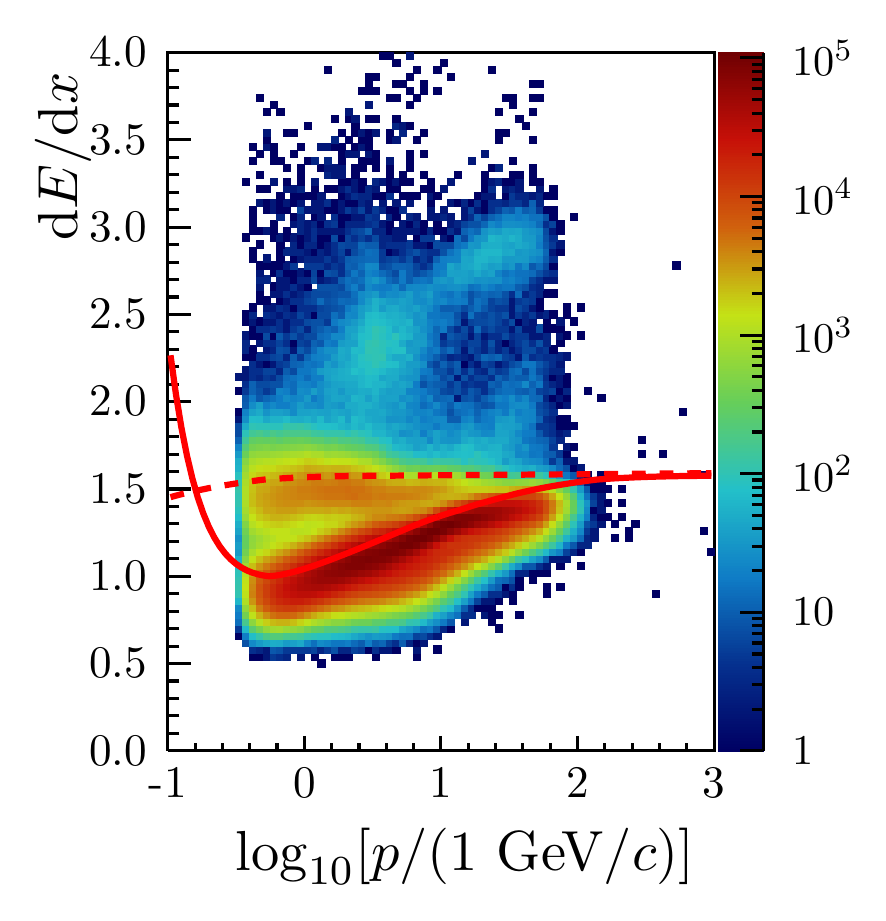}
    \includegraphics[width=0.45\textwidth]{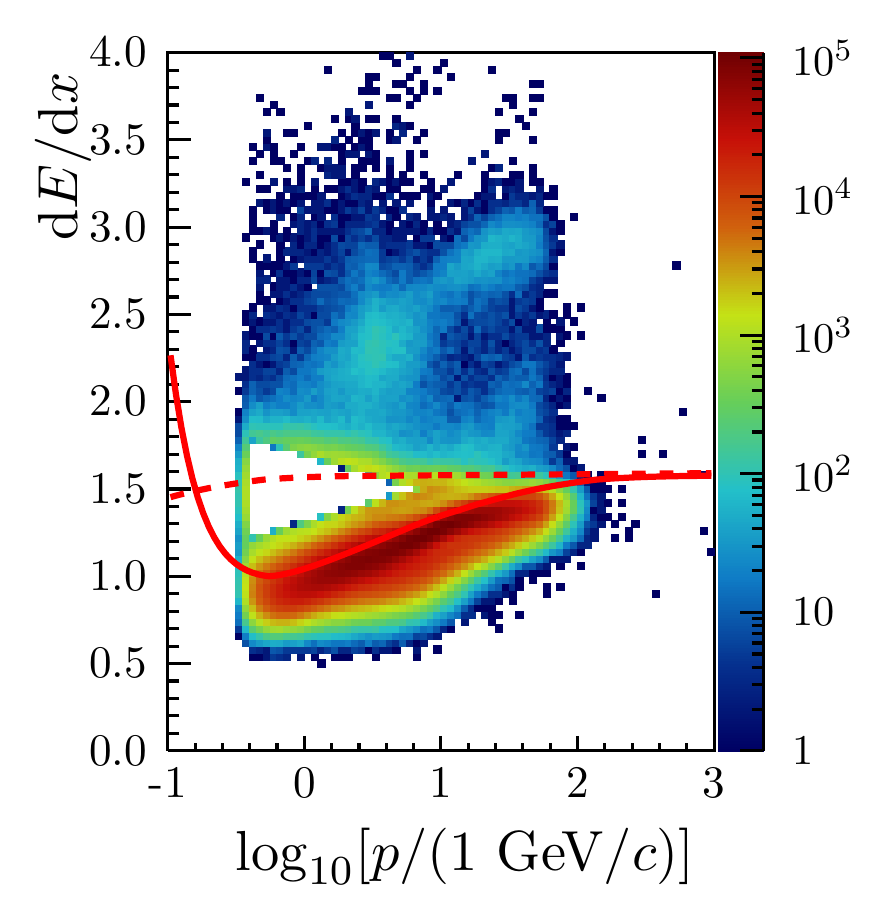}\\
    \textcolor{kRed}{\dashedLine}~e$^{-}$ \hspace{.5cm} \textcolor{kRed}{\solidLine}~\piNeg}
\caption{2D histograms of specific energy loss d$E$/dx versus momentum in Ar+Sc interactions at 150\AGeVc before (\textit{left}) and after (\textit{right}) electron exclusion. The Bethe-Bloch functions for electrons and negatively charged pions are plotted by dashed and solid lines, respectively.}
\label{fig:electronCut}
\end{figure}

In order to select tracks of primary charged hadrons and to reduce the contamination of tracks from secondary interactions and weak decays, the following track selection criteria were applied:

\begin {enumerate}[(i)]
    \item track momentum fit at the interaction vertex should have converged,
    \item fitted \coordinate{x} component of track momentum is negative. This selection minimizes the angle between the track trajectory and the TPC pad direction for the chosen magnetic field direction, reducing uncertainties of the reconstructed cluster position, energy deposition and track parameters,
    \item total number of reconstructed points on the track should be greater than~30,
    \item sum of the number of reconstructed points in VTPC-1 and VTPC-2 should be greater than~15,
    \item the distance between the track extrapolated to the interaction plane and the interaction point (impact parameter) should be smaller than 4~cm in the horizontal (bending) plane and 2~cm in the vertical (drift) plane,
    \item electron tracks were excluded by a cut on the particle energy loss d$E$/dx in the TPCs (see Fig.~\ref{fig:electronCut}).
\end {enumerate}

The analysis was performed in $(y,p_\text{T})$ and $(y,m_\text{T}-m_{\pi})$ bins. The bin sizes were selected taking into account the statistical uncertainties and the resolution of the momentum reconstruction~\cite{Abgrall:2013pp_pim}. Corrections as well as statistical and systematic uncertainties were calculated for each bin.

\subsection{Corrections}
\label{sec:corrections}

\begin{figure}[!htbp]
    \centering
    \includegraphics[width=0.45\textwidth]{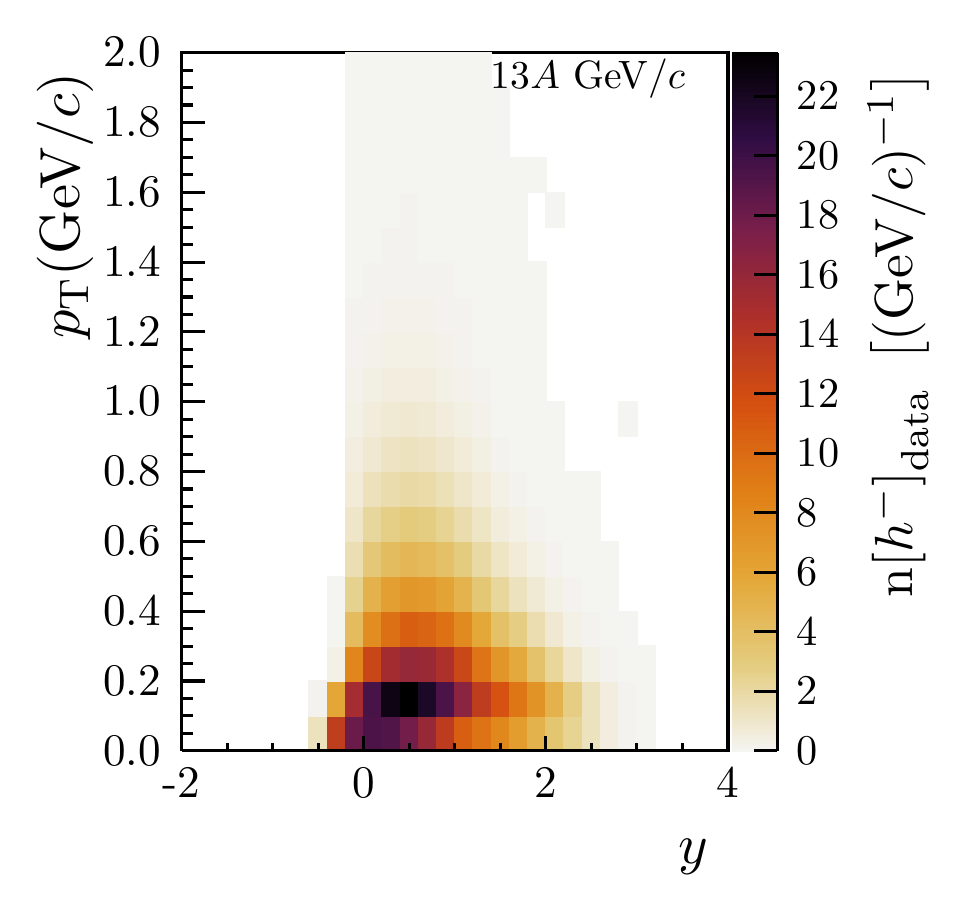}
    \includegraphics[width=0.45\textwidth]{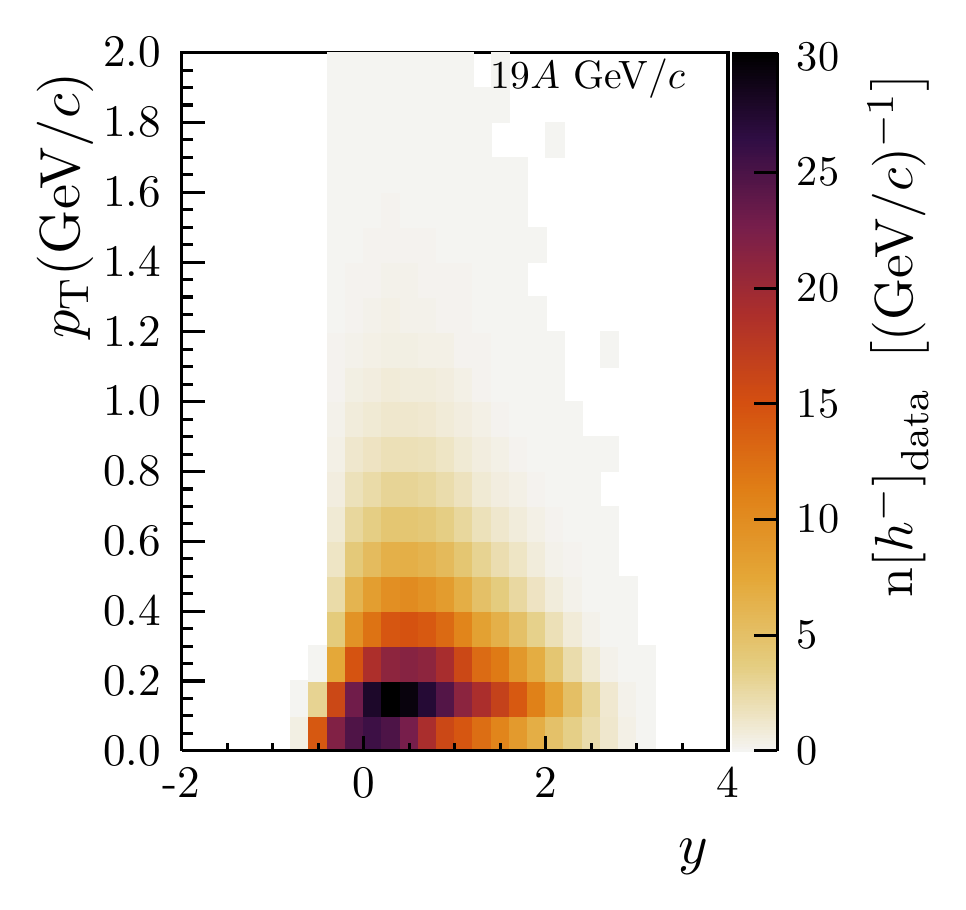}\\
    \includegraphics[width=0.45\textwidth]{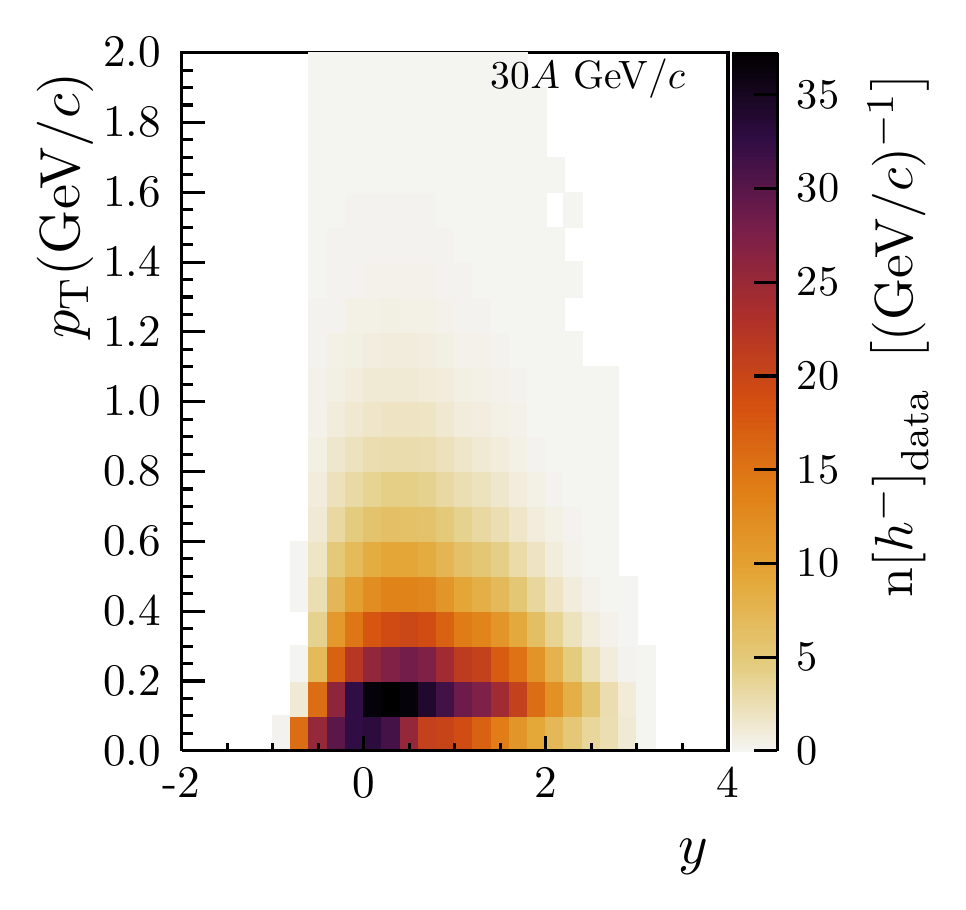}
    \includegraphics[width=0.45\textwidth]{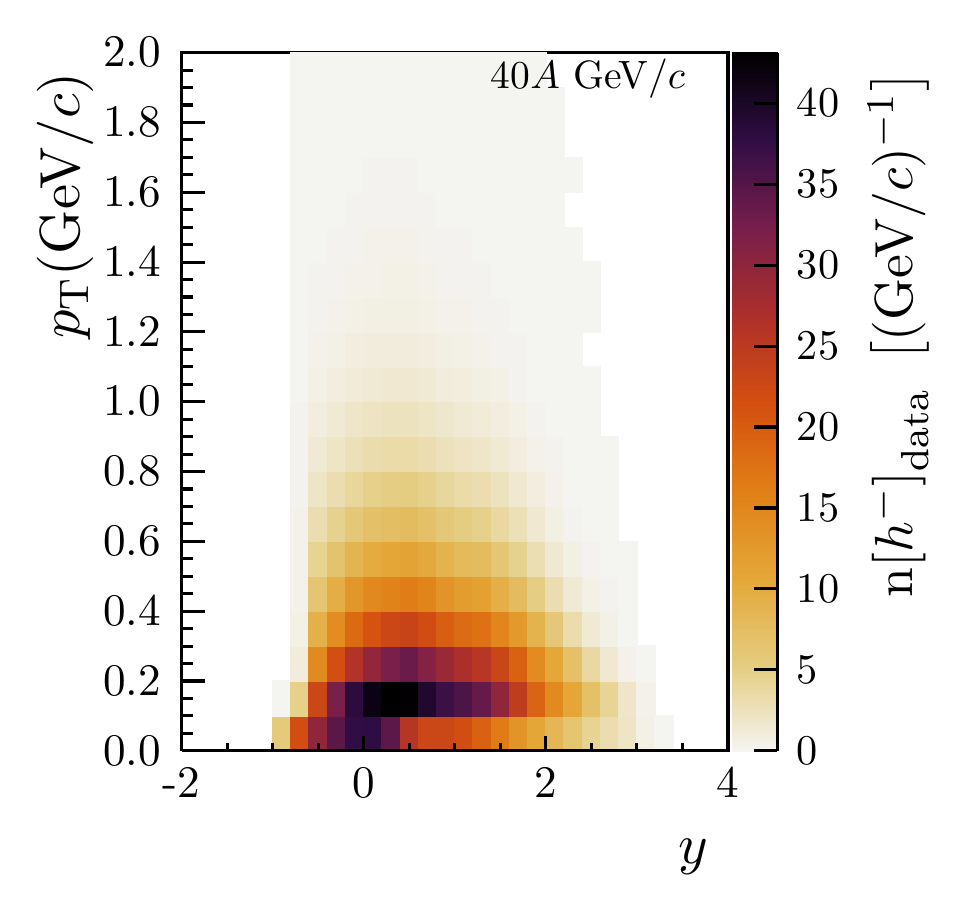}\\
    \includegraphics[width=0.45\textwidth]{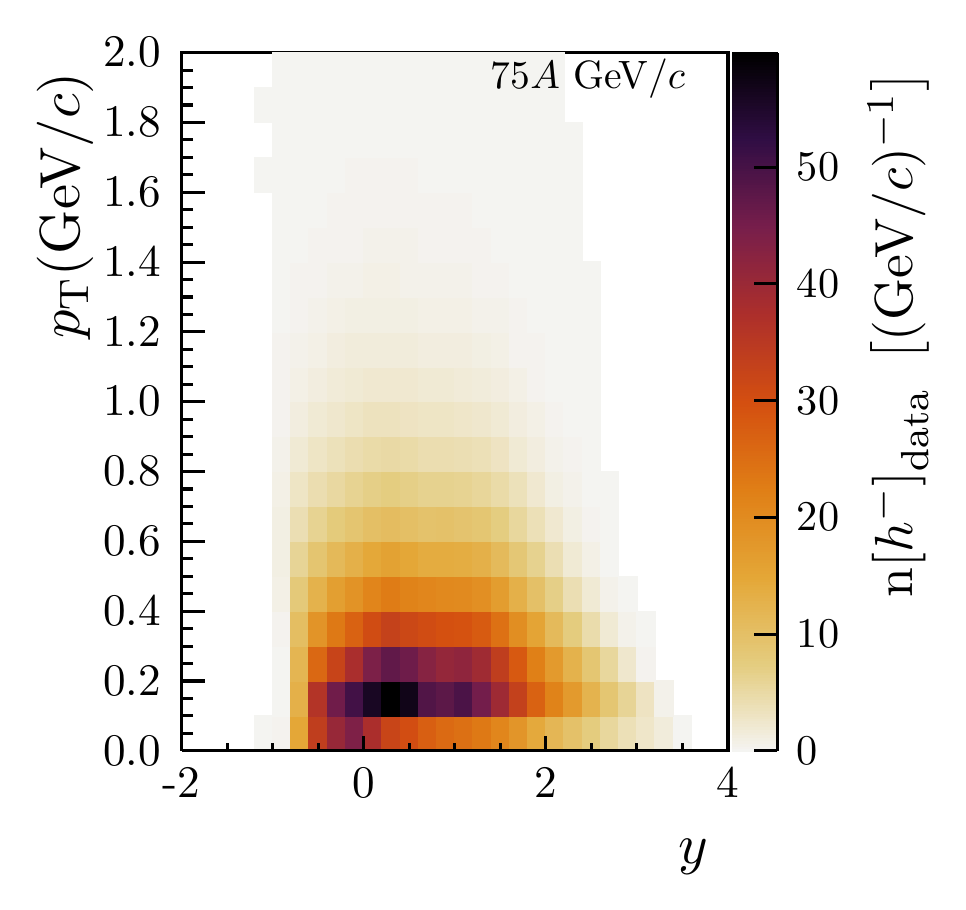}
    \includegraphics[width=0.45\textwidth]{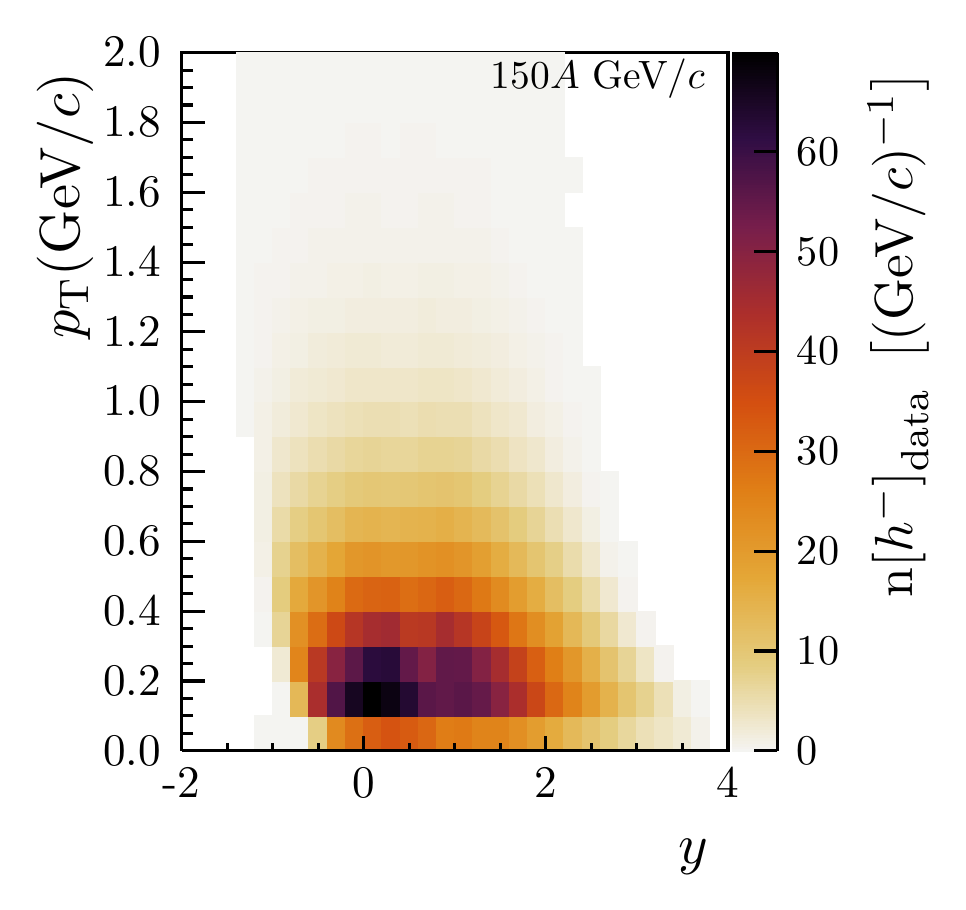}\\
    \caption{Uncorrected double-differential spectra $n[h^-]^\text{raw}/\Delta y /\Delta p_\text{T}$ of negatively charged hadrons produced in the 5\%  Ar+Sc collisions with the smallest $E_{PSD}$ energy at beam momenta of 13$A$, 19$A$, 30$A$, 40$A$, 75$A$ and 150\AGeVc.}
    \label{fig:data_dndydpt}
\end{figure}

Uncorrected yields of negatively charged hadrons per event after all event and track cuts $n[h^-]^\text{raw}$ divided by bin dimensions are shown in Fig.~\ref{fig:data_dndydpt}. In order to determine the mean multiplicity of \textit{primary} $\pi^-$~mesons produced in \textit{central} Ar+Sc collisions a set of corrections was applied to the extracted raw yields. The main biasing effects are detector acceptance, loss of events due to the cut on reconstructed vertex position, track selection cuts, reconstruction efficiency, contributions of particles from weak decays (feed-down), and contribution of \textit{primary} hadrons other than negatively charged pions (mostly $K^-$ mesons). Contamination from events occurring outside the target was negligible.

A simulation of the \NASixtyOne detector is used to correct the data for acceptance, reconstruction efficiency, feed-down and contamination from re-interactions of produced particles. Only Ar+Sc interactions in the target material were simulated and reconstructed. The \Epos model~\cite{Werner:2005jf, Pierog:2009zt, Pierog:2018} was selected to generate the primary interactions. A \GeantThree based program chain was used to track particles through the spectrometer, generate decays and secondary interactions and simulate the detector response (for more detail see Ref.~\cite{Abgrall:2013pp_pim}). Simulated events were then reconstructed using the standard \NASixtyOne reconstruction chain and reconstructed tracks were matched to the simulated particles based on the cluster positions. Hadrons which were not produced in the primary interaction can amount to a significant fraction of the selected track sample. Thus a careful effort was undertaken to evaluate and subtract this contribution. 

Since \Epos provides only an approximate description of the measurements of particle production in Ar+Sc collisions, a data-based effort was made {to improve the estimate of contamination from $\pi^-$ wrongly accepted as coming from the \textit{primary} interaction. Yields of misidentified Kaons were estimated from preliminary results of \NASixtyOne (see Ref.~\cite{Gazdzicki:2692088}) on $K^-$ production and the contribution of $\pi^-$~from decays of hyperons was estimated from published results of other experiments (see Ref.~\cite{Naskret:2020}). The relative effect of such tuning of the yields of negatively charged hadrons was below 5\% for majority of the bins and did not exceed 7\% for all the beam momenta.}

Backward rapidity bins with relative statistical uncertainties exceeding 20\% in case of the higher beam momenta (150$A$ and 75\AGeVc) and 30\% in case of the lower beam momenta (40$A$, 30$A$, 19$A$ and 13\AGeVc) were not used since they suffer from limited backward rapidity acceptance of the detector.

The correction factor $c_{yp_\text{T}}$, based on the event and detector simulation was calculated for each \y~and \pt~bin as:
\begin{equation}
	c_{yp_\text{T}} =  n[\pi^-]^\mathrm{MC}_\mathrm{gen}~/~n[h^-]^\mathrm{MC}_\mathrm{sel}    ,
	\label{eq:correctiongeo}
\end{equation}
where $ n[h^-]^\mathrm{MC}_\mathrm{sel} $ is the mean multiplicity of reconstructed  negatively charged particles after the event and track selection criteria and $n[\pi^-]^\mathrm{MC}_\mathrm{gen} $ is the mean multiplicity of \textit{primary} negatively charged pions from the \textit{centrality} selected Ar+Sc collisions generated by the \Epos model.

The corrected multiplicities were then calculated as:
\begin{equation}
	n[\pi^-]^\text{corr} = c_{yp_\text{T}} \cdot n[h^-]^\text{raw}.
	\label{eq:corrdndydpt}
\end{equation}
Double differential distributions \dndydpT~of per event multiplicities are then given by:
\begin{equation}
	\frac{\text{d}^{2}n}{\text{d}y\text{d}p_\text{T}}  =  \frac{1}{\Delta y  \cdot \Delta p_{\text{T}}} n[\pi^-]^\text{corr},
	\label{eq:dndydpt}
\end{equation}
where $n[\pi^-]^{corr}$ are the corrected per event multiplicities for $\pi^-$ in the ($y$, \pt) bins with size $\Delta y$ and $\Delta p_{\text{T}}$. The distributions $\frac{\text{d}^{2}n}{\text{d}y\text{d}m_{\text{T}}}$ were calculated with an analogous formula.

\subsection{Statistical uncertainties}

Statistical uncertainties of the yields receive contributions from the finite statistics of both the data and the correction factors derived from the simulations. The contribution from the statistical uncertainty of the data is much larger than that from the correction factors $c_{yp_\text{T}}$ which was therefore neglected. The statistical uncertainty of the data was calculated assuming a Poisson probability distribution for the number of entries in each \y, \pt~bin. 

\subsection{Systematic uncertainties}

Systematic uncertainties presented in this paper were calculated taking into account contributions from the following effects:
\begin{enumerate}[(i)]
    \item Possible biases which were not corrected for. These are:
    \begin{enumerate}
        \item a possible bias due to the \dEdx~cut applied to reject electron tracks,
        \item a possible bias due to the removal of events with off-time beam particles close in time to the trigger particle.
    \end{enumerate}
    Their magnitude was estimated by varying the values of the corresponding cut. The values of the selected \dEdx~band around the Bethe-Bloch function was changed by $\pm 0.01$~\dEdx~units (where 1 corresponds to a minimum ionizing particle, and 0.04 is a typical width of the \dEdx~distribution for \piNeg), and the rejection time window was changed to $\pm 3~\mu$s and $\pm 5~\mu$s. The systematic uncertainty was estimated as half of the maximum absolute difference between $h^{-}$ multiplicities when varying the cut values.
    \item Uncertainty of the correction for the track selection cuts used for data and Monte Carlo data selection were estimated by removing the impact parameter cut and varying the minimum number of required points by $\pm 3$. The observed changes suggest the potential bias is around 1\%.
    \item Uncertainty of the correction for contamination of the primary \piNeg~mesons by daughters of decays and re-interactions. It was estimated from simulations using the \Epos model where the production rates of parents were adjusted to extrapolations of published data (see Ref.~\cite{Naskret:2020}). The systematic uncertainty was estimated as $15\%$ of the correction value.
    \item Uncertainty of correction for the contamination of particles other than \piNeg~in negatively charged hadrons $h^{-}$ spectrum. The value of the uncertainty was assumed as 15\% of the simulated contribution of $K^-$, $\Sigma^{-}$ and \pbar to the total number of negatively charged hadrons.
\end{enumerate}

Values of $\sigma_\text{sys}$ are listed in the table \ref{tab:piMultiplicity}. The total systematic uncertainty was calculated by adding in quadrature the individual contributions. Note that systematic biases in different bins are correlated, whereas statistical fluctuations are independent.

\begin{figure*}[!htbp]
  \centering
        \centering
   \includegraphics[width=0.4\textwidth]{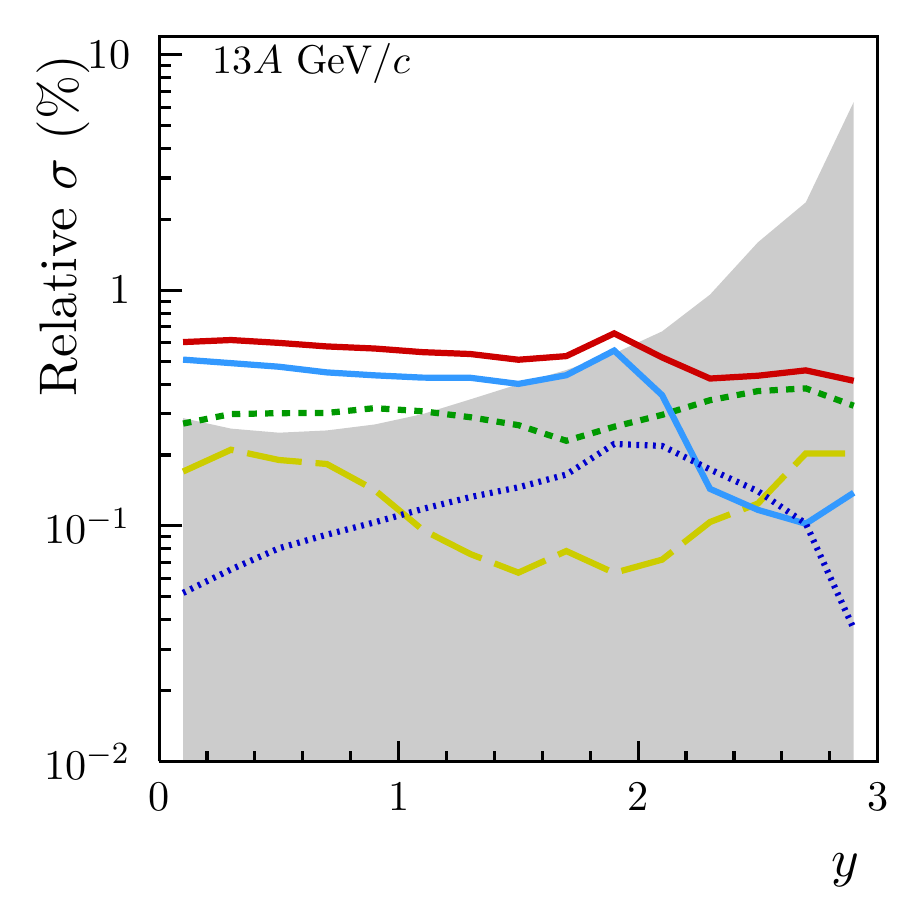}
   \includegraphics[width=0.4\textwidth]{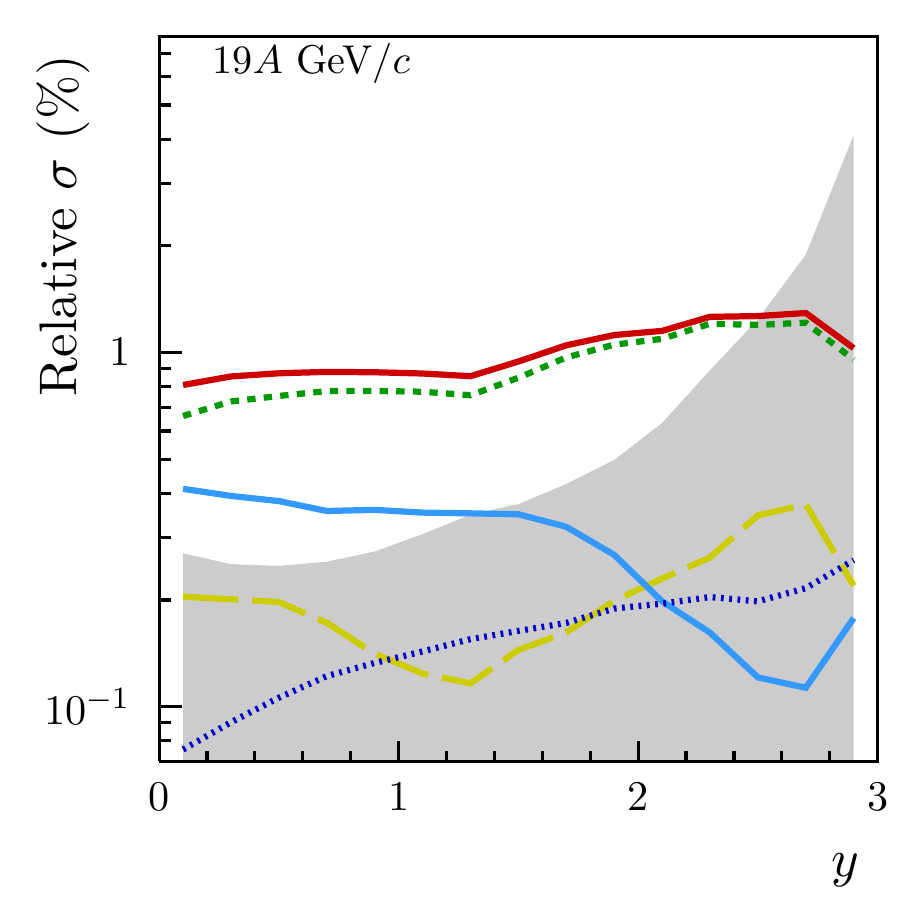}\\
   \includegraphics[width=0.4\textwidth]{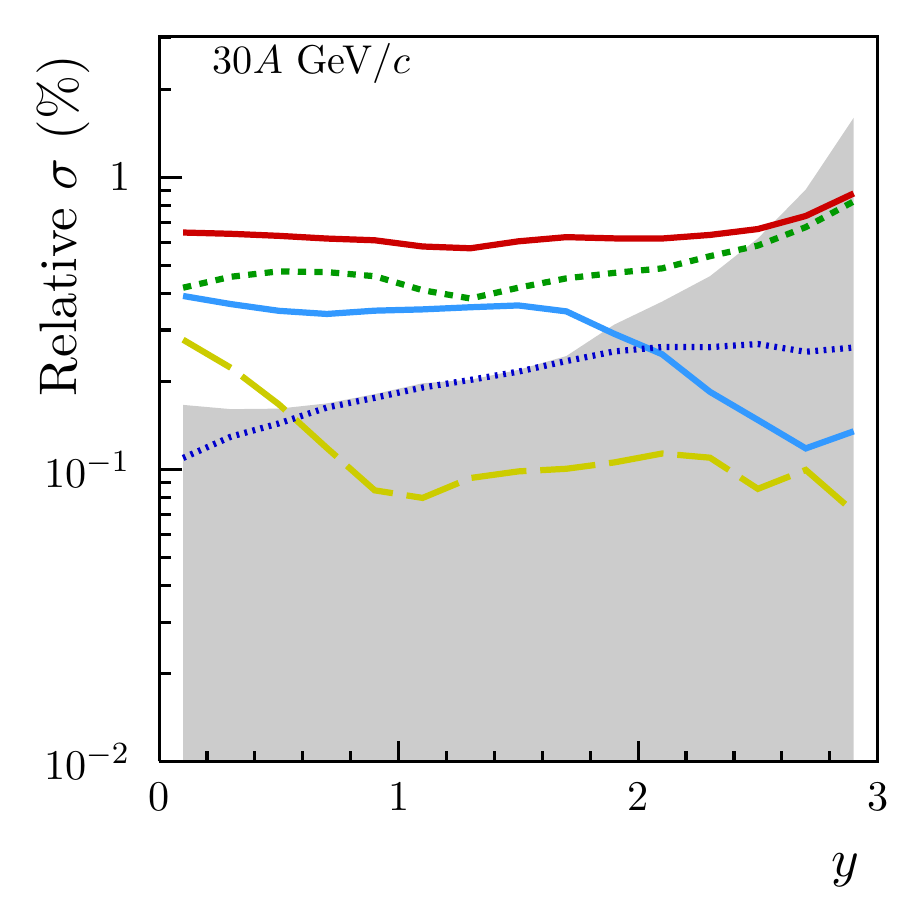}
   \includegraphics[width=0.4\textwidth]{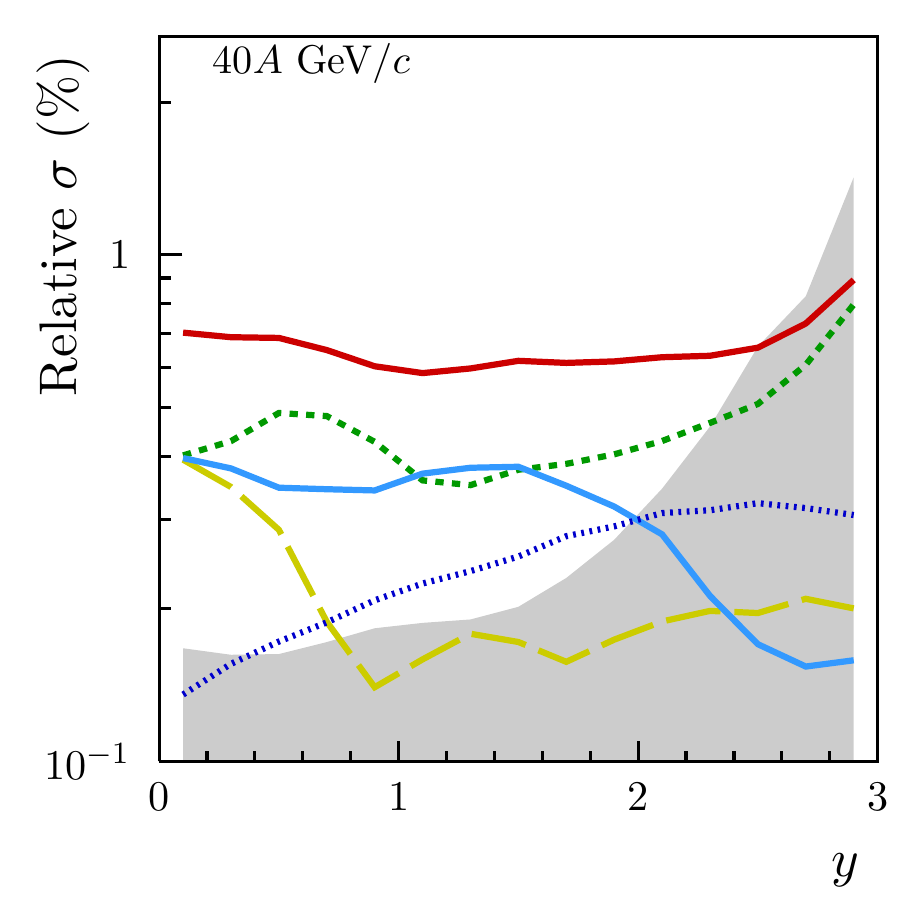}\\
   \includegraphics[width=0.4\textwidth]{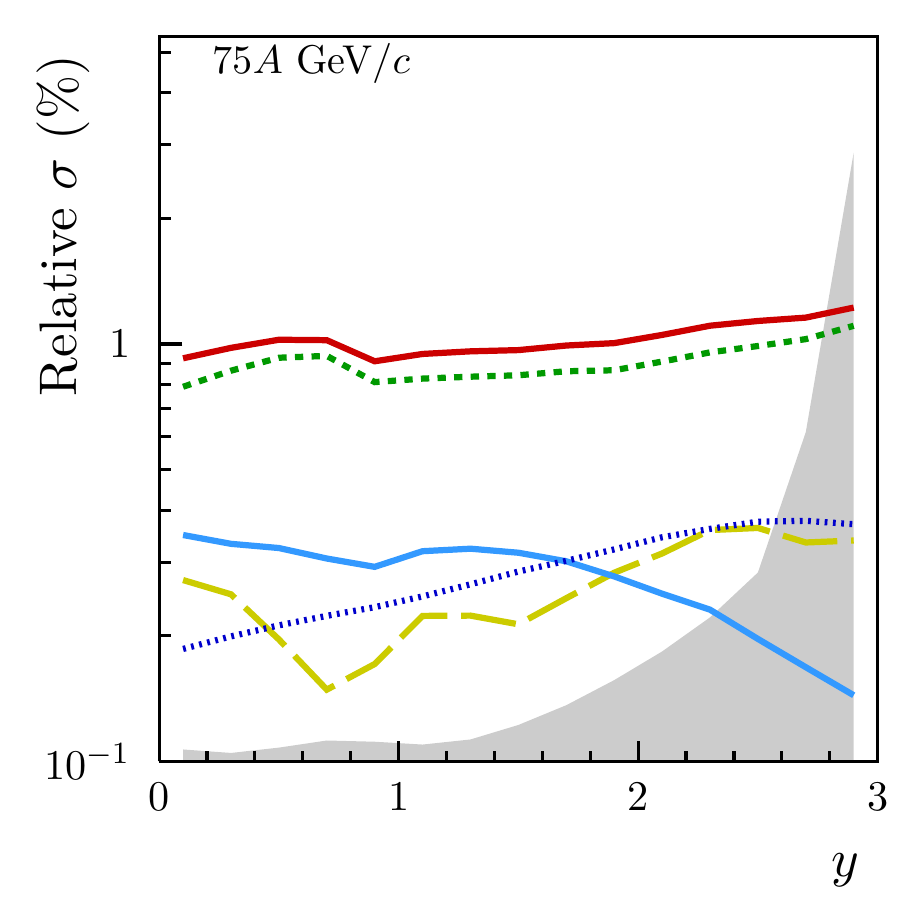}
   \includegraphics[width=0.4\textwidth]{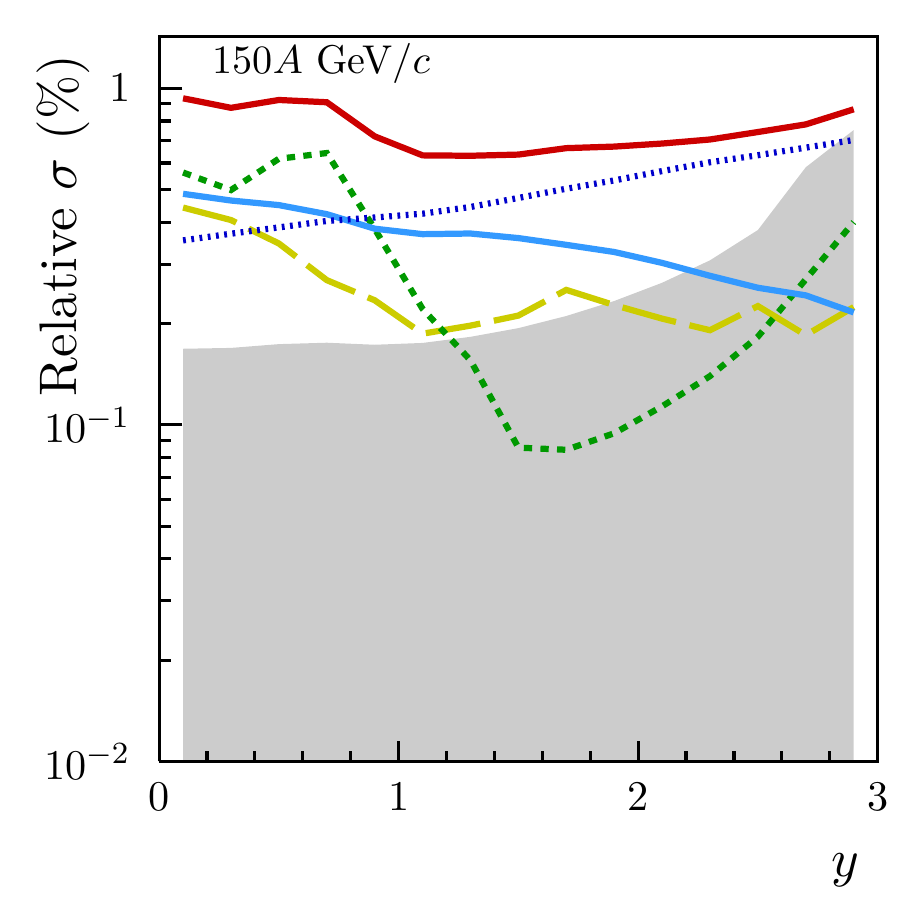}\\
   {\textcolor{kGray}\SquareSolid} $\sigma_\text{stat}$\hspace{.5cm}
   {\textcolor{kRed+2}\solidLine} $\sigma_\text{sys}$\hspace{.5cm}\hspace{.5cm}
   {\textcolor{kYellow+1}\looseDashedLine} $\sigma_\text{i}$\hspace{.5cm}
   {\textcolor{kGreen+2}\dashedLine} $\sigma_\text{ii}$\hspace{.5cm}\hspace{.5cm}
   {\textcolor{kAzure+1}\solidLine} $\sigma_\text{iii}$\hspace{.5cm}
   {\textcolor{kBlue+2}\dottedLine} $\sigma_\text{iv}$\\
 \caption{Statistical and systematic uncertainties for all six beam momenta as a function of rapidity \y. Statistical uncertainties are shown by gray shaded area, systematic uncertainties by curves referring to electron rejection and off-time events (i), track selection cuts (ii), contamination by decay daughters (iii), and contamination by primary mesons other than $\pi^{-}$ (iv), see text for details.}
 \label{fig:sys_uncertainties}
\end{figure*}

Statistical and systematic uncertainties for all six beam momenta are shown as a function of rapidity \y~in Fig.~\ref{fig:sys_uncertainties}.

\section{Experimental results}

This section presents results on negatively charged pion spectra at 13$A$, 19$A$, 30$A$, 40$A$, 75$A$ and 150\AGeVc beam momentum in the 5\% most \textit{central} ${}^{40}$Ar+${}^{45}$Sc collisions  with statistical and systematic uncertainties. The spectra refer to pions produced by strong interaction processes and in electromagnetic decays of produced hadrons. Comparisons of the new measurements of spectra, their parameters and mean multiplicities of $\pi^-$ mesons in \textit{central} Ar+Sc collisions with predictions of the \EposLong~\cite{Werner:2005jf, Pierog:2009zt, Pierog:2018}, \Urqmd~\cite{Bass:1998ca,Bleicher:1999xi} and \Hijing~\cite{Hijing:1991} models are presented. In the model calculations connected with spectra the selection of the 5\% most \textit{central} collisions was based on the number of projectile spectator nucleons.

\subsection{Double-differential ($y$, \pt) and ($y$, $m_{\text{T}}-m_{\pi}$) yields}

\begin{figure}[!htbp]
    \centering
    \includegraphics[width=0.45\textwidth]{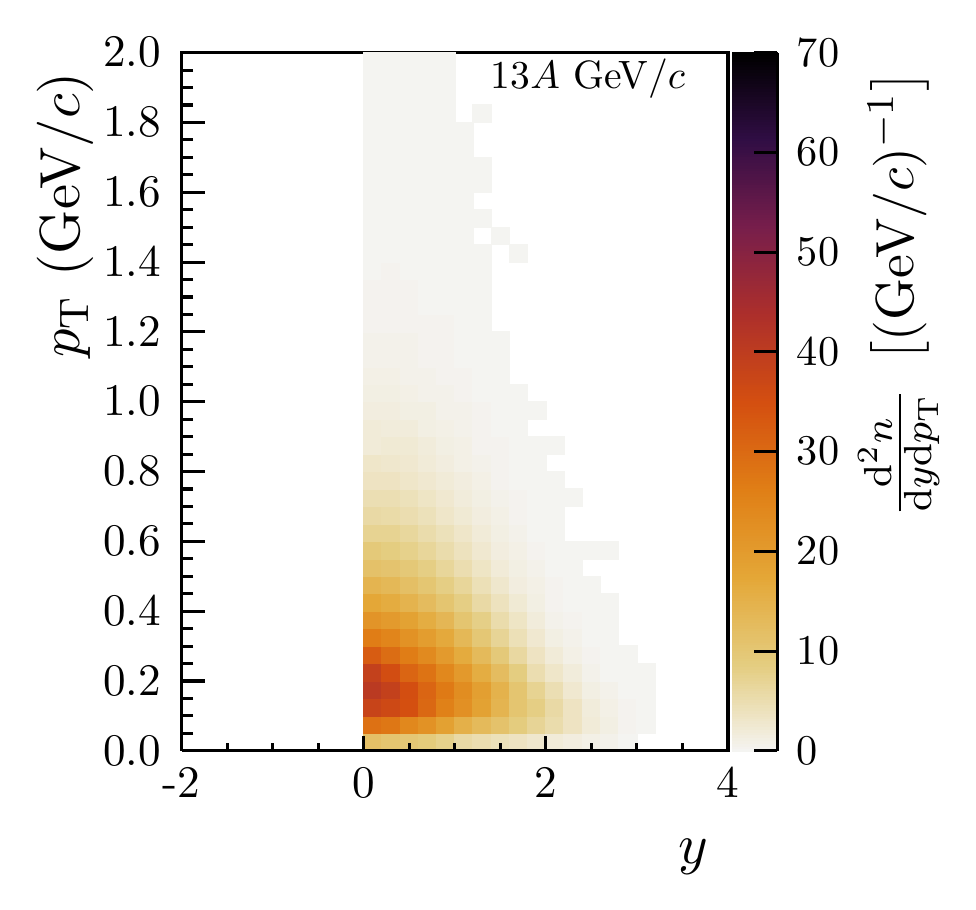}
    \includegraphics[width=0.45\textwidth]{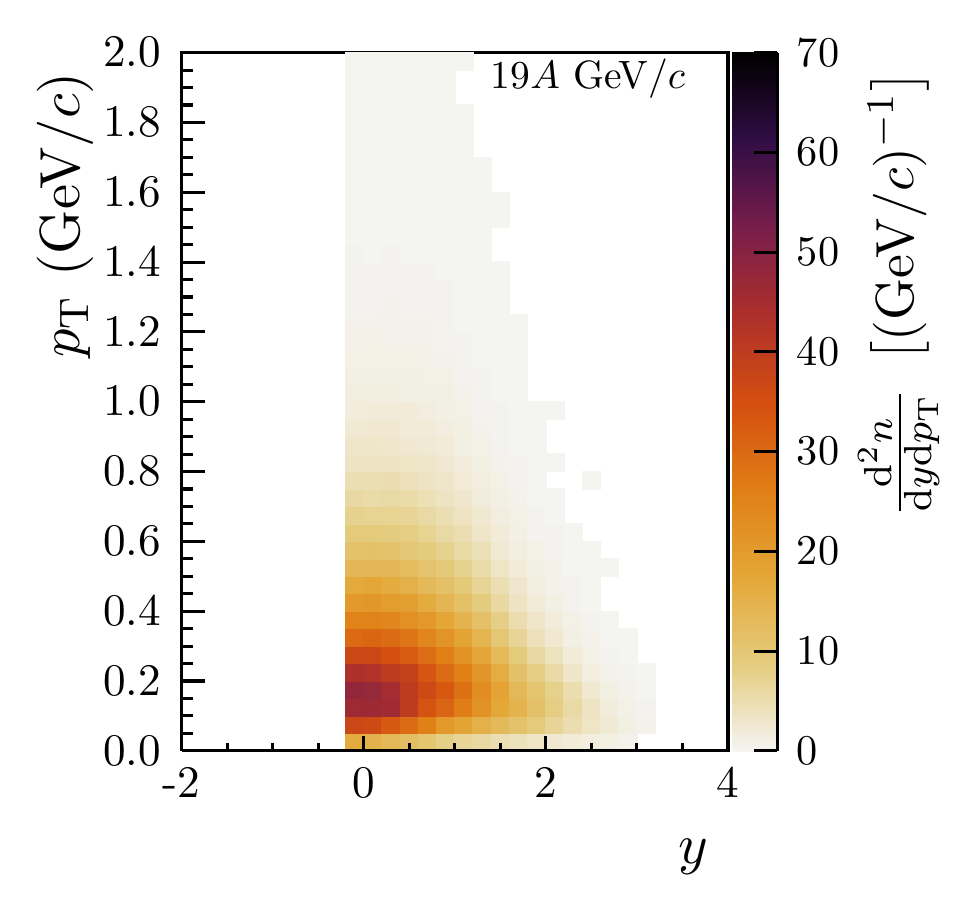}\\
    \includegraphics[width=0.45\textwidth]{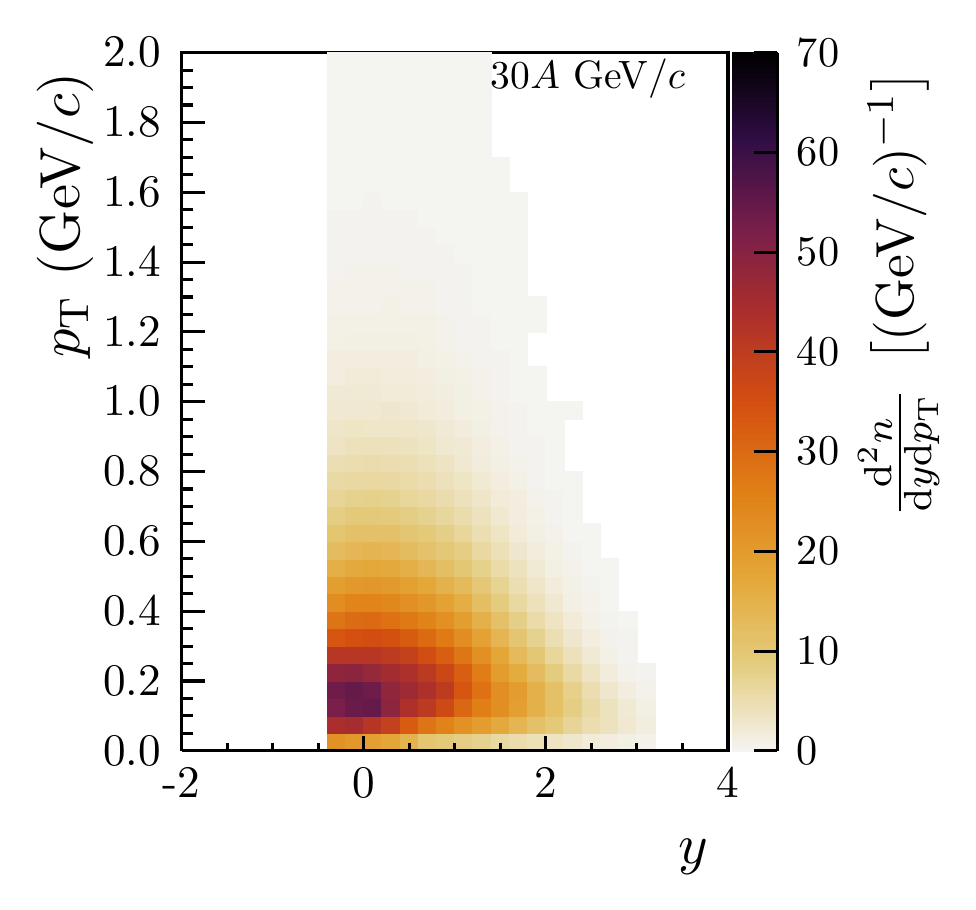}
    \includegraphics[width=0.45\textwidth]{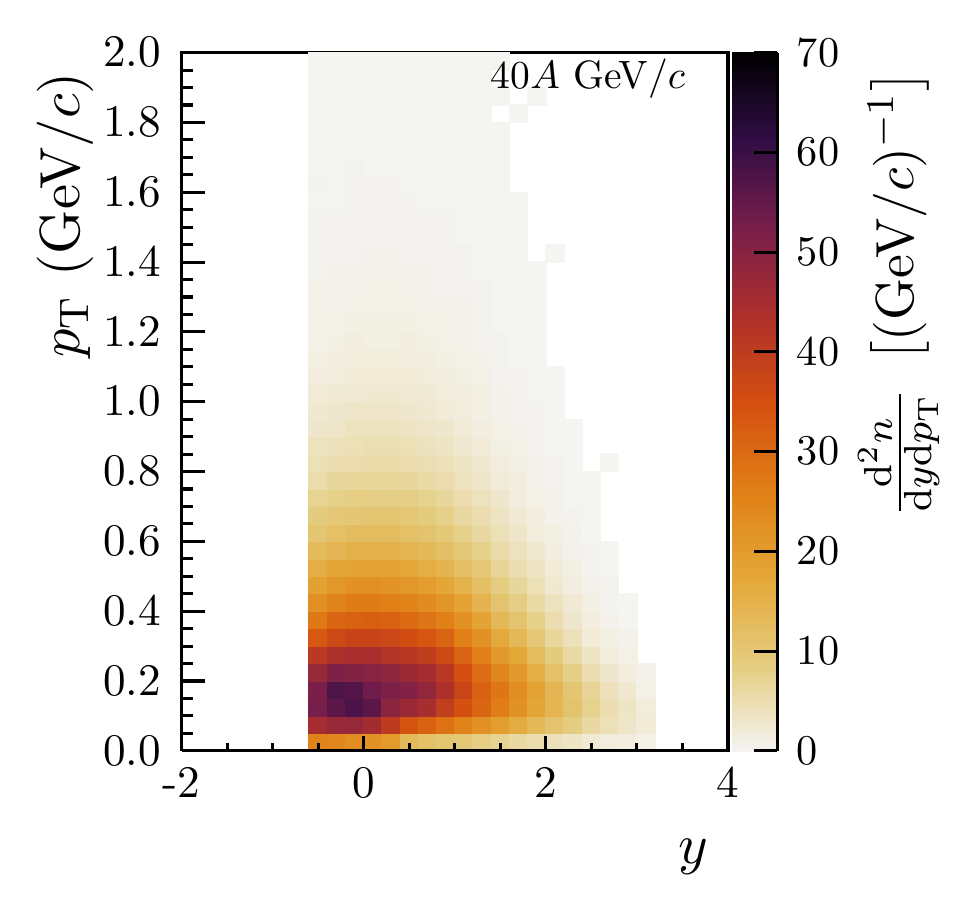}\\
    \includegraphics[width=0.45\textwidth]{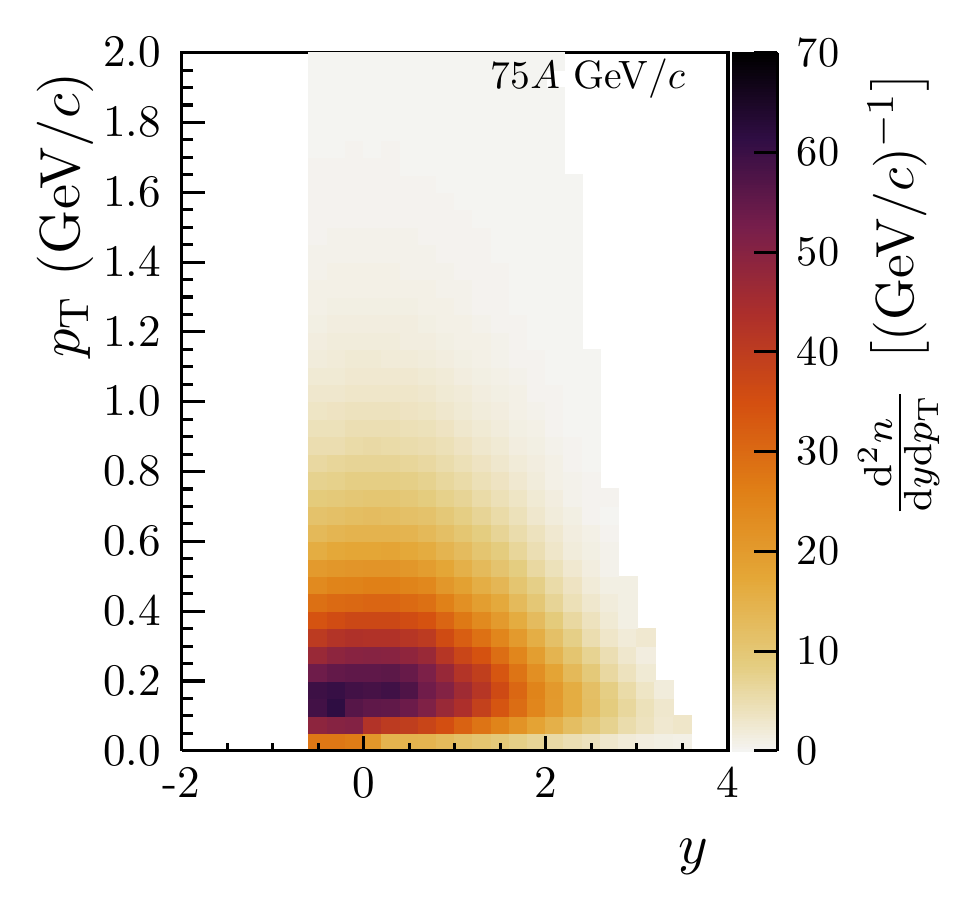}
    \includegraphics[width=0.45\textwidth]{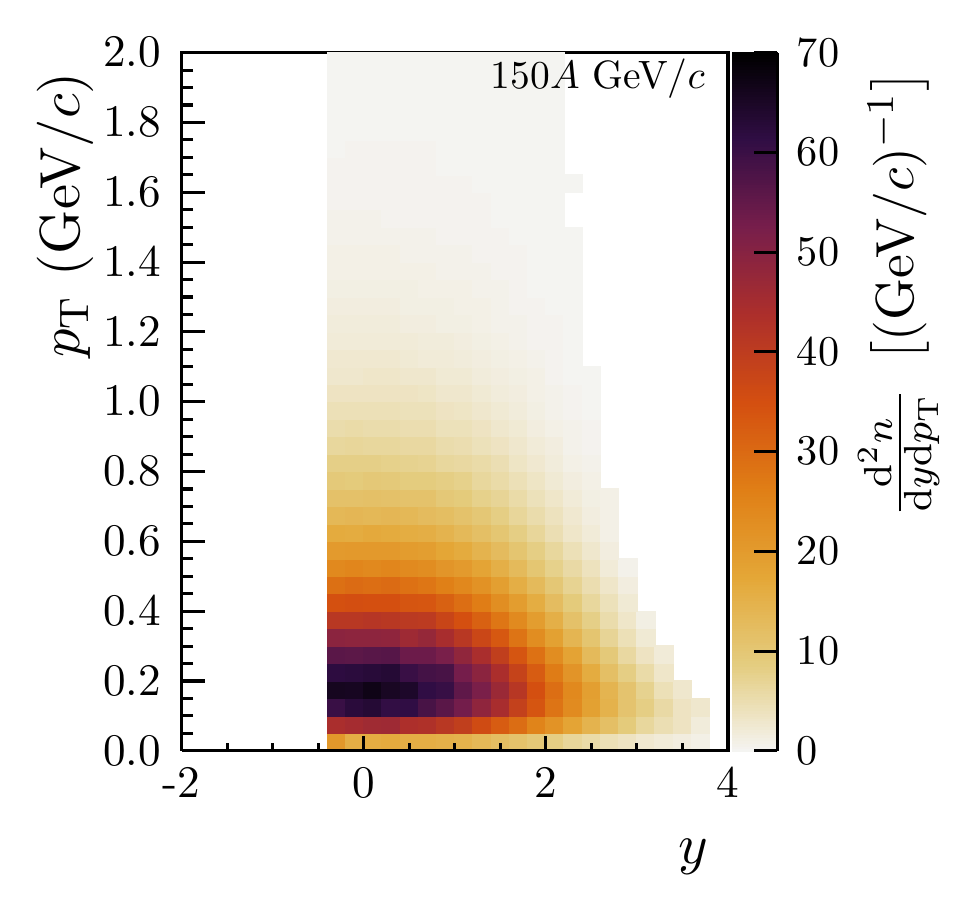}\\
    \caption{Corrected double-differential spectra \dndydpT~of negatively charged pions produced in the 5\% most \textit{central} Ar+Sc collisions at beam momenta of 13$A$, 19$A$, 30$A$, 40$A$, 75$A$ and 150$A$ \GeVc.}
    \label{fig:dndydpt}
\end{figure}

Figure~\ref{fig:dndydpt} shows fully corrected double-differential $(y,p_\text{T})$ distributions $\frac{\text{d}^2n}{\text{d}y\text{d}p_\text{T}}$ of $\pi^{-}$ measured in \textit{central} Ar+Sc collisions and illustrates the wide phase space acceptance of the detector. The distributions $\frac{\text{d}^{2}n}{\text{d}y\text{d}m_\text{T}}$ were calculated using an analogous procedure. From these results spectra of transverse momentum $p_\text{T}$, transverse mass $m_\text{T}-m_{\pi}$, rapidity $y$, as well as total multiplicities $\langle \pi^{-}\rangle$ were derived.

\subsection{Transverse momentum distributions}

\begin{figure*}[!htbp]
	\centering
	\includegraphics[width=0.4\textwidth]{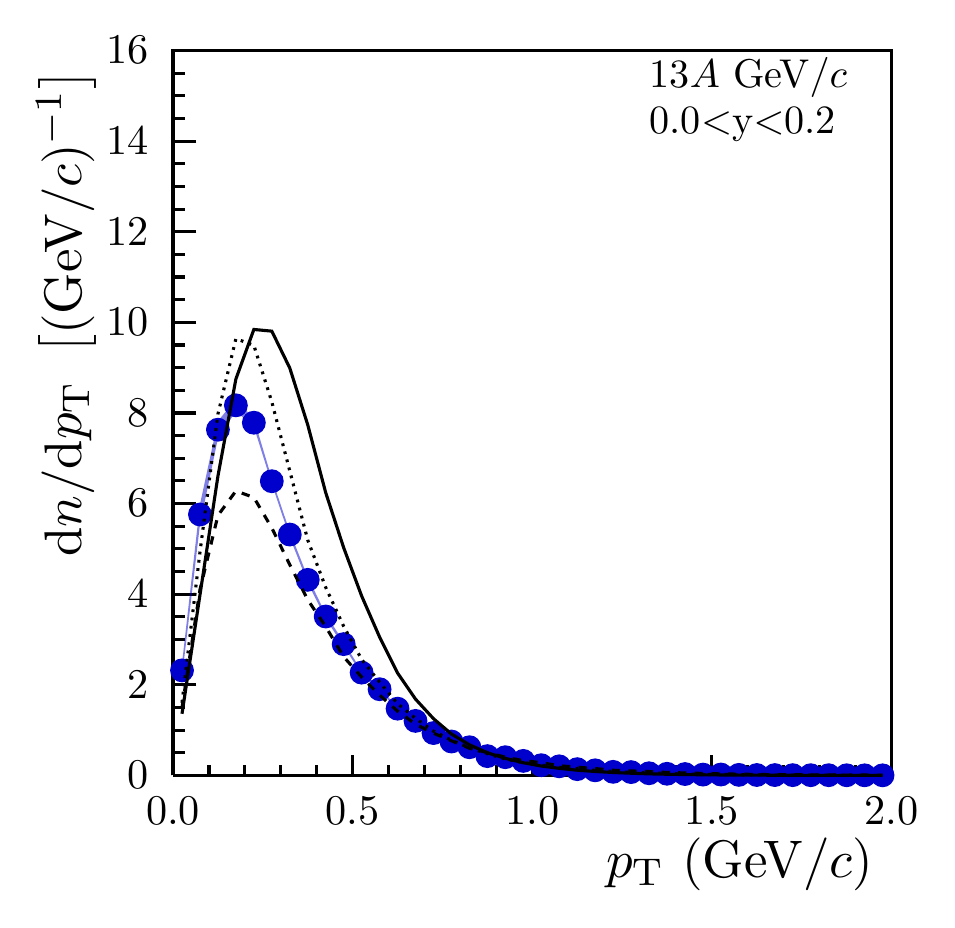}
	\includegraphics[width=0.4\textwidth]{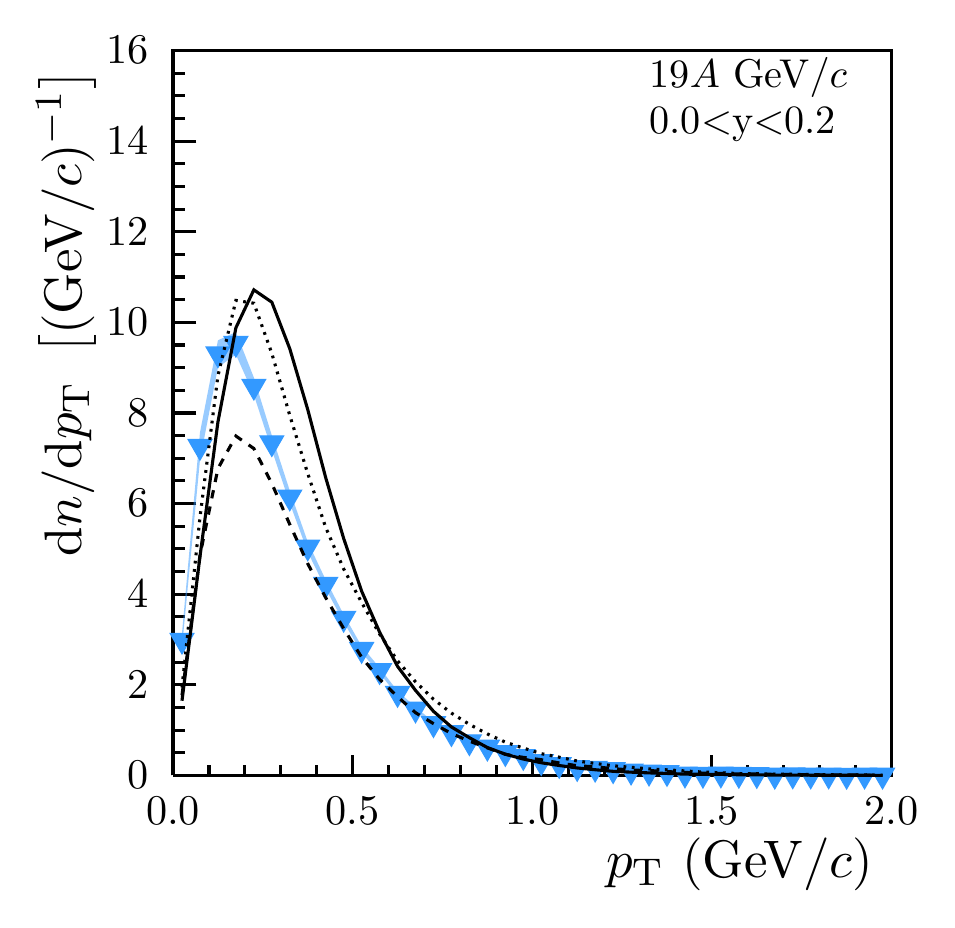}\\
	\includegraphics[width=0.4\textwidth]{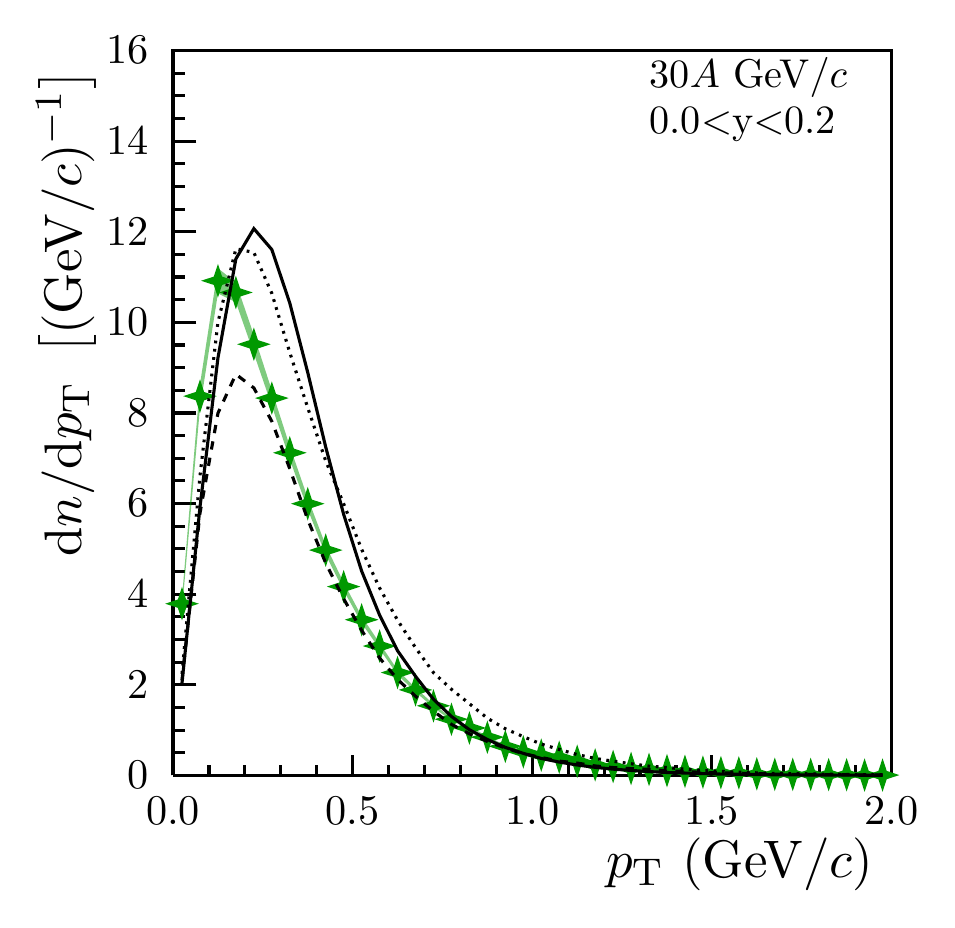}
	\includegraphics[width=0.4\textwidth]{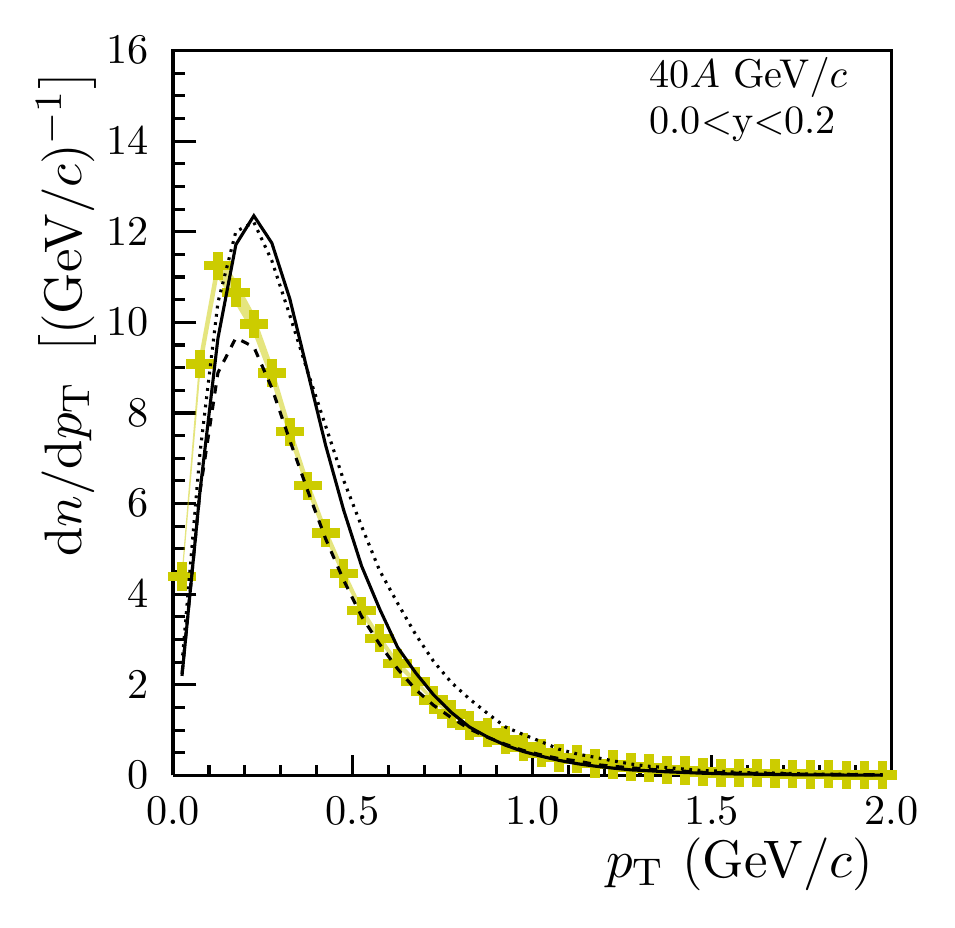}\\
	\includegraphics[width=0.4\textwidth]{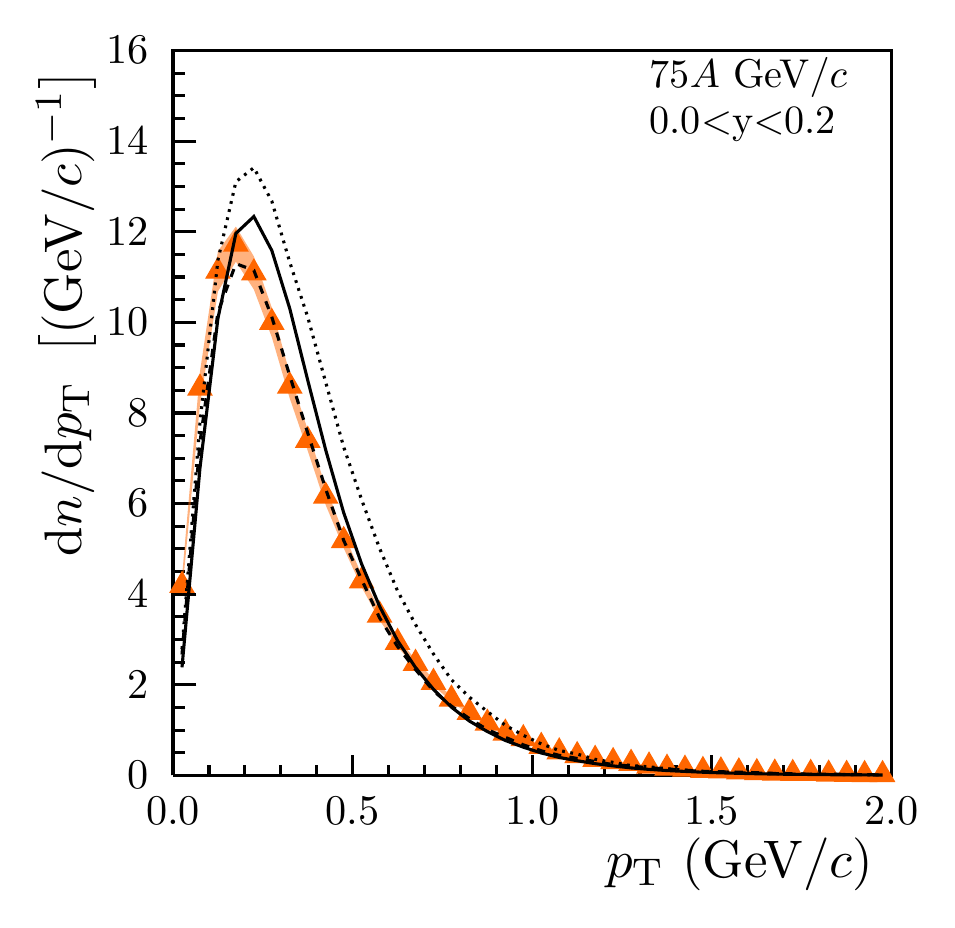}
	\includegraphics[width=0.4\textwidth]{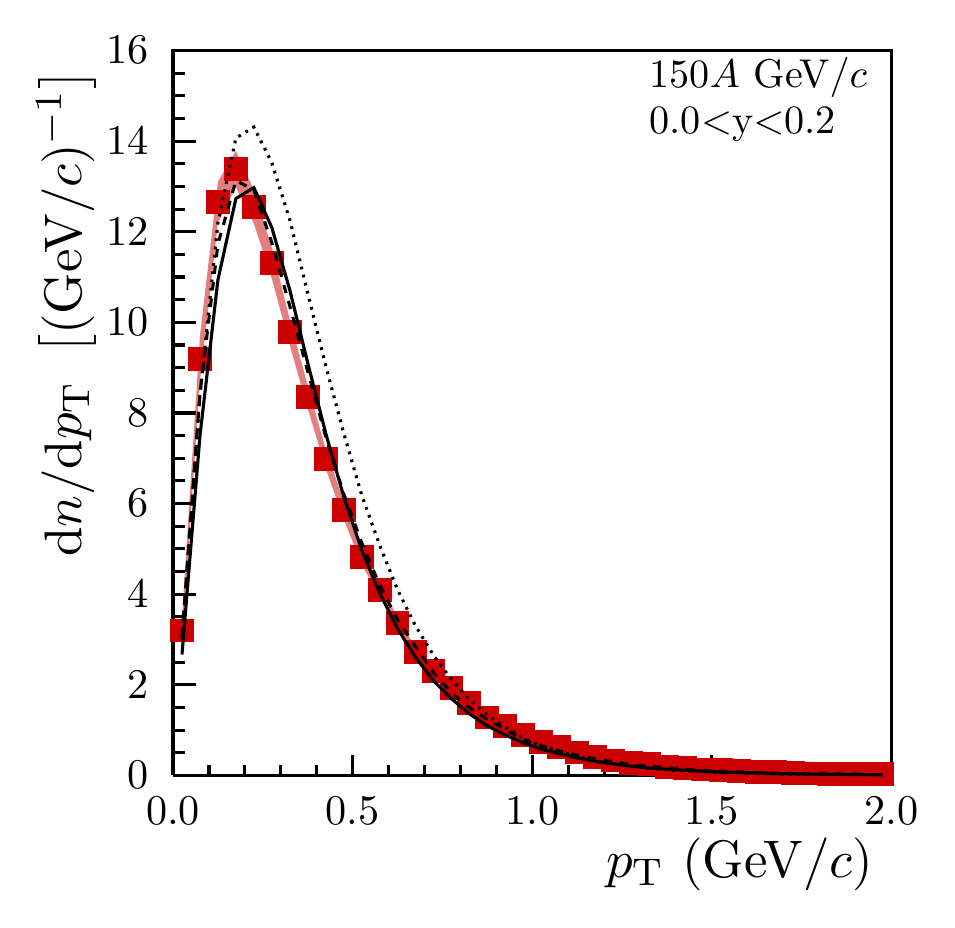}\\
	\textcolor{kBlack}{\solidLine} \Epos~\cite{Werner:2005jf, Pierog:2009zt, Pierog:2018}\hspace{1cm} \textcolor{kBlack}{\dashedLine} \Urqmd~\cite{Bass:1998ca,Bleicher:1999xi}\hspace{1cm} \textcolor{kBlack}{\dottedLine} \Hijing~\cite{Hijing:1991}
	\caption{Transverse momentum distributions \dndptLong~at mid-rapidity for all six beam momenta. Predictions of \Epos, \Urqmd and \Hijing models are shown by curves of different line styles. Statistical uncertainties are smaller than the marker size and systematic uncertainties are indicated by shaded bands.}
	\label{fig:dnDPTComparison}
\end{figure*}

Figure~\ref{fig:dnDPTComparison} shows measured transverse momentum $p_\text{T}$ spectra at mid-rapidity for all six beam momenta. The results are compared with \Epos, \Urqmd and \Hijing model calculations. The spectra differ significantly from the models' predictions, especially for lower beam momenta, where none of the models can describe the data well. For higher beam momenta \Epos and \Urqmd describe data with reasonable accuracy.

\clearpage

\subsection{Transverse mass distributions}

\begin{figure}[!htbp]
  \centering
        \begin{minipage}[m]{0.7\textwidth}
        \includegraphics[width=0.95\textwidth]{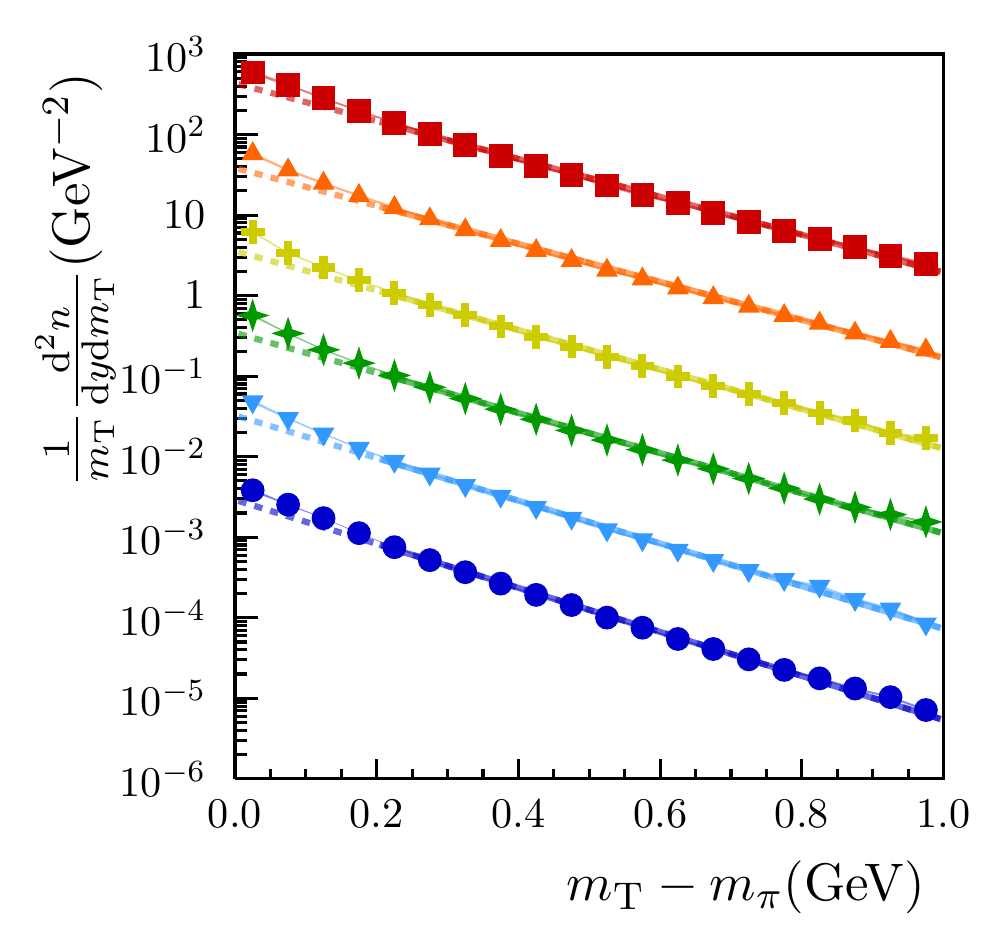}
        \end{minipage}
        \begin{minipage}[m]{0.25\textwidth}
        \normalsize
        \begin{itemize}
	        \item[\tiny\textcolor{color150}{\markerOneFifty}] 150\AGeVc
	        \item[\tiny\textcolor{color75}{\markerSeventyFife}] 75\AGeVc $\times 10^{-1}$
	        \item[\tiny\textcolor{color40}{\markerFourty}] 40\AGeVc $\times 10^{-2}$
	        \item[\tiny\textcolor{color30}{\markerThirty}] 30\AGeVc $\times 10^{-3}$
	        \item[\tiny\textcolor{color19}{\markerNineteen}] 19\AGeVc $\times 10^{-4}$
	        \item[\tiny\textcolor{color13}{\markerThirteen}] 13\AGeVc $\times 10^{-5}$
		\end{itemize}
        \end{minipage}
  \caption{Transverse mass spectra at mid-rapidity ($0<y<0.2$). The fitted exponential function is indicated by solid lines in the fit range $0.24<m_\text{T}-m_{\pi}<0.72$~\GeV and dashed lines outside the fit range. The data points for different beam momenta were scaled for better readability. Statistical uncertainties are smaller than the marker size. Systematic uncertainties are not plotted.}
  \label{fig:mTDist}
\end{figure}

Spectra of transverse mass $m_\text{T}-m_{\pi}$ at mid-rapidity ($0<y<0.2$) are shown in Fig.~\ref{fig:mTDist}. For further comparisons a function
\begin{equation}
 \frac{\text{d}n}{\text{d}m_\text{T}}=A \cdot m_\text{T} \cdot \exp\left(-\frac{m_\text{T}}{T}\right)
\label{eq:mT_expfit}
\end{equation}
was fitted in the range $0.24<m_\text{T}-m_{\pi}<0.72$~\GeV where no strong contributions from resonance decays and radial flow are expected. The fitted parameters were the normalization $A$ and the inverse slope parameter $T$.  The results of the fits are indicated by lines in Fig.~\ref{fig:mTDist}. The deviation of the measurements from the exponential shape for larger values of $m_\text{T}-m_{\pi}$ is indicative of collective transverse flow~\cite{Alt:2007uj} whereas the excess below the fit range can be explained by the contribution of resonance decay products to the $\pi^-$ spectrum.

\begin{figure}[!htbp]
	\centering
	\begin{minipage}[m]{0.7\textwidth}
		\includegraphics[width=0.95\textwidth]{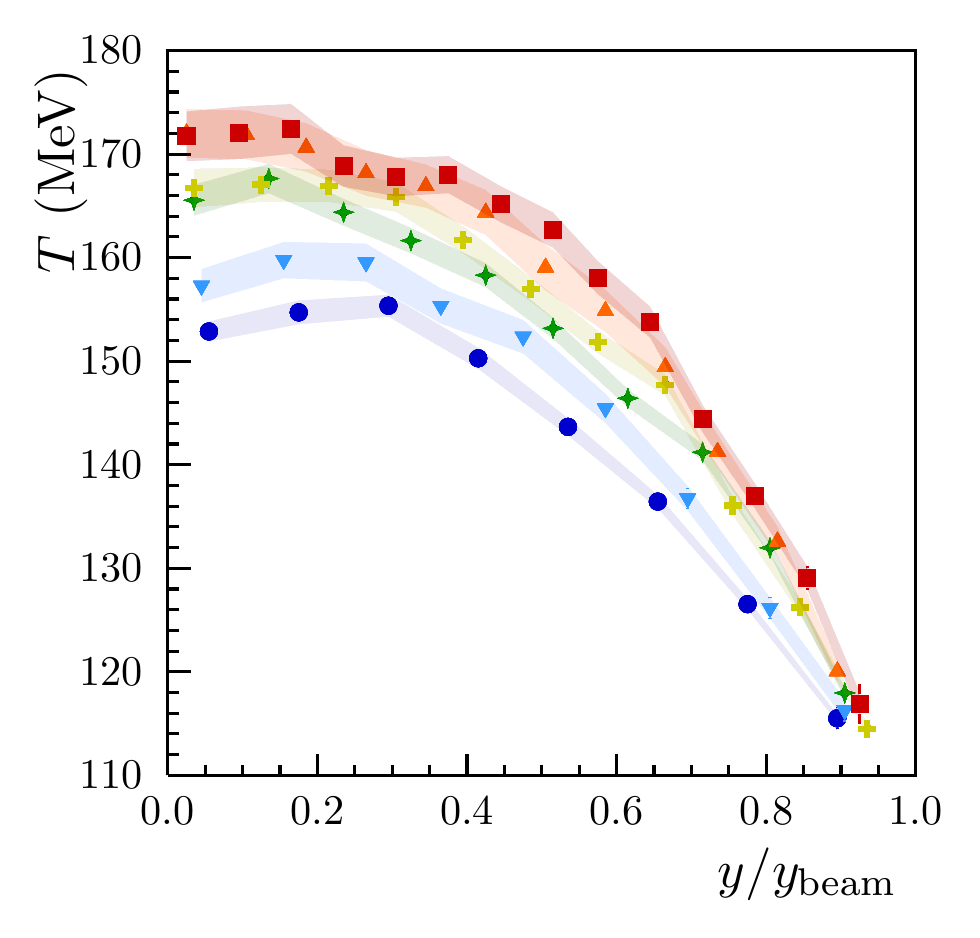}
	\end{minipage}
	\begin{minipage}[m]{0.25\textwidth}
		\normalsize
		\begin{itemize}
			\item[\tiny\textcolor{color150}{\markerOneFifty}] 150\AGeVc
			\item[\tiny\textcolor{color75}{\markerSeventyFife}] 75\AGeVc
			\item[\tiny\textcolor{color40}{\markerFourty}] 40\AGeVc
			\item[\tiny\textcolor{color30}{\markerThirty}] 30\AGeVc
			\item[\tiny\textcolor{color19}{\markerNineteen}] 19\AGeVc
			\item[\tiny\textcolor{color13}{\markerThirteen}] 13\AGeVc
		\end{itemize}
	\end{minipage}
	\caption{The inverse slope parameter $T$ of the transverse mass spectra as a function of rapidity divided by the beam rapidity. The fit range is $0.24<m_\text{T}-m_{\pi}<0.72$~\GeV. Statistical uncertainties are usually smaller than the marker size and systematic uncertainties are indicated by shaded bands.}
	\label{fig:invSlope}
\end{figure}

Figure~\ref{fig:invSlope} presents the dependence of the inverse slope parameter $T$ on the rapidity for the different beam energies. One finds the well known decrease towards larger rapidities. Moreover, an increase of the values of $T$ by about 20 MeV is seen from 13$A$ to 150\AGeVc beam momentum.

\begin{figure}[!htbp]
	\centering
	\begin{minipage}[m]{\textwidth}
		\includegraphics[width=0.45\textwidth]{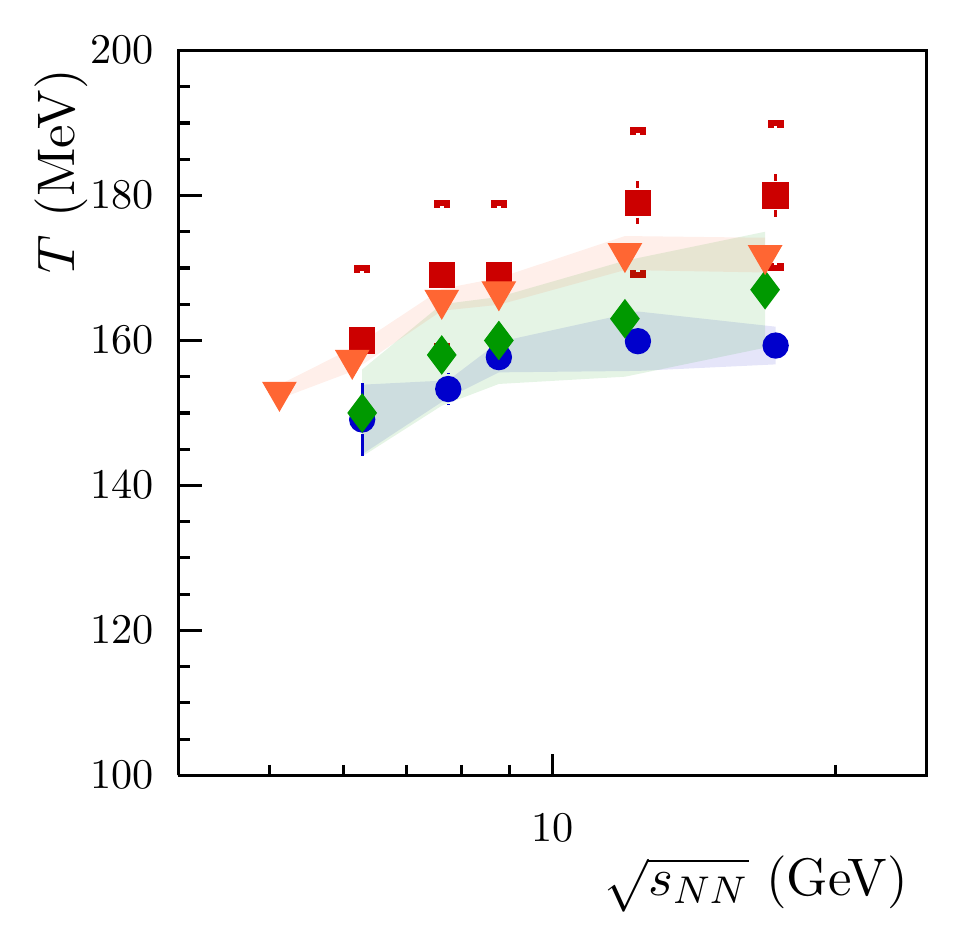}
		\includegraphics[width=0.45\textwidth]{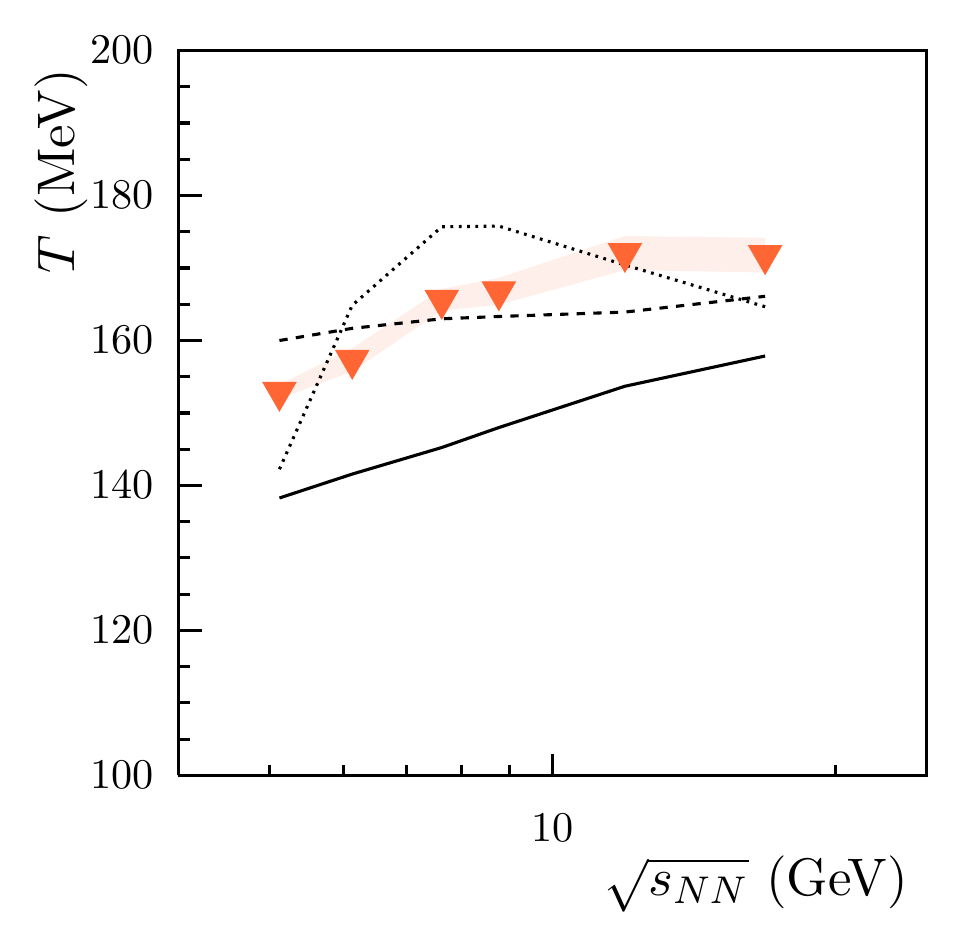}
	\end{minipage}
\begin{minipage}[m]{0.45\textwidth}
\begin{center}
\begin{varwidth}{\textwidth}
	\centering
	\normalsize
	\begin{itemize}
		\item[\footnotesize\textcolor{colorArSc}{\markerArSc}] Ar+Sc
		\item[\scriptsize\textcolor{colorBeBe}{\markerBeBe}] Be+Be~\cite{Acharya:2020cyb}
		\item[\footnotesize\textcolor{colorPP}{\markerPP}] \pp~\cite{Abgrall:2013pp_pim}
		\item[\footnotesize\textcolor{colorPbPb}{\markerPbPb}] Pb+Pb (NA49~\cite{Afanasiev:2002mx,Alt:2007aa})
	\end{itemize}
\end{varwidth}
\end{center}
\end{minipage}
\begin{minipage}[m]{0.45\textwidth}
\begin{center}
\begin{varwidth}{\textwidth}
	\normalsize
	\begin{itemize}
		\item[\footnotesize\textcolor{colorArSc}{\markerArSc}] Ar+Sc
		\item[\huge\textcolor{kBlack}{\dottedLine}] Ar+Sc (\Hijing~\cite{Hijing:1991})
		\item[\huge\textcolor{kBlack}{\dashedLine}] Ar+Sc (\Urqmd~\cite{Bass:1998ca,Bleicher:1999xi})
		\item[\huge\textcolor{kBlack}{\solidLine}] Ar+Sc (\Epos~\cite{Werner:2005jf, Pierog:2009zt, Pierog:2018})
	\end{itemize}
\end{varwidth}
\end{center}
\end{minipage}
	\caption{\textit{Left:} inverse slope parameter $T$ at mid-rapidity ($0<y<0.2$) fitted in the range $0.24<m_\text{T}-m_{\pi}<0.72$~\GeV plotted against the collision energy per nucleon together with measurements for inelastic \pp, \textit{central} Be+Be and central Pb+Pb collisions. Statistical uncertainties are shown as vertical bars (often smaller than the marker size) and systematic uncertainties are indicated by shaded bands (or caps for Pb+Pb collisions). \textit{Right:} comparison of the results for Ar+Sc collisions as presented in the left plot with \Epos, \Urqmd and \Hijing model calculations (black curves).}
	\label{fig:inverseSlopeVsSNN}
\end{figure}

Inverse slope parameters fitted at mid-rapidity ($0<y<0.2$) in the range $0.24<m_\text{T}-m_{\pi}<0.72$~\GeV are plotted versus $\sqrt{s_{NN}}$ for inelastic \pp, 5\% most \textit{central} Ar+Sc, \textit{central} Be+Be and central Pb+Pb collisions in Fig.~\ref{fig:inverseSlopeVsSNN} (\textit{left}). As seen from the plot, the values increase significantly (10 - 15~MeV) for all three reactions. The new Ar+Sc results are close to those for Pb+Pb reactions but about 15~MeV higher than for \pp and Be+Be reactions. Furthermore, Fig.~\ref{fig:inverseSlopeVsSNN} (\textit{right}) shows that \Epos model predictions for Ar+Sc collisions are lower than the \NASixtyOne measurements and stay even below the measurements in inelastic \pp interactions. Predictions of \Urqmd exhibit a non-monotonic behavior. They lie lower than measurements at low beam momenta and higher at high beam momenta. \Hijing shows a concave behavior and unsatisfactory agreement with measurements.

\begin{figure}[!htbp]
	\centering
	\begin{minipage}[m]{0.5\textwidth}
		\includegraphics[width=0.95\textwidth]{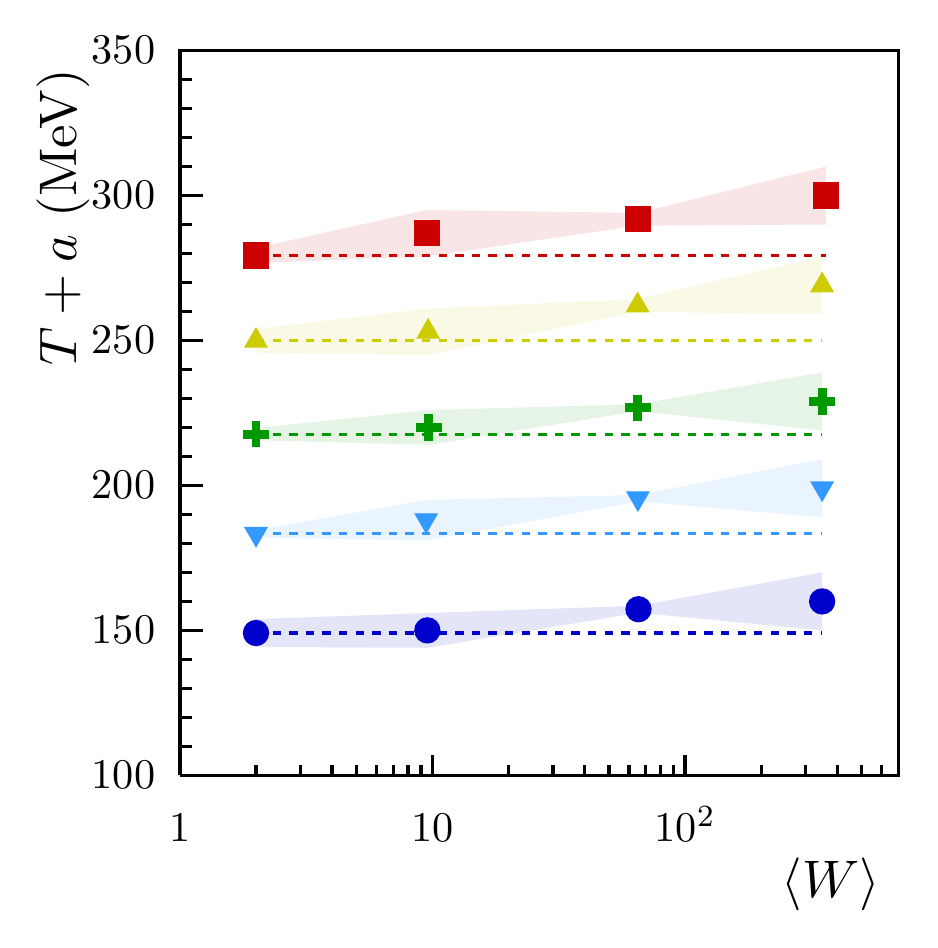}
	\end{minipage}
	\begin{minipage}[m]{0.35\textwidth}
		\normalsize
		\begin{itemize}
			\item[\tiny\textcolor{color150Comparison}{\markerOneFiftyComparison}] 150\AGeVc, $a=120$ MeV
			\item[\tiny\textcolor{color75Comparison}{\markerSeventyFifeComparison}] 75\AGeVc, $a=90$ MeV
			\item[\tiny\textcolor{color40Comparison}{\markerFourtyComparison}] 40\AGeVc, $a=60$ MeV
			\item[\tiny\textcolor{color30Comparison}{\markerThirtyComparison}] 30\AGeVc, $a=30$ MeV
			\item[\tiny\textcolor{color19Comparison}{\markerNineteenComparison}] 19\AGeVc, $a=0$ MeV
		\end{itemize}
	\end{minipage}
	\caption{The inverse slope parameter $T$ versus the mean number of wounded nucleons $\langle W\rangle$ in \textit{central} Ar+Sc collisions at beam momenta from 19$A$ to 150\AGeVc. Statistical uncertainties are smaller than the marker size. Systematic uncertainties are shown by shaded bands. For better visibility different energies are offset vertically.}
	\label{fig:inverseSlopeVsW}
\end{figure}

Figure~\ref{fig:inverseSlopeVsW} presents the inverse slope parameter $T$ plotted versus the number of wounded nucleons $\langle W\rangle$ which are a measure of the initial volume of the collision system. Although the uncertainties are large, the measurements show a modest monotonic rise with increasing system size for all the beam momenta.

\begin{figure}[!htbp]
\centering
\begin{minipage}[m]{\textwidth}
	\includegraphics[width=0.45\textwidth]{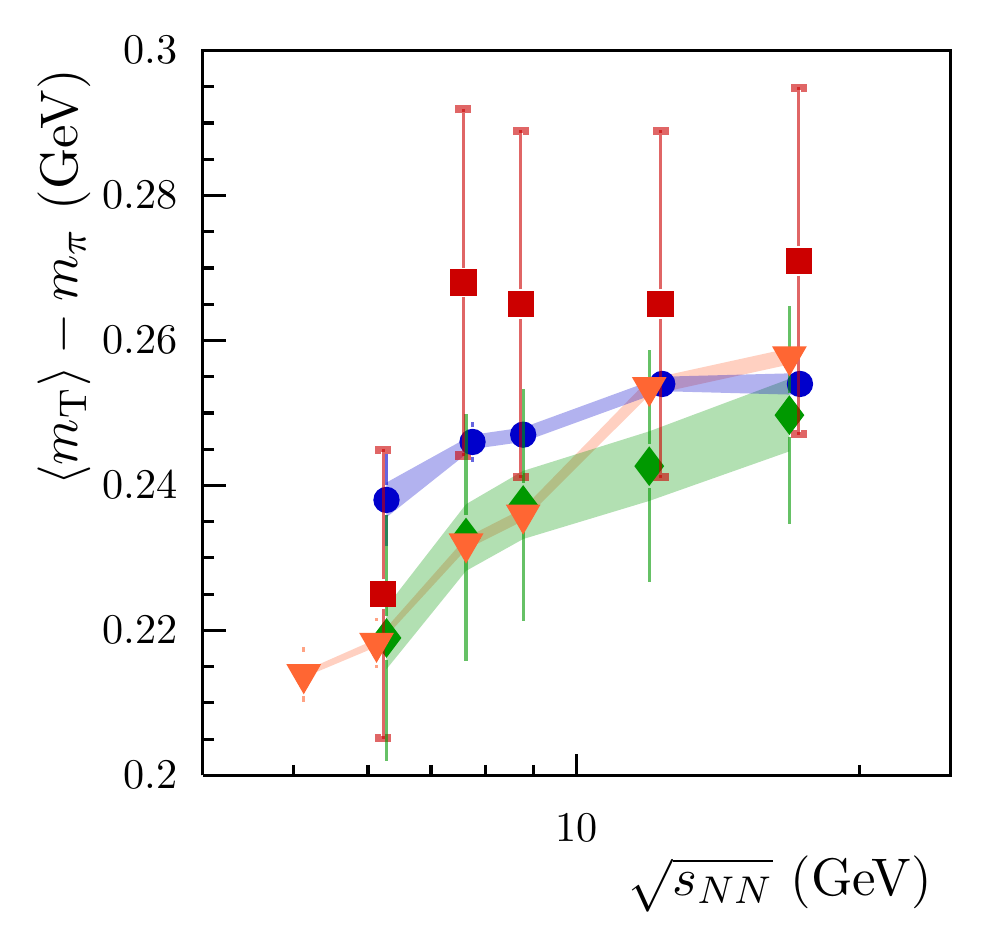}
	\includegraphics[width=0.45\textwidth]{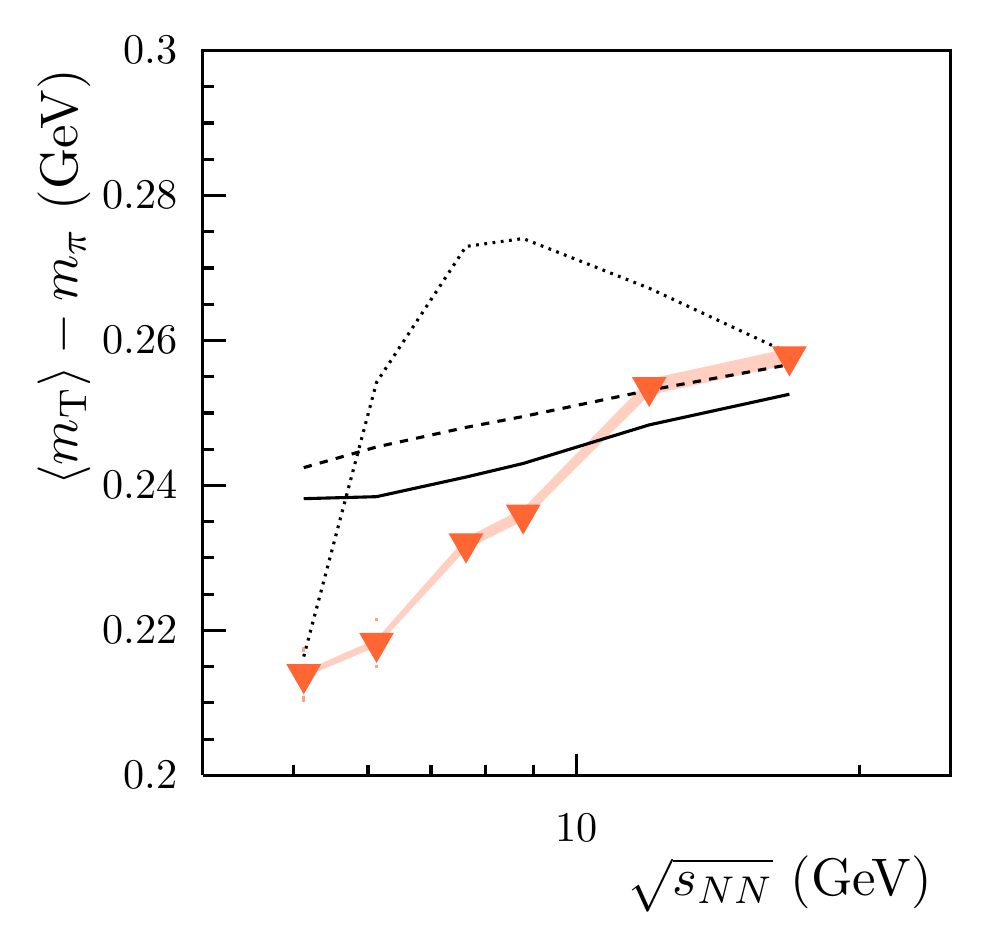}
\end{minipage}
\begin{minipage}[m]{0.45\textwidth}
\begin{center}
\begin{varwidth}{\textwidth}
	\centering
	\normalsize
	\begin{itemize}
		\item[\footnotesize\textcolor{colorArSc}{\markerArSc}] Ar+Sc
		\item[\scriptsize\textcolor{colorBeBe}{\markerBeBe}] Be+Be~\cite{Acharya:2020cyb}
		\item[\footnotesize\textcolor{colorPP}{\markerPP}] \pp~\cite{Abgrall:2013pp_pim}
		\item[\footnotesize\textcolor{colorPbPb}{\markerPbPb}] Pb+Pb (NA49~\cite{Afanasiev:2002mx,Alt:2007aa})
	\end{itemize}
\end{varwidth}
\end{center}
\end{minipage}
\begin{minipage}[m]{0.45\textwidth}
\begin{center}
\begin{varwidth}{\textwidth}
	\normalsize
	\begin{itemize}
		\item[\footnotesize\textcolor{colorArSc}{\markerArSc}] Ar+Sc
		\item[\huge\textcolor{kBlack}{\dottedLine}] Ar+Sc (\Hijing~\cite{Hijing:1991})
		\item[\huge\textcolor{kBlack}{\dashedLine}] Ar+Sc (\Urqmd~\cite{Bass:1998ca,Bleicher:1999xi})
		\item[\huge\textcolor{kBlack}{\solidLine}] Ar+Sc (\Epos~\cite{Werner:2005jf, Pierog:2009zt, Pierog:2018})
	\end{itemize}
\end{varwidth}
\end{center}
\end{minipage}
\caption{\textit{Left:} average transverse mass $\langle m_\text{T}\rangle-m_{\pi}$ at mid-rapidity ($0<y<0.2$) versus the collision energy. The results are compared with the corresponding data on inelastic \pp, \textit{central} Be+Be and central Pb+Pb collisions. Statistical uncertainties are shown as vertical bars (occasionally smaller than the marker size and with caps for Pb+Pb) and systematic uncertainties are indicated by shaded bands. \textit{Right:} comparison of the results for Ar+Sc collisions as presented in the left plot with \Epos, \Urqmd and \Hijing model calculations (black curves).}
\label{fig:mTMeanComparisonSystems}
\end{figure}

As the distributions are not strictly exponential it may be better to characterize them by their average values $\langle m_\text{T}\rangle-m_{\pi}$. These were calculated by summing the $m_\text{T}-m_{\pi}$-weighted distributions and adding an extrapolation for the region $m_\text{T}-m_{\pi} > 1.2$~\GeV based on the exponential fits Eq.~\ref{eq:mT_expfit}. The results are plotted at mid-rapidity versus the collision energy in Fig.~\ref{fig:mTMeanComparisonSystems}. The differences between \pp and \textit{central} Be+Be, Ar+Sc and central Pb+Pb reactions are compatible within their uncertainties. Ar+Sc, Be+Be and \pp measurements show a rise with increasing collision energy which is more pronounced for Ar+Sc. Due to the large uncertainties for the Pb+Pb data a significant discrimination between rise and constancy is not possible for this reaction. \Epos and \Urqmd model predictions show a slope that is more similar to \pp interactions and less steep than for Ar+Sc with \Urqmd covering two highest collision energies. \Hijing shows a concave behavior with extreme points matching the measurements.

\begin{figure}[!htbp]
  \centering
        \begin{minipage}[t]{0.45\textwidth}
            \includegraphics[width=0.95\textwidth]{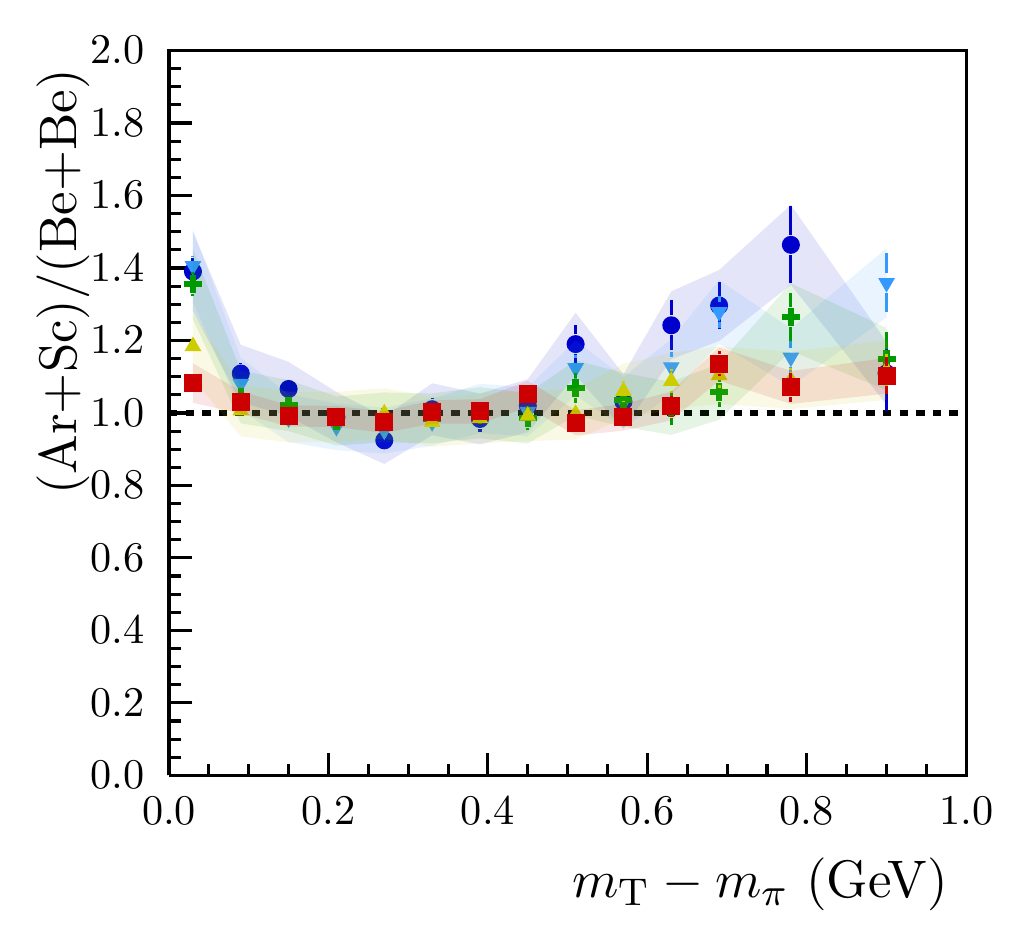}\\
            \includegraphics[width=0.95\textwidth]{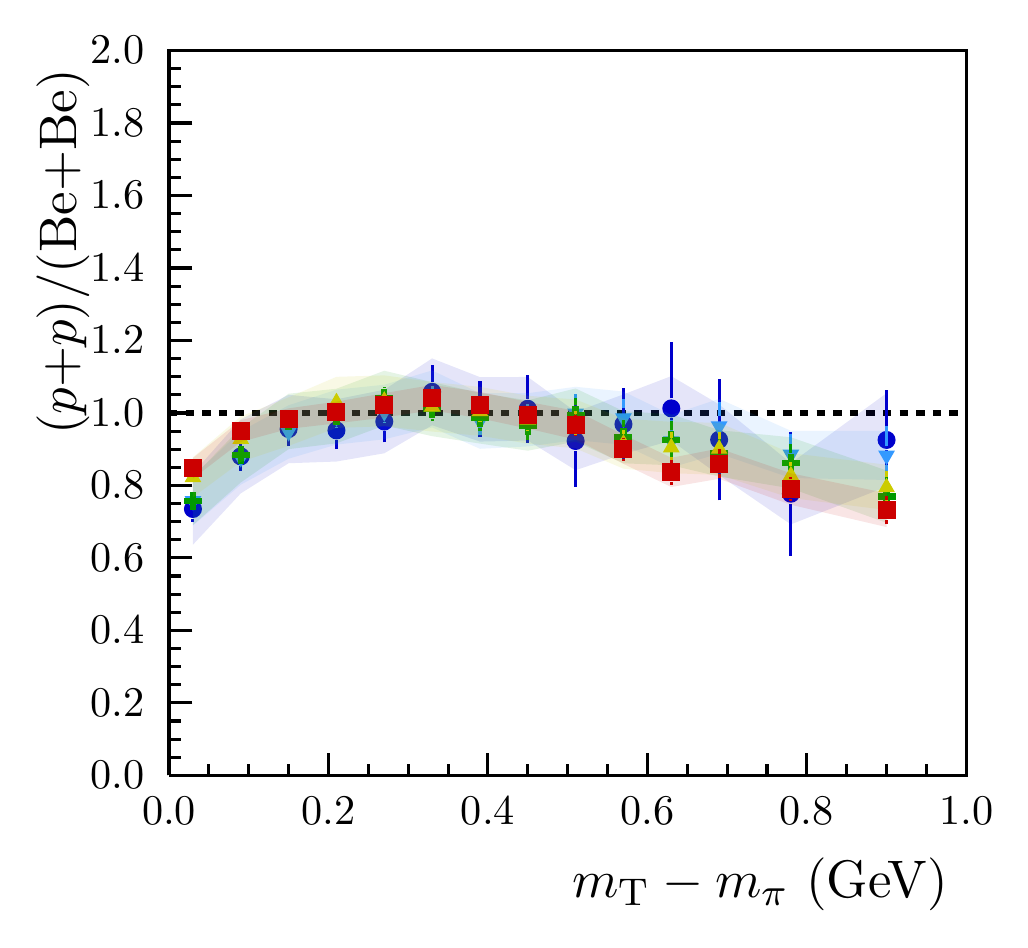}
        \end{minipage}
        \begin{minipage}[t]{0.45\textwidth}
        	\includegraphics[width=0.95\textwidth]{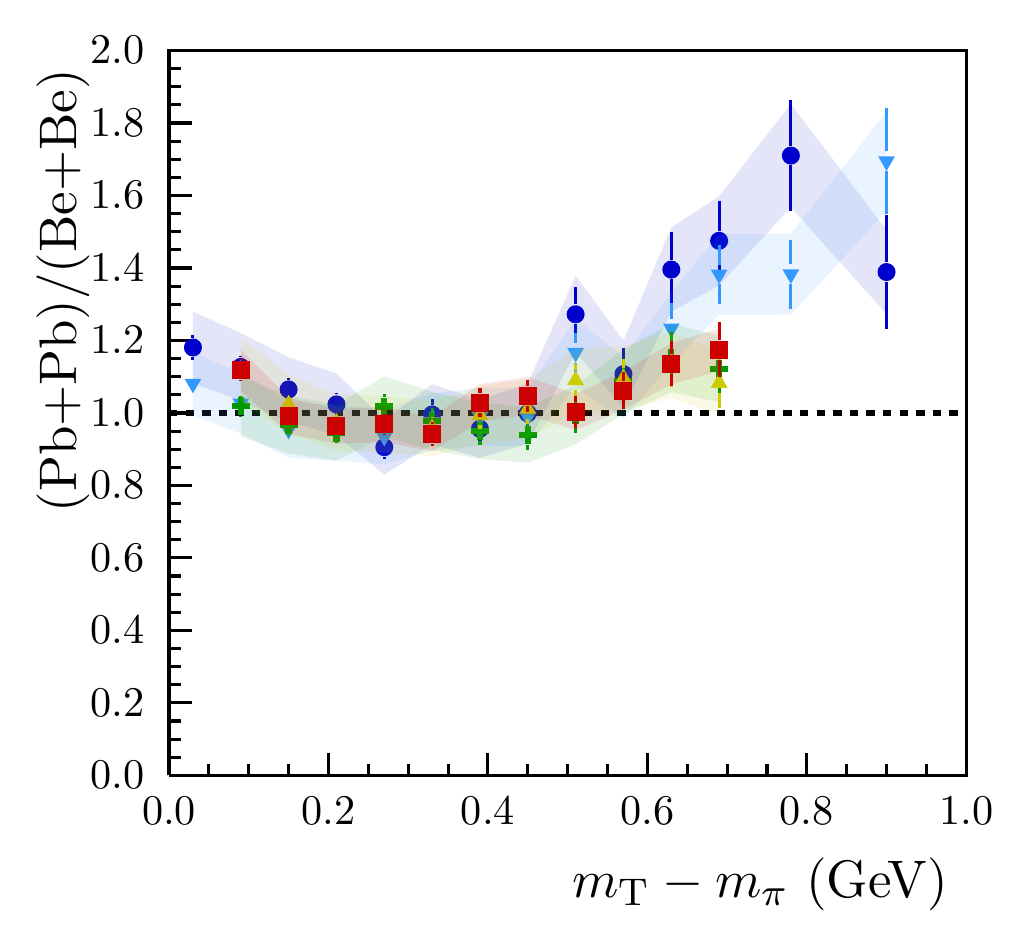}\\
            \normalsize
            \begin{itemize}
                \item[\tiny\textcolor{color150Comparison}{\markerOneFiftyComparison}] 150\AGeVc
	    		\item[\tiny\textcolor{color75Comparison}{\markerSeventyFifeComparison}] 75\AGeVc
			    \item[\tiny\textcolor{color40Comparison}{\markerFourtyComparison}] 40\AGeVc
			    \item[\tiny\textcolor{color30Comparison}{\markerThirtyComparison}] 30\AGeVc
			    \item[\tiny\textcolor{color19Comparison}{\markerNineteenComparison}] 19\AGeVc
            \end{itemize}
       \end{minipage}
	\caption{The ratio of transverse mass spectra of $\pi^{-}$ mesons at mid-rapidity: \textit{central}~Ar+Sc to \textit{central}~Be+Be, central Pb+Pb to \textit{central}~Be+Be and inelastic \pp to \textit{central}~Be+Be collisions. Statistical uncertainties are shown as vertical bars and systematic uncertainties are indicated by shaded bands. Data on Pb+Pb, \pp and Be+Be were taken from Refs.~\cite{Afanasiev:2002mx,Alt:2007aa},~\cite{Abgrall:2013pp_pim} and~\cite{Acharya:2020cyb}, respectively.}
\label{fig:mTComparison}
\end{figure}

In order to compare the detailed features of the spectra, the ratios of the scaled differential yields from \pp, Ar+Sc and Pb+Pb reactions to isospin symmetric Be+Be reference are plotted in Fig.~\ref{fig:mTComparison}. From the three panels of the figure one may conclude that compared to inelastic \pp collisions nucleus+nucleus interactions show a slightly concave behavior and a significant enhancement at small values of $m_\text{T}-m_{\pi}$. One clearly observes a hardening of the spectra at high values of $m_\text{T}-m_{\pi}$ and an increased peak at low $m_\text{T}-m_{\pi}$ for Ar+Sc and Pb+Pb collisions, most likely due to radial expansion flow and decays of strongly decaying resonance states, respectively.

\clearpage

\subsection{Rapidity distributions and mean multiplicities}

The \NASixtyOne experimental apparatus is characterized by large, but limited acceptance. In order to compute the rapidity distribution \dndyLong~and mean multiplicity, one needs to extrapolate the measured data to unmeasured regions.

\begin{figure*}[!htbp]
	\centering
	\includegraphics[width=0.39\textwidth]{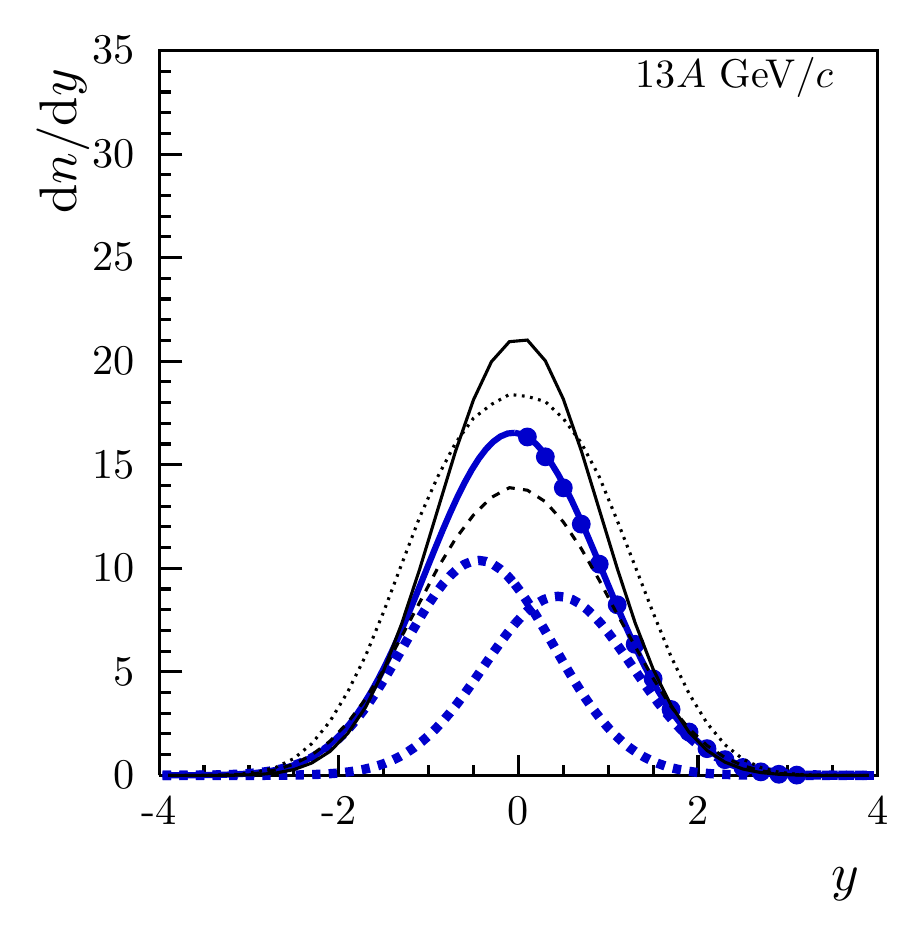}
	\includegraphics[width=0.39\textwidth]{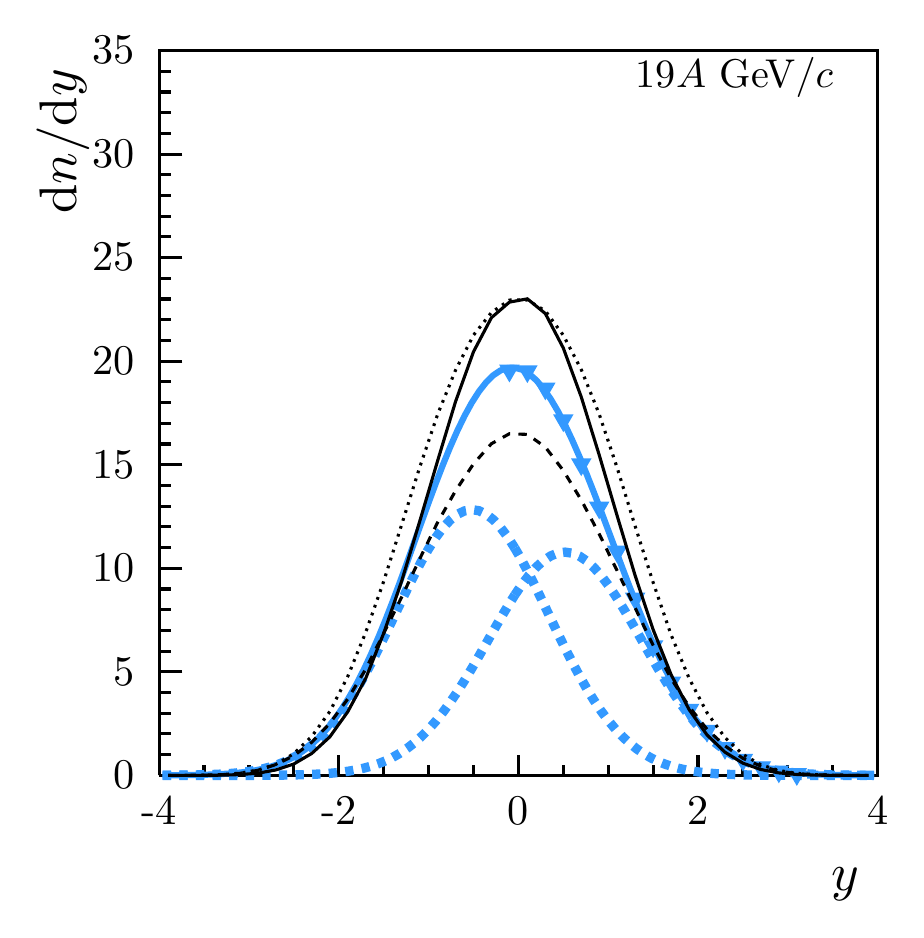}\\
	\includegraphics[width=0.39\textwidth]{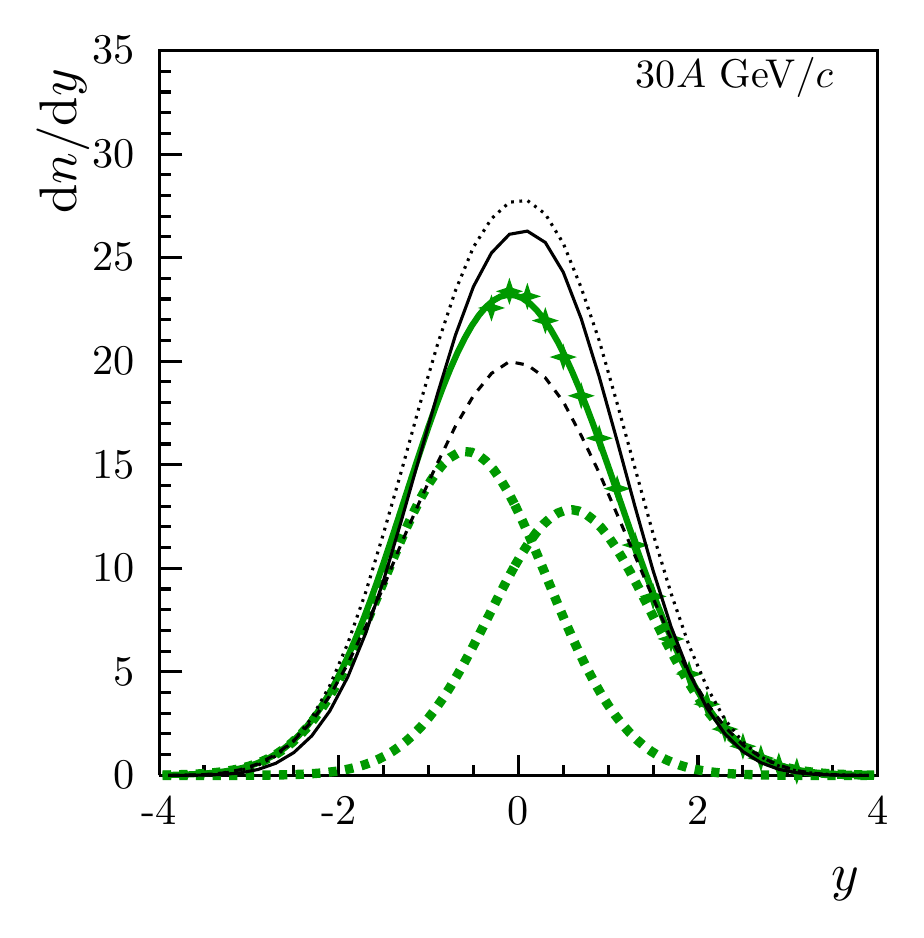}
	\includegraphics[width=0.39\textwidth]{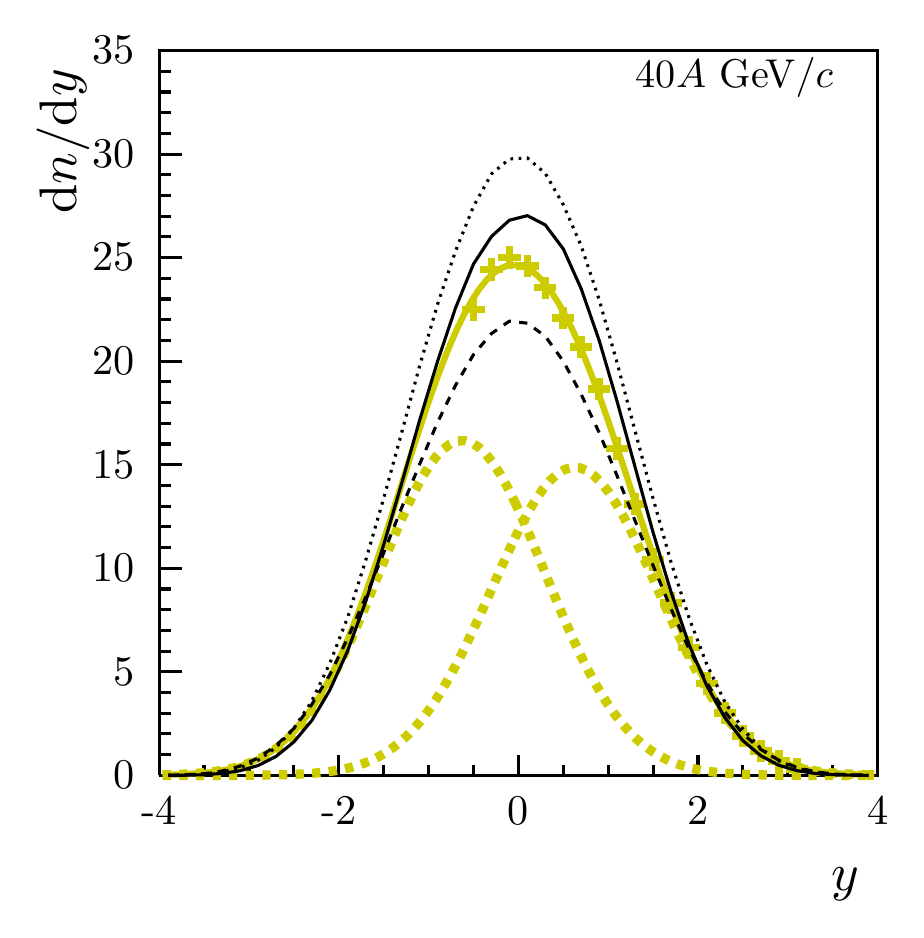}\\
	\includegraphics[width=0.39\textwidth]{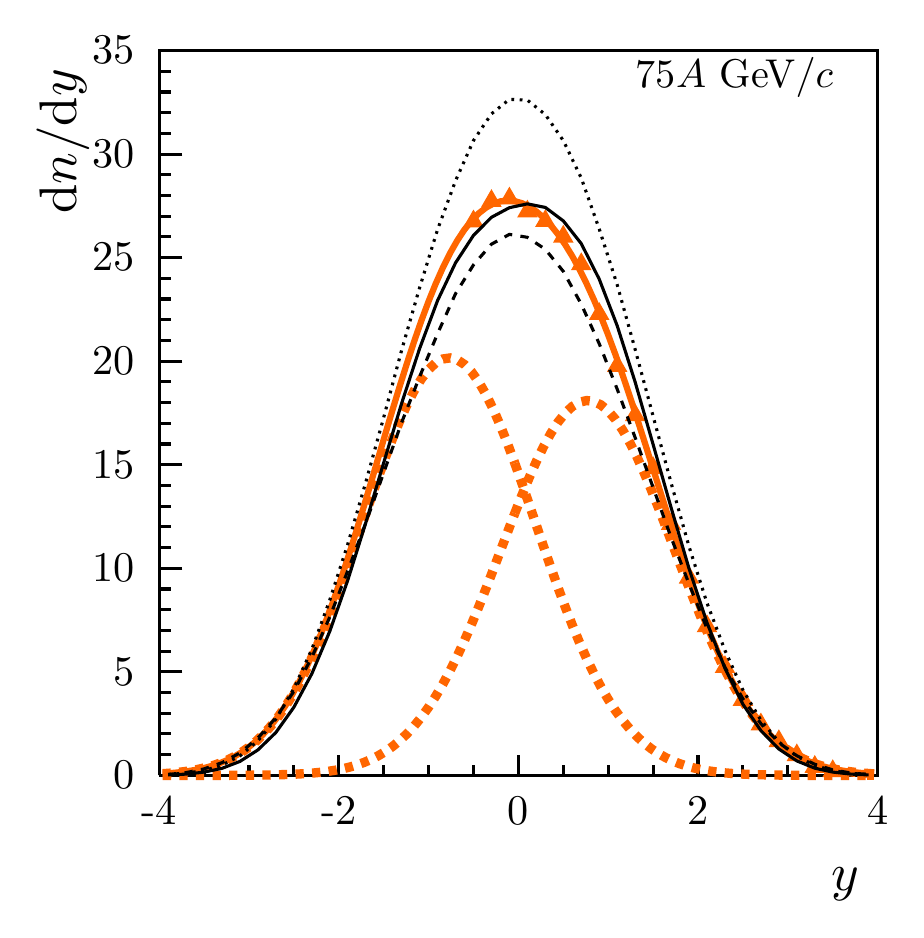}
	\includegraphics[width=0.39\textwidth]{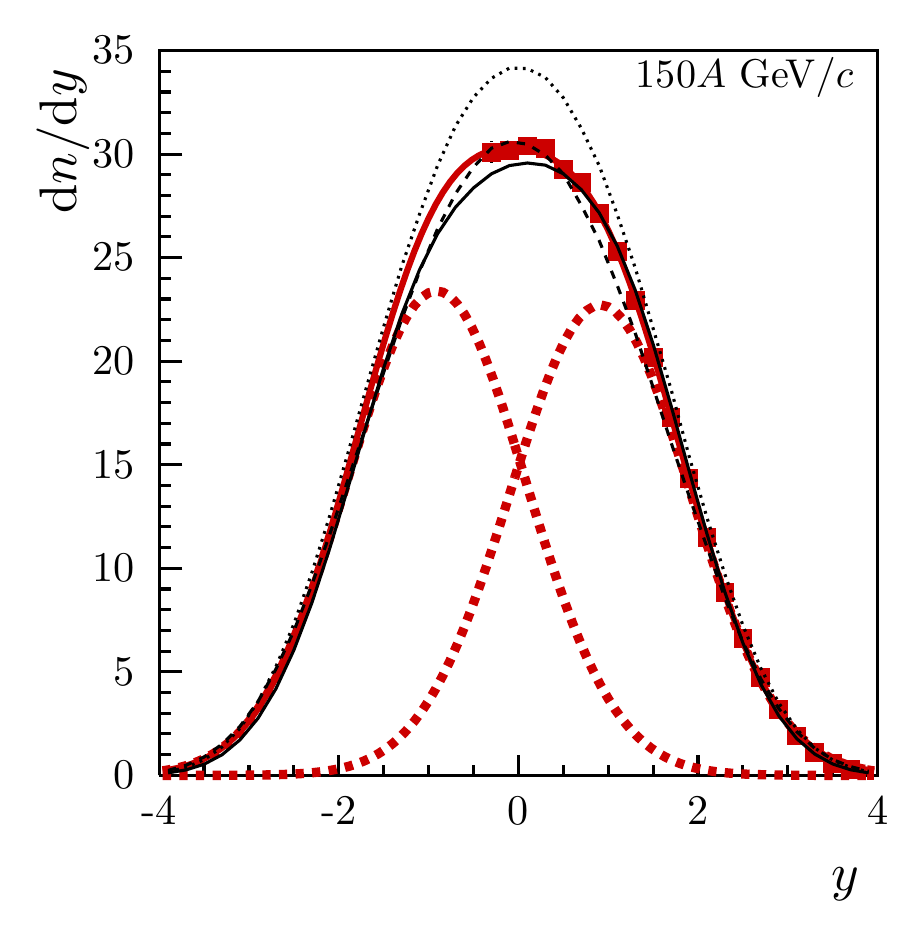}\\
	\textcolor{kBlack}{\solidLine} \Epos~\cite{Werner:2005jf, Pierog:2009zt, Pierog:2018}\hspace{1cm} \textcolor{kBlack}{\dashedLine} \Urqmd\cite{Bass:1998ca,Bleicher:1999xi}\hspace{1cm} \textcolor{kBlack}{\dottedLine} \Hijing~\cite{Hijing:1991}
	\caption{Rapidity distributions \dndyLong~for all six beam momenta obtained by $p_\text{T}$ integration. Points included in the fit of the distribution are shown by full markers. The solid coloured curve is the result of  a fit to the data points using two Gaussians which are indicated by the dashed coloured curves. Predictions of \Epos, \Urqmd and \Hijing models are shown by black curves. All uncertainties are smaller than the symbol size.}
	\label{fig:extrapolation}
\end{figure*}

\begin{figure}[!htbp]
  \centering
        \begin{minipage}[m]{0.7\textwidth}
        \includegraphics[width=0.98\textwidth]{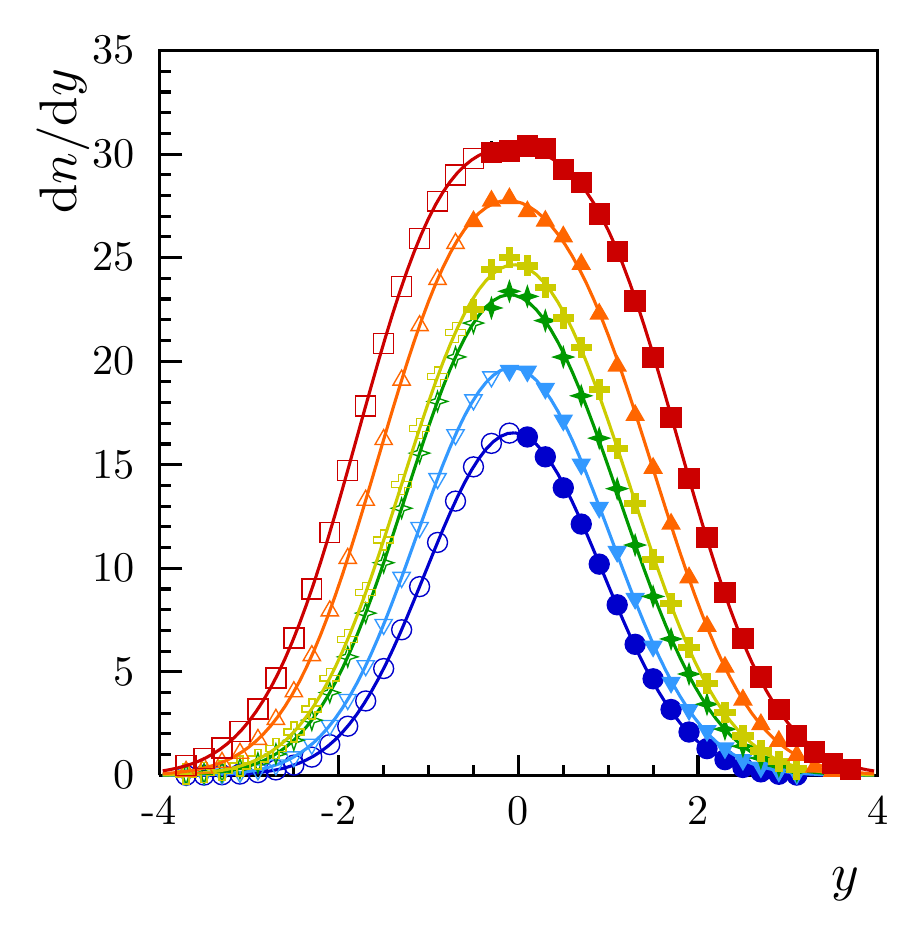}
        \end{minipage}
        \begin{minipage}[m]{0.25\textwidth}
                \normalsize
          \begin{itemize}
	    \item[\tiny\textcolor{color150}{\markerOneFifty}] 150\AGeVc
	    \item[\tiny\textcolor{color75}{\markerSeventyFife}] 75\AGeVc
	    \item[\tiny\textcolor{color40}{\markerFourty}] 40\AGeVc
	    \item[\tiny\textcolor{color30}{\markerThirty}] 30\AGeVc
	    \item[\tiny\textcolor{color19}{\markerNineteen}] 19\AGeVc
	    \item[\tiny\textcolor{color13}{\markerThirteen}] 13\AGeVc
                \end{itemize}
        \end{minipage}
  \caption{Rapidity distributions \dndyLong~for all six beam momenta. Measured points are shown by full markers, values extrapolated by the fit function (see text) are plotted by open markers. All uncertainties are smaller than the symbol size.}
  \label{fig:rapDist}
\end{figure}

First, the \pt~distributions in each rapidity bin were extrapolated from the edge of acceptance to \mbox{\pt = 2~\GeVc}, using the exponential form
\begin{equation}          
  f(p_\text{T})=C \cdot p_\text{T} \cdot \exp \left( \frac{-\sqrt{(c p_\text{T})^2+m_{\pi}^2}}{T}\right)   ,
\end{equation}
where $C$ and $T$ are fit parameters. To obtain \dndyLong, the measured \pt~data bins  are summed and the integral of the extrapolated curve is added:
\begin{equation}
  \frac{\text{d}n}{\text{d}y}=\sum_0^{p_\text{T}^\text{max}}\text{d}p_\text{T}\left(\frac{\text{d}^{2}n}{\text{d}y\text{d}p_\text{T}}\right)_\text{measured}+
  \int_{p_\text{T}^\text{max}}^2f(p_\text{T})\text{d}p_\text{T}   .
\end{equation}
The results are shown by the solid data points in Fig.~\ref{fig:extrapolation} together with \Epos, \Urqmd and \Hijing model predictions. Figure~\ref{fig:rapDist} shows the results for all beam momenta combined into one plot.

In a second step the rapidity spectra are extrapolated to the missing rapidity acceptance. The event trigger requires small energy emitted into the projectile spectator region but puts no constraints on the target spectators. This together with the different number of nucleons in the Ar and Sc nucleus might cause some forward-backward asymmetry of the rapidity distribution. Therefore, the sum of two symmetrically displaced Gaussians -- related to projectile and target contributions -- was fitted to the distributions:
\begin{equation}
\label{eq:asymm}
 g(y)=\frac{A_\text{0}A_\text{rel}}{\sigma_{0}\sqrt{2\pi}}\exp\left(-\frac{(y-y_0)^2}{2\sigma_{0}^2}\right)
      +\frac{A_0}{\sigma_{0}\sqrt{2\pi}}\exp\left(-\frac{(y+y_0)^2}{2\sigma_{0}^2}\right)~,
\end{equation}

where $A_0$ and $A_\text{rel}$ are absolute and relative amplitudes, $y_0$ is the displacement from center-of-mass rapidity and $\sigma_{0}$ is the common width. The fitted two Gaussians are plotted as dashed colored curves in Fig.~\ref{fig:extrapolation}. The figure shows that the asymmetry between the amplitudes of the two Gaussians increases with decreasing beam momentum from 0.97 to 0.84 between 150$A$ and 19$A$~\GeVc beam momentum. This contrasts with the behavior observed for Be+Be collisions~\cite{Acharya:2020cyb}. At 13$A$~\GeVc there is no acceptance for $y < 0$ and the value 0.83 had to be determined from extrapolation of results at higher energies. The resulting values of the fitted parameters are presented in Table~\ref{tab:gauss_params}. The r.m.s. widths of the rapidity distributions $\sigma_{\text{d}n/\text{d}y}$ differ by less than 5\% from the widths approximated by a single Gaussian function. 

\begin{table*}[!htbp]
 \caption{Parameters $A_\text{rel}$, $y_0$ of the double Gaussian fit and r.m.s. width $\sigma_{\text{d}n/\text{d}y}$ of the rapidity distribution together with their statistical and systematic uncertainties.}
 \vspace{0.5cm}
 \centering
 \footnotesize
 \begin{tabular}{l|cccccc}
  Momentum (\AGeVc) & 13 & 19 & 30 & 40 & 75 & 150\\
    \hline
    $A_\text{rel}$ & 0.833 & 0.840 & 0.820 & 0.920 & 0.898 & 0.971\\
	$\sigma_\text{stat}(A_\text{rel})$ & 0.0024 & 0.0110 & 0.0046 & 0.0038 & 0.0022 & 0.0052\\
	$\sigma_\text{sys}(A_\text{rel})$ & 0.2062 & 0.2275 & 0.1632 & 0.1284 & 0.1575 & 0.1988\\
    \hline
	$y_0$ & 0.441 & 0.518 & 0.583 & 0.627 & 0.772 & 0.913\\
	$\sigma_\text{stat}(y_0)$ & 0.0293 & 0.0079 & 0.0038 & 0.0030 & 0.0014 & 0.0020\\
	$\sigma_\text{sys}(y_0)$ & 0.3402 & 0.2587 & 0.2415 & 0.2201 & 0.2737 & 0.3356\\
    \hline
	$\sigma_{dn/dy}$ & 0.940 & 0.999 & 1.077 & 1.114 & 1.232 & 1.351\\
	$\sigma_\text{stat}(\sigma_{\text{d}n/\text{d}y})$ & 0.0156 & 0.0061 & 0.0033 & 0.0028 & 0.0016 & 0.0024\\
	$\sigma_\text{sys}(\sigma_{\text{d}n/\text{d}y})$ & 0.0891 & 0.0824 & 0.0868 & 0.0944 & 0.1026 & 0.1350\\
 \end{tabular}
 \label{tab:gauss_params}
\end{table*}

The total mean $\pi^-$ multiplicity was calculated using the formula:
\begin{equation}
 \langle \pi^- \rangle = \int_{-4.0}^{y_{\text{min}}}g(y)\text{d}y + 
      \sum_{y_{\text{min}}}^{y_{\text{max}}} \text{d}y\left(\frac{\text{d}n}{\text{d}y}\right)
      +\int_{y_{\text{max}}}^{4.0} g(y)\text{d}y   ,
\end{equation}
where $y_{\text{min}}$ to $y_{\text{max}}$ is the interval of measurements for \dndyLong. The results are presented in Table~\ref{tab:piMultiplicity}. Statistical uncertainties $\sigma_{\text{stat}}(\langle \pi^{-} \rangle)$ were obtained by propagating the statistical uncertainties of the $\frac{\text{d}^{2}n}{\text{d}y\text{d}p_{\text{T}}}$ spectra.

\begin{table*}[!htbp]
 \caption{Mean $\pi^-$ multiplicities in the 5\% most \textit{central} Ar+Sc collisions with statistical and systematic uncertainties.}
 \vspace{0.5cm}
 \centering
 \footnotesize
 \begin{tabular}{l|cccccc}
  Momentum (\AGeVc) & 13 & 19 & 30 & 40 & 75 & 150\\
  \hline
  $\langle\pi^-\rangle$ & 39.6 & 50.6 & 64.6 & 71.7 & 92.0 & 114.9\\
  $\sigma_{\text{stat}}(\langle \pi^{-} \rangle)$ & 0.041 & 0.222 & 0.121 & 0.100 & 0.084 & 0.216\\
  $\sigma_{\text{sys}}(\langle \pi^{-} \rangle)$ & 7.3 & 4.6 & 5.3 & 5.7 & 6.7 & 12.8\\
 \end{tabular}
 \label{tab:piMultiplicity}
\end{table*}

The systematic uncertainty connected with the extrapolation procedure was estimated varying the parametrization of the rapidity distribution. In particular, widths and the positions of projectile and target Gaussians were assumed to be independent, i.e. separate parameters $y_{0,\text{proj}}$, $y_{0,\text{targ}}$ and $\sigma_{0,\text{proj}}$, $\sigma_{0,\text{targ}}$ were fitted. The uncertainty of each fitting parameter and the mean $\pi^{-}$ multiplicity was calculated as a standard deviation from the value calculated for the standard values of all parameters. The definition of $\sigma_{\text{d}n/\text{d}y}$ was generalized to take into account that the width and the shift of the projectile and target Gaussians can be different and is given by the formula:

\begin{equation*}
    \sigma_{\text{d}n/\text{d}y} = \sqrt{\left(\frac{y_{0,\text{proj}}+y_{0,\text{targ}}}{2}\right)^2+\left(\frac{\sigma_{0,\text{proj}}+\sigma_{0,\text{targ}}}{2}\right)^2}~.
\end{equation*}

\begin{figure}[!htbp]
\centering
\begin{minipage}[m]{\textwidth}
	\includegraphics[width=0.45\textwidth]{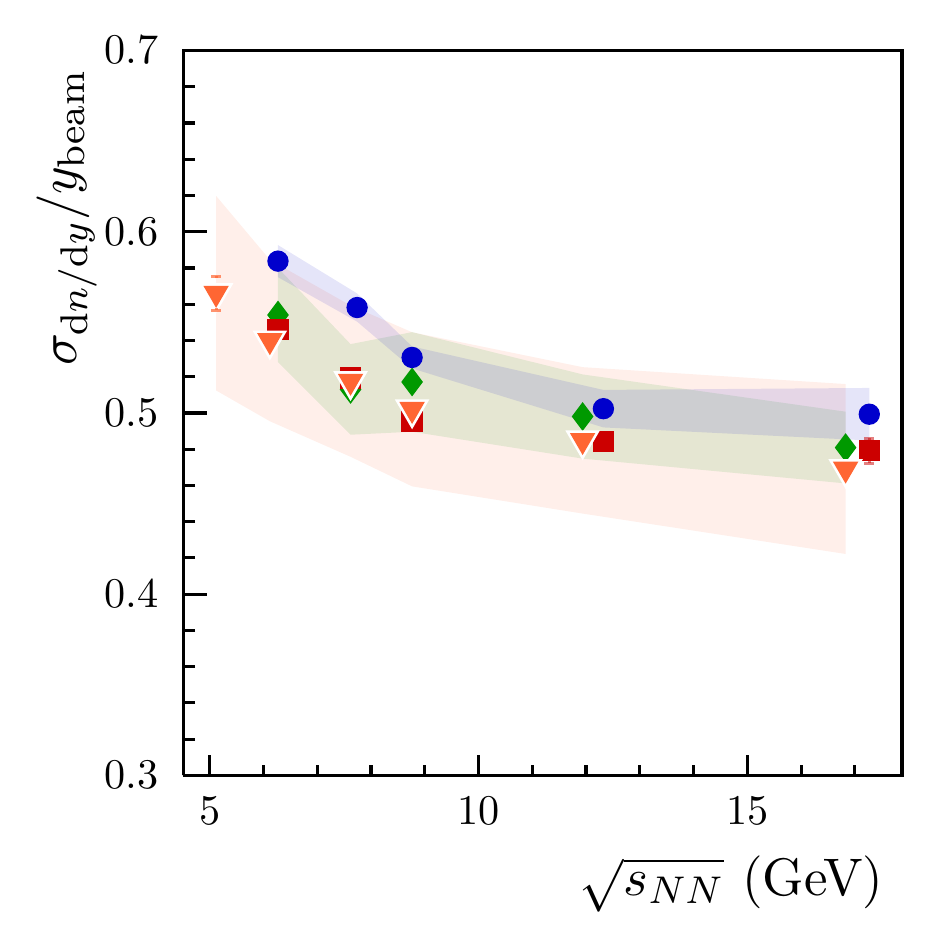}
	\includegraphics[width=0.45\textwidth]{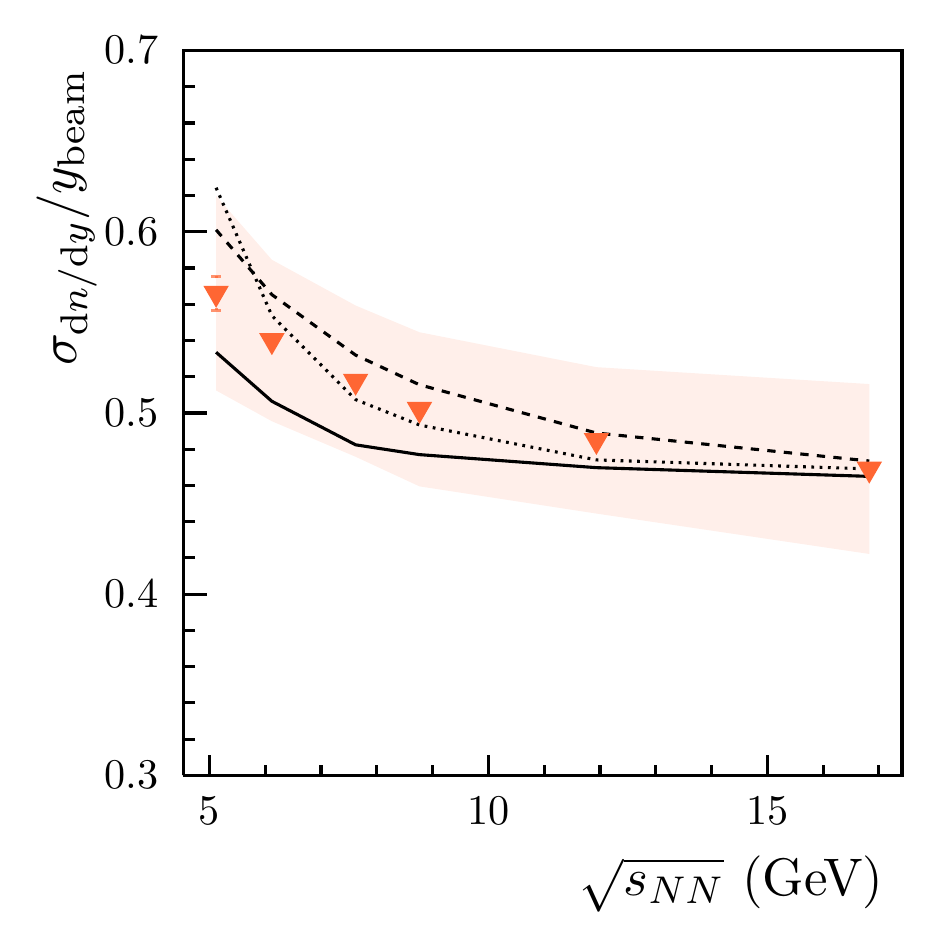}
\end{minipage}
\begin{minipage}[m]{0.45\textwidth}
\begin{center}
\begin{varwidth}{\textwidth}
	\centering
	\normalsize
	\begin{itemize}
		\item[\footnotesize\textcolor{colorArSc}{\markerArSc}] Ar+Sc
		\item[\scriptsize\textcolor{colorBeBe}{\markerBeBe}] Be+Be~\cite{Acharya:2020cyb}
		\item[\footnotesize\textcolor{colorPP}{\markerPP}] \NN~\cite{Abgrall:2013pp_pim, Aduszkiewicz:2017sei}
		\item[\footnotesize\textcolor{colorPbPb}{\markerPbPb}] Pb+Pb (NA49~\cite{Afanasiev:2002mx,Alt:2007aa})
	\end{itemize}
\end{varwidth}
\end{center}
\end{minipage}
\begin{minipage}[m]{0.45\textwidth}
\begin{center}
\begin{varwidth}{\textwidth}
	\normalsize
	\begin{itemize}
		\item[\footnotesize\textcolor{colorArSc}{\markerArSc}] Ar+Sc
		\item[\huge\textcolor{kBlack}{\dashedLine}] Ar+Sc (\Urqmd~\cite{Bass:1998ca,Bleicher:1999xi})
		\item[\huge\textcolor{kBlack}{\dottedLine}] Ar+Sc (\Hijing~\cite{Hijing:1991})
		\item[\huge\textcolor{kBlack}{\solidLine}] Ar+Sc (\Epos~\cite{Werner:2005jf, Pierog:2009zt, Pierog:2018})
	\end{itemize}
\end{varwidth}
\end{center}
\end{minipage}
	\caption{\textit{Left:} the width $\sigma_{\text{d}n/\text{d}y}$ of the rapidity distributions of negatively charged pions to the beam rapidity $y_\text{beam}$ in inelastic \NN interactions and in \textit{central} Ar+Sc, Be+Be and central Pb+Pb collisions as a function of the center of mass energy $\sqrt{s_{NN}}$. Ar+Sc, \NN and Be+Be measurements are presented with statistical (vertical bars, often smaller than marker size) and systematic (shaded band) uncertainty, whereas Pb+Pb with statistical uncertainty only. \textit{Right:} Comparison of the results for Ar+Sc collisions as shown in the left plot with \Epos, \Urqmd and \Hijing model calculations (black curves).}
	\label{fig:sigmaVsSNN}
\end{figure}

\begin{figure}[!htbp]
	\centering
	\begin{minipage}[m]{0.55\textwidth}
		\includegraphics[width=0.95\textwidth]{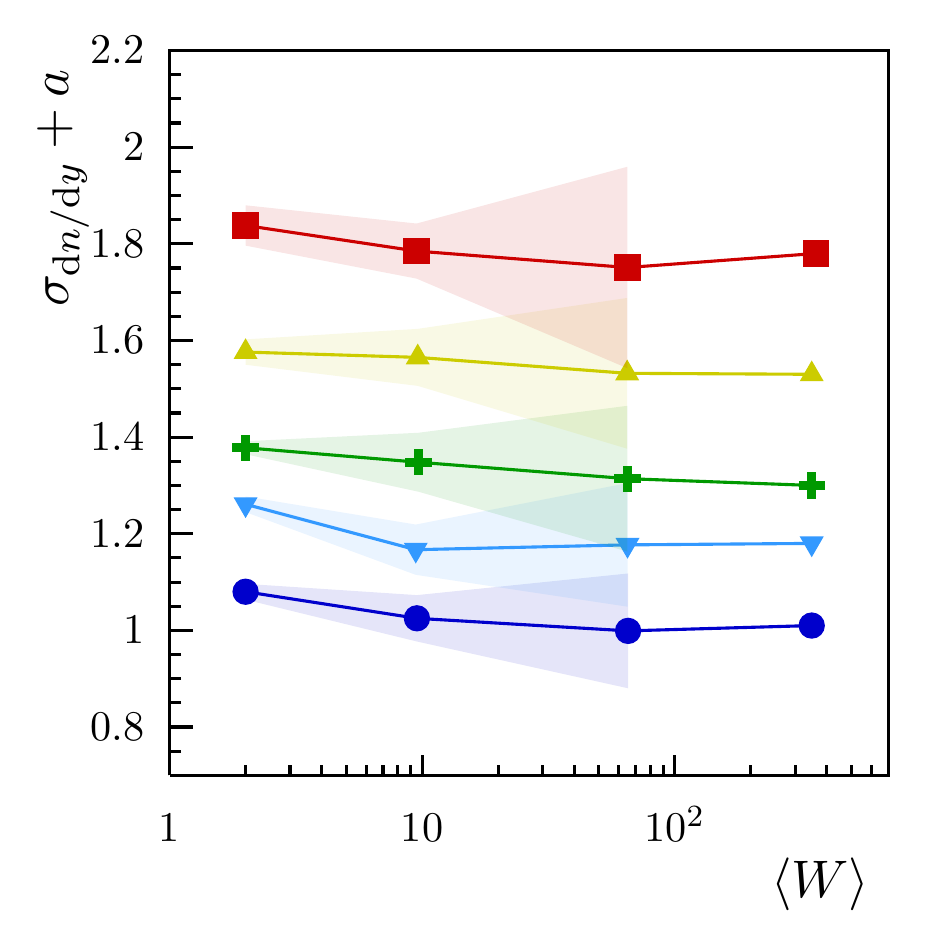}
	\end{minipage}
	\begin{minipage}[m]{0.35\textwidth}
		\normalsize
		\begin{itemize}
			\item[\tiny\textcolor{color150Comparison}{\markerOneFiftyComparison}] 150\AGeVc, $a=0.4$
			\item[\tiny\textcolor{color75Comparison}{\markerSeventyFifeComparison}] 75\AGeVc, $a=0.3$
			\item[\tiny\textcolor{color40Comparison}{\markerFourtyComparison}] 40\AGeVc, $a=0.2$
			\item[\tiny\textcolor{color30Comparison}{\markerThirtyComparison}] 30\AGeVc, $a=0.1$
			\item[\tiny\textcolor{color19Comparison}{\markerNineteenComparison}] 19\AGeVc, $a=0.0$
		\end{itemize}
	\end{minipage}
	\caption{The widths $\sigma_{\text{d}n/\text{d}y}$ of the rapidity distributions of negatively charged pions versus the mean number of wounded nucleons $\langle W\rangle$ for beam momenta from 19$A$ to 150\AGeVc. The data points for different beam momenta were shifted for better readability. Ar+Sc, \NN and Be+Be measurements are presented with statistical uncertainty as vertical bars (often smaller than marker size) and systematic uncertainty as a shaded band. For Pb+Pb statistical uncertainty only was published.}
	\label{fig:sigmaVsW}
\end{figure}

Experimental data on the width $\sigma_{\text{d}n/\text{d}y}$ of the rapidity distributions of $\pi^-$ mesons produced in central nucleus-nucleus collisions and inelastic nucleon-nucleon interactions as function of the collision energy are presented in Fig.~\ref{fig:sigmaVsSNN}. Since the \pp collision system is not isospin symmetric the isospin average $(\pi^{-}+\pi^{+})/2$ was plotted for comparison. These results are referred to as results for nucleon-nucleon (\NN) collisions~\cite{Gazdzicki:1995zs}. One observes that the width normalized to the beam rapidity decreases slowly with increasing collision energy. When correcting the \pp data for the isospin asymmetry one finds a monotonic decrease of $\sigma_{\text{d}n/\text{d}y}$ with decreasing number of nucleons in the colliding nuclei. One also observes that the measured values differ little within the SPS energy range. \Epos, \Urqmd and \Hijing model predictions fit into the measurements' systematic uncertainties band. Figure~\ref{fig:sigmaVsW} shows the width $\sigma_{\text{d}n/\text{d}y}$ of the rapidity distributions plotted versus the mean number of wounded nucleons $\langle W\rangle$. The results seem to be independent of $\langle W\rangle$ from \NN to Pb+Pb collisions for all studied collision energies.

\clearpage

\subsection{Mean multiplicities}

\begin{figure}[!htbp]
\centering
\begin{minipage}[m]{\textwidth}
	\includegraphics[width=0.45\textwidth]{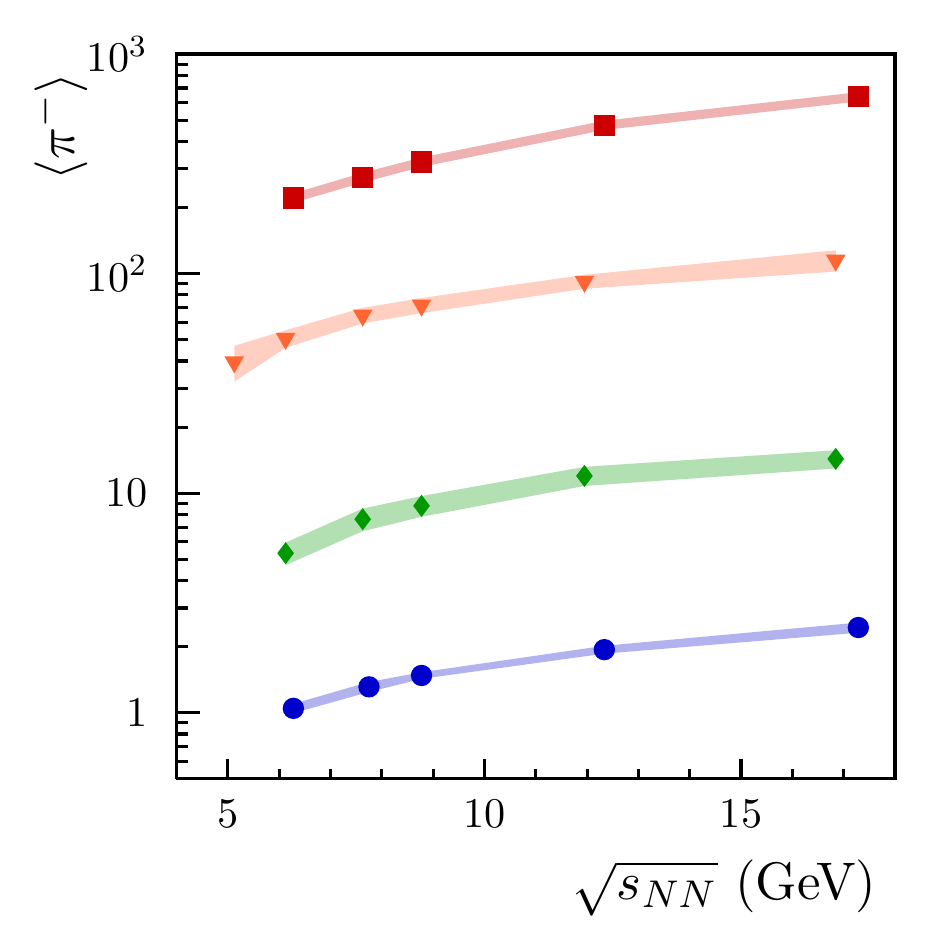}
	\includegraphics[width=0.45\textwidth]{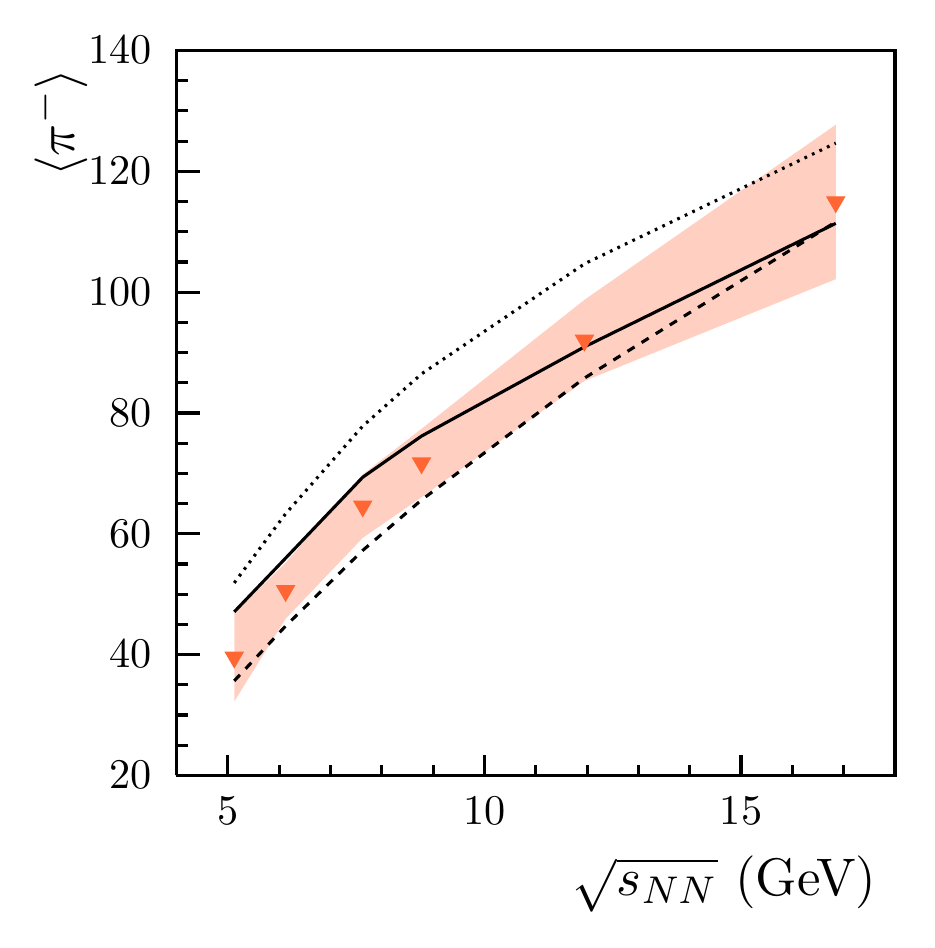}
\end{minipage}
\begin{minipage}[m]{0.45\textwidth}
\begin{center}
\begin{varwidth}{\textwidth}
	\centering
	\normalsize
	\begin{itemize}
		\item[\footnotesize\textcolor{colorArSc}{\markerArSc}] Ar+Sc
		\item[\scriptsize\textcolor{colorBeBe}{\markerBeBe}] Be+Be~\cite{Acharya:2020cyb}
		\item[\footnotesize\textcolor{colorPP}{\markerPP}] \pp~\cite{Abgrall:2013pp_pim}
		\item[\footnotesize\textcolor{colorPbPb}{\markerPbPb}] Pb+Pb (NA49~\cite{Afanasiev:2002mx,Alt:2007aa})
	\end{itemize}
\end{varwidth}
\end{center}
\end{minipage}
\begin{minipage}[m]{0.45\textwidth}
\begin{center}
\begin{varwidth}{\textwidth}
	\normalsize
	\begin{itemize}
		\item[\footnotesize\textcolor{colorArSc}{\markerArSc}] Ar+Sc
		\item[\huge\textcolor{kBlack}{\dottedLine}] Ar+Sc (\Hijing~\cite{Hijing:1991})
		\item[\huge\textcolor{kBlack}{\solidLine}] Ar+Sc (\Epos~\cite{Werner:2005jf, Pierog:2009zt, Pierog:2018})
		\item[\huge\textcolor{kBlack}{\dashedLine}] Ar+Sc (\Urqmd~\cite{Bass:1998ca,Bleicher:1999xi})
	\end{itemize}
\end{varwidth}
\end{center}
\end{minipage}
	\caption{\textit{Left:} The mean multiplicity of negatively charged pions in inelastic \pp interactions and in \textit{central} Ar+Sc, Be+Be and central Pb+Pb collisions as a function of center of mass collision energy. Statistical uncertainties of the data points are smaller than the marker size. The systematic uncertainties are indicated by shaded bands. 
	\textit{Right:} Comparison of the results for Ar+Sc collisions as shown in the left plot with \Epos, \Urqmd and \Hijing model calculations (black curves). }
	\label{fig:PiToWVsSNN}
\end{figure}

The mean multiplicity of $\pi^-$ mesons is plotted versus the center-of-mass energy in Fig.~\ref{fig:PiToWVsSNN} for \pp and \textit{central} Be+Be, Ar+Sc and central Pb+Pb collisions. As shown by the curves, the predictions of the \Epos and \Urqmd models are within the uncertainties of the measurements. \Hijing predictions are systematically higher than
the measurements.

The Ar+Sc system is approximately isospin symmetric. The $\langle\pi^-\rangle/\langle\pi^+\rangle$ ratio calculated within \EposLong was found to change from 0.954 to 0.984 between 13$A$ and 150$A$~\GeVc beam momentum. Based on these results one calculates mean multiplicity of $\pi=\pi^{+}+\pi^{-}+\pi^{0}$, $\langle \pi\rangle$, as:

\begin{equation} 
\langle\pi\rangle_{\text{Ar+Sc}}= 1.5 \cdot (\langle\pi^-\rangle + \langle\pi^+\rangle) =
      1.5 \cdot (1 + c_\text{isospin}) \cdot \langle\pi^-\rangle~,
\end{equation}

where $c_\text{isospin}=\langle\pi^-\rangle/\langle\pi^+\rangle$.

\begin{figure}[!htbp]
	\centering
	\begin{minipage}[m]{0.55\textwidth}
		\includegraphics[width=0.95\textwidth]{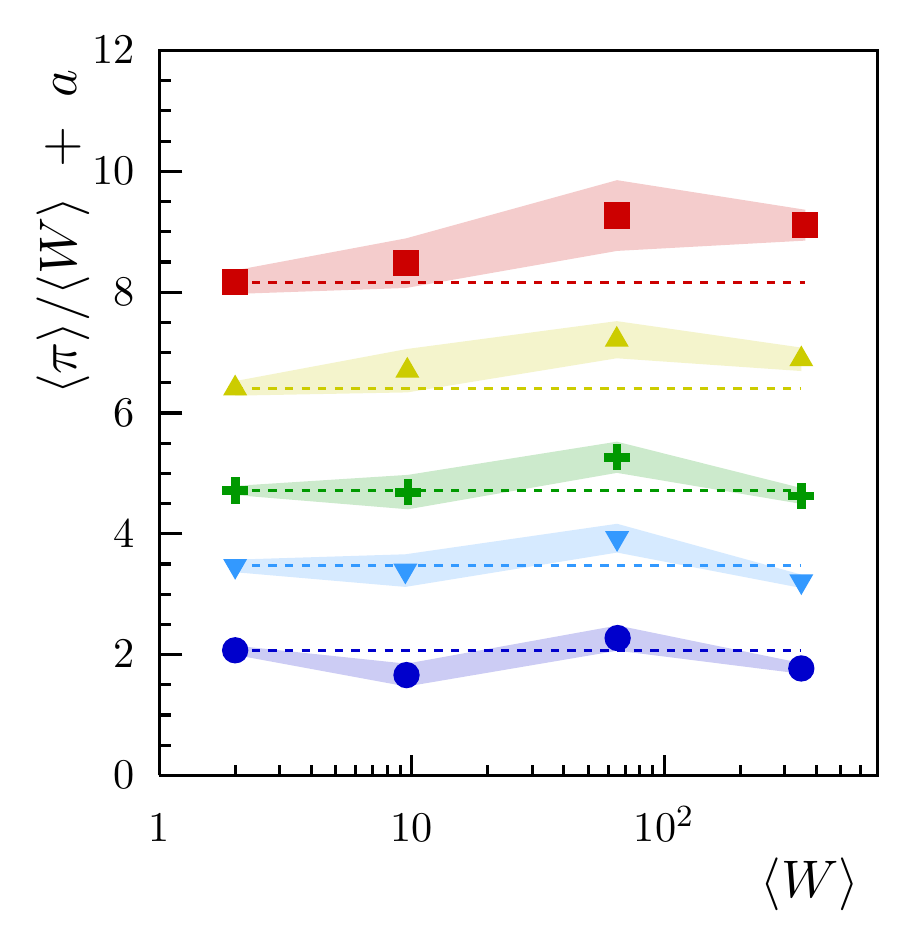}
	\end{minipage}
	\begin{minipage}[m]{0.35\textwidth}
		\normalsize
		\begin{itemize}
			\item[\tiny\textcolor{color150Comparison}{\markerOneFiftyComparison}] 150\AGeVc, $a=4$
			\item[\tiny\textcolor{color75Comparison}{\markerSeventyFifeComparison}] 75\AGeVc, $a=3$,
			\item[\tiny\textcolor{color40Comparison}{\markerFourtyComparison}] 40\AGeVc, $a=2$
			\item[\tiny\textcolor{color30Comparison}{\markerThirtyComparison}] 30\AGeVc, $a=1$
			\item[\tiny\textcolor{color19Comparison}{\markerNineteenComparison}] 19\AGeVc, $a=0$
		\end{itemize}
	\end{minipage}
	\caption{Ratio of the mean pion multiplicity $\langle \pi \rangle$ over the mean number of wounded nucleons $\langle W\rangle$ plotted versus the collision system size for beam momenta from 19$A$ to 150\AGeVc. The data points for different beam momenta were shifted for better readability. Statistical uncertainties are marked with vertical bars and are smaller than marker size. Systematic uncertainties are marked with shaded bands.}
	\label{fig:PiToWVsW}
\end{figure}

Figure~\ref{fig:PiToWVsW} shows the ratios of $\langle \pi \rangle$\footnote{For \pp interactions the figure shows isospin symmetrized values~\cite{Abgrall:2013pp_pim}} over the mean number of wounded nucleons $\langle W\rangle$ plotted versus the collision system size. In general the measurements are close to the expectations of the wounded nucleon model which are shown as horizontal dashed lines. A trend of increase for $\langle W\rangle$ can be seen at higher beam momenta of 75$A$ and 150\AGeVc in Ar+Sc and Pb+Pb reactions. Such an increase is not evident for lower beam momenta.

\section{Relevance of the results to the onset of deconfinement}

The speed of sound in the dense matter produced in the collisions was predicted to show a minimum around
the collision energy of the onset of deconfinement. This paper studies this energy dependence for 
\textit{central} Ar+Sc collisions.

The Landau hydrodynamical model of high energy collisions~\cite{Landau:1953,Belenkij:1956cd} predicts 
rapidity distributions of Gaussian shapes. In fact this prediction is approximately confirmed by the experimental data, see Ref.~\cite{Blume:2005ru} and references therein. Moreover, the collision energy dependence of the width was derived by Shuryak~\cite{Shuryak:1972zq} from the same model under simplifying assumptions and reads:

\begin{equation}
\label{eq:speedOfSound}
\sigma^{2} = \frac{8}{3} \cdot \frac{c_s^2}{1-c^4_s} \cdot \ln {\left(\frac{\sqrt{s_{NN}}}{2 m_p}\right)},
\end{equation}
where $c_s$ denotes the speed of sound, and $c^2_s = 1/3$ for an ideal gas of massless particles.

\begin{figure*}[!htbp]
    \centering
    \begin{minipage}[m]{0.6\textwidth}
        \includegraphics[width=\textwidth]{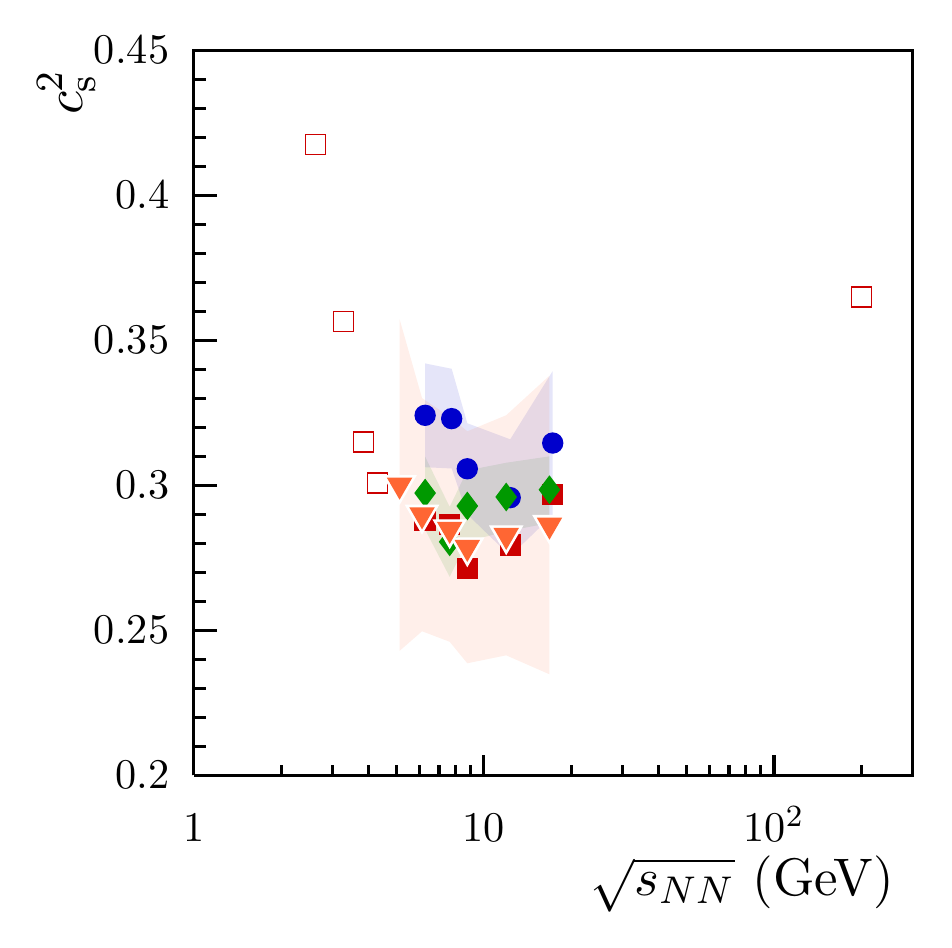}
    \end{minipage}
    \begin{minipage}{0.35\textwidth}
        \normalsize
        \begin{itemize}
            \item[\scriptsize\textcolor{colorArSc}{\markerArSc}] Ar+Sc
            \item[\scriptsize\textcolor{colorBeBe}{\markerBeBe}] Be+Be~\cite{Acharya:2020cyb}
            \item[\scriptsize\textcolor{colorNN}{\markerNN}] \NN~\cite{Abgrall:2013pp_pim}
            \item[\scriptsize\textcolor{colorPbPb}{\markerPbPb}] Pb+Pb~\cite{Afanasiev:2002mx,Alt:2007aa}
            \item[\scriptsize\textcolor{colorAuAu}{\markerAuAu}] Au+Au~\cite{Klay:2003zf, Bearden:2004yx}
        \end{itemize}
    \end{minipage}
 \caption{The speed of sound as a function of beam energy as extracted from the data using Eq.~\ref{eq:speedOfSound}. Statistical uncertainties are marked with a vertical bar (usually smaller than the bin size) and the systematic uncertainties as a shaded area. Only statistical uncertainties were available for Pb+Pb and for Au+Au measurements systematic uncertainties were negligible.}
 \label{fig:sigmaDnDy}
\end{figure*}

By inverting Eq.~\ref{eq:speedOfSound} one can express $c^2_\text{s}$ in the medium as a function of the measured width of the rapidity distribution. The sound velocities extracted from the data on central Pb+Pb collisions, in combination with results from AGS and RHIC on central Au+Au collisions, cover a wide energy range. Here, the sound velocity exhibits a clear minimum~\cite{Bleicher:2005tb,Petersen:2006mp} (usually called the softest point) at $\sqrt{s_{NN}} \approx 10$~GeV consistent with the reported onset of deconfinement~\cite{Afanasiev:2002mx,Alt:2007aa}. The energy dependence of the sound velocities extracted from the new measurement are presented in Fig.~\ref{fig:sigmaDnDy}. The energy range covered by \NASixtyOne for results from \textit{central} Ar+Sc, Be+Be collisions and inelastic \NN reactions is too limited to allow a significant conclusion about a possible minimum.

Pions are the most copiously produced hadrons ($\approx90\%$) in collisions of nucleons and nuclei at SPS energies. Their multiplicity is closely related to the entropy produced in such interactions~\cite{Landau:1953,vHove:1982}. Since the number of degrees of freedom is higher for the quark-gluon plasma than for confined matter, it is expected that the entropy density of the produced final state at given temperature should also be higher in the first case. Therefore, the entropy and information regarding the state of matter formed in the early stage of a collision should be reflected in the number of produced pions normalized to the volume of the system. This intuitive argument was quantified in the Statistical Model of the Early Stage (SMES)~\cite{Gazdzicki:2010iv}. The increase with collision energy of the mean number of produced pions $\langle \pi \rangle$, normalized to the number of wounded nucleons $\langle W \rangle$~\cite{Bialas:1976ed} is expected to be linear when plotted against the Fermi energy measure
\begin{equation}
	F=\left[\left(\sqrt{s_{NN}}-2m_{N}\right)^3/\sqrt{s_{NN}}\right]^{1/4}   ,
\end{equation}
where $\sqrt{s_{NN}}$ is center-of-mass collision energy. The rate of increase depends on  the number of degrees of freedom in the system, $g$, as $g^{1/4}$.

\begin{figure}[!htbp]
    \centering
    \begin{minipage}{0.75\textwidth}
        \centering
        \includegraphics[width=0.98\textwidth]{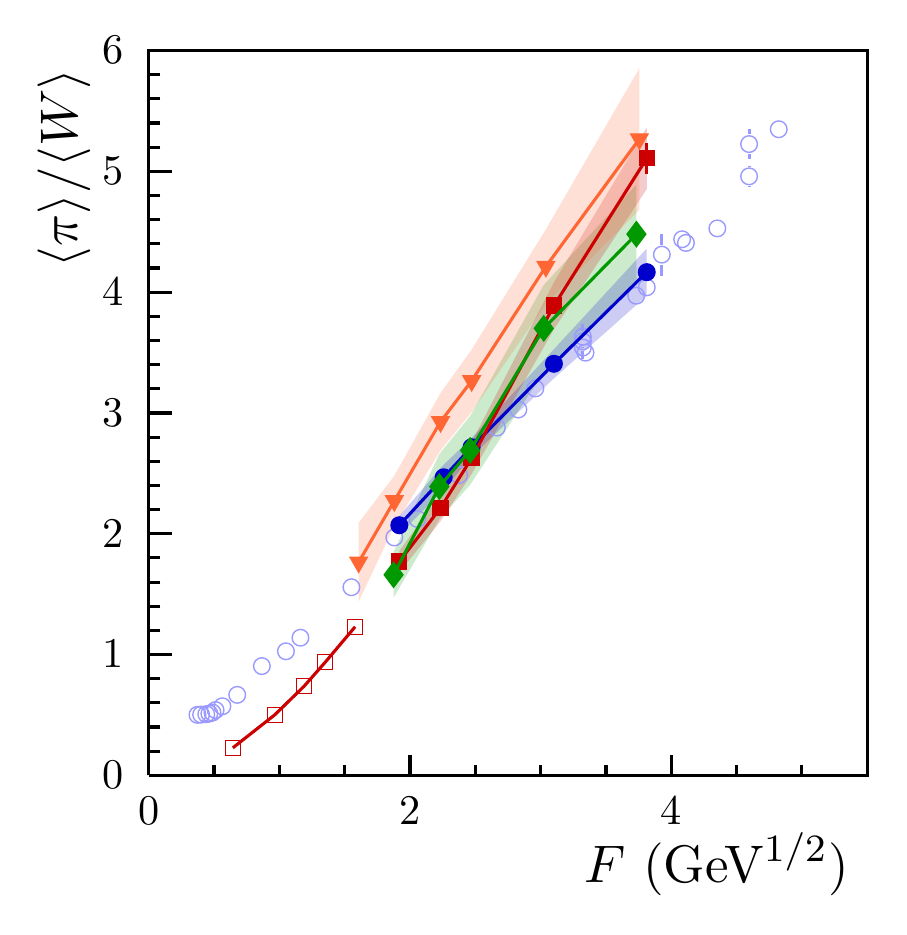}
    \end{minipage}
    \begin{minipage}{0.22\textwidth}
        \normalsize \NASixtyOne
        \begin{itemize}
            \item[\scriptsize\textcolor{colorArSc}{\markerArSc}] Ar+Sc
            \item[\scriptsize\textcolor{colorBeBe}{\markerBeBe}] Be+Be~\cite{Acharya:2020cyb}
            \item[\scriptsize\textcolor{colorNN}{\markerNN}] \NN~\cite{Abgrall:2013pp_pim}
        \end{itemize}
        NA49
        \begin{itemize}
            \item[\scriptsize\textcolor{colorPbPb}{\markerPbPb}] Pb+Pb~\cite{Afanasiev:2002mx,Alt:2007aa}
        \end{itemize}
        AGS
        \begin{itemize}
            \item[\scriptsize\textcolor{colorAuAu}{\markerAuAu}] Au+Au~\cite{Klay:2003zf}
        \end{itemize}
        WORLD
        \begin{itemize}
            \item[\scriptsize\textcolor{colorNNWorld}{\markerNNWorld}] \NN~\cite{Afanasiev:2002mx}
        \end{itemize}
    \end{minipage}
    \caption{The "kink" plot showing the ratio of pion multiplicity $\langle\pi\rangle$ to number of wounded nucleons $\langle W\rangle$ versus the Fermi energy variable $F \approx \sqrt[\uproot{2}4]{s_{NN}}$. Published results for inelastic nucleon-nucleon reactions and central nucleus-nucleus collisions are compared.}
    \label{fig:kinkNew}
\end{figure}

The new \NASixtyOne results are presented in Fig.\ref{tab:piMultiplicity}. These, together with available measurements from other experiments are presented in Fig.~\ref{fig:kinkNew}\footnote{For \pp interactions the figure shows isospin symmetrized values denoted as \NN~\cite{Abgrall:2013pp_pim}}. The uncertainty connected with the calculation of the number of wounded nucleons $\langle W\rangle$ was studied using different MC models. This indicated a variation of $\langle \pi\rangle/\langle W\rangle$ of up to 6\%. This source of uncertainty was not included in the systematic uncertainties plotted in Fig.~\ref{fig:kinkNew}. The world data on \NN and central Pb+Pb (Au+Au) collisions established a well-known picture -- the ''kink'' plot. The results on \NN interactions increase linearly with $F$, whereas the slope of the Pb+Pb results increase by about 30\% in the low SPS beam energy range (at $\approx$ 30\AGeV). The suppression of pion yield in Pb+Pb collisions at low collision energies was attributed to pion absorption in the evolving fireball~\cite{Stock:1986,Roehr:1994}. The increase of the ratio $\langle\pi\rangle/\langle W\rangle$ can be related to activation of additional quark-gluon degrees of freedom.

The NA61/SHINE results on \NN interactions agree well with the world data. The results on Be+Be collisions are mostly between measurements from \NN and Pb+Pb collisions. The new data on Ar+Sc collisions seem to be systematically higher than the results for \NN, Be+Be and Pb+Pb collisions at the lower energies. They are close to the Pb+Pb results at the highest energies. There appears to be a systematic steepening of the rate of increase of this ratio with energy between light and heavy collision systems. This behavior suggests in the statistical scenario a systematic increase of the effective number of degrees of freedom. 

Interestingly $\langle \pi \rangle / \langle W \rangle$ for Ar+Sc reactions equals that for inelastic \NN reactions at low SPS energies whereas it is closer to that for Pb+Pb reactions at high SPS energies. Moreover, a suppression of the pion yield per wounded nucleon was observed in central Pb+Pb collisions compared to inelastic \NN reactions at low energies which was attributed to pion absorption in the evolving fireball \cite{Stock:1986,Roehr:1994}. This effect is not found for the intermediate size Ar+Sc system in which a smaller less dense fireball is created.

\section{Summary and outlook}

Spectra and mean multiplicities of $\pi^-$ produced in the 5\% most \textit{central} Ar+Sc collisions were measured by the \NASixtyOne experiment at the CERN SPS for beam momenta of 13$A$, 19$A$, 30$A$, 40$A$, 75$A$ and 150\AGeVc using the $h^-$ method. The results represent the first measurements on pion production in an intermediate size collision system at SPS energies.

Energy and system size dependence of parameters of measured distributions -- mean transverse mass, the inverse slope parameter of transverse mass spectra, width of the rapidity distribution and mean multiplicity -- were presented and discussed.

The inverse slope parameter of the transverse mass distribution increases with increasing system size and collision energy. Width of the rapidity distribution is independent of the system size and increases with collision energy. The mean multiplicity at high energies increases faster with the system size than expected from the Wounded Nucleon Model. The rate of the increase with collision energy is faster in Ar+Sc and Pb+Pb collisions than in \NN interactions. The new measurements were compared to predictions of \EposLong~\cite{Werner:2005jf, Pierog:2009zt, Pierog:2018}, \UrqmdLong~\cite{Bass:1998ca,Bleicher:1999xi} and \Hijing~\cite{Hijing:1991} models. None of them provides a consistent description of the new \NASixtyOne measurements in Ar+Sc collisions. 

The new results on \emph{central} Ar+Sc collisions   were discussed in the context of the signatures of the onset of deconfinement. The velocity of sound extracted form the width of
rapidity distribution is consistent with results for central Pb+Pb collisions as well as  Be+Be and \emph{N+N} interactions. Measurements in a broader energy range are needed to conclude on a possible minimum of the sound velocity in Ar+Sc collisions. 
The ratio of mean pion multiplicity to the number of wounded nucleons and its collision energy dependence at the highest SPS energies are close to the ones for central Pb+Pb collisions and higher than the corresponding results for \emph{N+N} and Be+Be interactions.
This behavior suggests an increase of the effective number of degrees of freedom already in
\emph{central} Ar+Sc collisions at the top SPS energies.

\clearpage

\section{Acknowledgments}
We would like to thank the CERN EP, BE, HSE and EN Departments for the
strong support of NA61/SHINE.

This work was supported by
the Hungarian Scientific Research Fund (grant NKFIH 123842\slash123959),
the Polish Ministry of Science
and Higher Education (grants 667\slash N-CERN\slash2010\slash0,
NN\,202\,48\,4339 and NN\,202\,23\,1837), the National Science Centre Poland(grants
2014\slash14\slash E\slash ST2\slash00018,
2014\slash15\slash B\slash ST2 \slash\- 02537 and
2015\slash18\slash M\slash ST2\slash00125,
2015\slash 19\slash N\slash ST2 \slash01689,
2016\slash23\slash B\slash ST2\slash00692,
DIR\slash WK\slash\- 2016\slash 2017\slash\- 10-1,
2017\slash\- 25\slash N\slash\- ST2\slash\- 02575,
2018\slash 30\slash A\slash ST2\slash 00226,
2018\slash 31\slash G\slash ST2\slash 03910,
2019\slash 32\slash T\slash ST2\slash 00432),
the Russian Science Foundation, grant 16-12-10176 and 17-72-20045,
the Russian Academy of Science and the
Russian Foundation for Basic Research (grants 08-02-00018, 09-02-00664
and 12-02-91503-CERN),
the Russian Foundation for Basic Research (RFBR) funding within the research project no. 18-02-40086,
the National Research Nuclear University MEPhI in the framework of the Russian Academic Excellence Project (contract No.\ 02.a03.21.0005, 27.08.2013),
the Ministry of Science and Higher Education of the Russian Federation, Project "Fundamental properties of elementary particles and cosmology" No 0723-2020-0041,
the European Union's Horizon 2020 research and innovation programme under grant agreement No. 871072,
the Ministry of Education, Culture, Sports,
Science and Tech\-no\-lo\-gy, Japan, Grant-in-Aid for Sci\-en\-ti\-fic
Research (grants 18071005, 19034011, 19740162, 20740160 and 20039012),
the German Research Foundation (grant GA\,1480/8-1), the
Bulgarian Nuclear Regulatory Agency and the Joint Institute for
Nuclear Research, Dubna (bilateral contract No. 4799-1-18\slash 20),
Bulgarian National Science Fund (grant DN08/11), Ministry of Education
and Science of the Republic of Serbia (grant OI171002), Swiss
Nationalfonds Foundation (grant 200020\-117913/1), ETH Research Grant
TH-01\,07-3 and the Fermi National Accelerator Laboratory (Fermilab), a U.S. Department of Energy, Office of Science, HEP User Facility managed by Fermi Research Alliance, LLC (FRA), acting under Contract No. DE-AC02-07CH11359 and the IN2P3-CNRS (France).

\bibliographystyle{unsrt}
%\bibliography{bibl.bib}
\bibliography{include/na61References}

\end{document}